\documentclass[fleqn,usenatbib]{mnras}

\usepackage{newtxtext,newtxmath}

\usepackage[T1]{fontenc}

\DeclareRobustCommand{\VAN}[3]{#2}
\let\VANthebibliography\thebibliography
\def\thebibliography{\DeclareRobustCommand{\VAN}[3]{##3}\VANthebibliography}

\usepackage{graphicx}	
\usepackage{amsmath}	
\usepackage{tablefootnote}
\usepackage{longtable}
\usepackage[dvipsnames]{xcolor}
\usepackage{float}

\title[Globular Clusters around Ultra-Diffuse Galaxies]{Implications for Galaxy Formation Models from Observations of Globular Clusters around Ultra-Diffuse Galaxies}
\author[Teymoor Saifollahi et al.]{Teymoor Saifollahi$^{1}$\thanks{E-mail: teymur.saif@gmail.com}, 
Dennis Zaritsky$^{2}$,
Ignacio Trujillo$^{3,4}$
Reynier F. Peletier$^{1}$,
Johan H. Knapen$^{3,4}$,
\newauthor
Nicola Amorisco$^{5}$,
Michael A. Beasley$^{3,4}$, 
Richard Donnerstein$^{2}$
\\\\
$^{1}$Kapteyn Astronomical Institute, University of Groningen, PO Box 800, 9700 AV Groningen, The Netherlands\\
$^{2}$Steward Observatory, University of Arizona, Tucson, AZ 85719, USA\\
$^{3}$Instituto de Astrof\'isica de Canarias, V\'ia L\'actea S/N, E-38205 La Laguna, Spain\\
$^{4}$Departamento de Astrof\'isica, Universidad de La Laguna, E-38206 La Laguna, Spain\\
$^{5}$Institute for Computational Cosmology, Department of Physics, Durham University, South Road, Durham DH1 3LE, UK\\
}

\date{Accepted XXX. Received YYY; in original form ZZZ}

\pubyear{2021}

\begin{document}
\label{firstpage}
\pagerange{\pageref{firstpage}--\pageref{lastpage}}
\maketitle

\begin{abstract}
We present an analysis of {\sl Hubble Space Telescope} observations of globular clusters (GCs) in six ultra-diffuse galaxies (UDGs) in the Coma cluster, a sample that represents UDGs with large effective radii ($R_{\rm e}$), and use the results to evaluate competing formation models. We eliminate two significant sources of systematic uncertainty in the determination of the number of GCs, $N_{\rm GC}$ by using sufficiently deep observations that (i) reach the turnover of the GC luminosity function and (ii) provide a sufficient number of GCs with which to measure the GC number radial distribution. We find that $N_{\rm GC}$ for these galaxies is on average $\sim$\,20, which implies an average total mass, $M_{\rm total}$, $\sim$\,$10^{11}$\,$M_{\odot}$ when applying the relation between $N_{\rm GC}$ and $M_{\rm total}$. This value of $N_{\rm GC}$ lies at the upper end of the range observed for dwarf galaxies of the same stellar mass and is roughly a factor of two larger than the mean. The GC luminosity function, radial profile and average colour are more consistent with those observed for dwarf galaxies than with those observed for the more massive ($L^*$) galaxies, while both the radial and azimuthal GC distributions closely follow those of the stars in the host galaxy. Finally, we discuss why our observations, specifically the GC number and GC distribution around these six UDGs, pose challenges for several of the currently favoured UDG formation models. 

\end{abstract}




\section{Introduction}

The increasing depth of astronomical surveys continues to reveal the existence of ever lower surface brightness galaxies (\citealp{Binggeli,bothun91,impey,vd15,2016MNRAS.456.1359F,2021A&A...654A..40T}), which due to their extreme nature can challenge galaxy formation models. In particular, a subclass of low surface brightness (LSB) galaxies with large effective radii ($R_{\rm e}$\,$>$\,1.5\,kpc) and low surface brightness ($\mu_{(0,g)}$\,$>$\,24.0\,mag arcsec$^{-2}$), dubbed ultra-diffuse galaxies (UDGs) by \citet{vd15}, has drawn attention both because of their large numbers and inferred properties.


The large population of UDGs in Coma and their survival in the cluster environment suggested large total masses, $M_{\rm total}$, despite their modest stellar masses \citep{vd15}. 
Further investigations and estimations of the total mass of UDGs using stellar kinematics (\citealp{vd19,chil,oliver2020,gannon,forbes1}), globular clusters (GCs) dynamics (\citealp{beasley2016,toloba}), H\,{\sc I} kinematics (\citealp{udgsfr1,leisman,manolis,pavel2020,poulain}), weak gravitational lensing (\citealp{sifon}), UDG abundance in clusters (\citealp{amorisco2016+}), scaling relations (\citealp{dennis}), and X-ray observations (\citealp{lee}) found that most UDGs are dark matter dominated objects with halo masses similar to those of dwarf galaxies and dynamical mass-to-light ratios that span a wide range between a few tens to a few thousands. The most massive known UDGs have $M_{\rm total}\,\sim 10^{11}$\,$M_{\odot}$, comparable to the most massive dwarf galaxies, such as the LMC (\citealp{erkal}). 

Because of the difficulty in obtaining direct mass estimates of UDGs, some authors have resorted to using the number of globular clusters, $N_{\rm GC}$, as a secondary mass estimator (\citealp{harris2013,beasley2016,harris2017}) for UDGs in the Coma (\citealp{beasley2016b,peng2016,amorisco2018,lim2018}), Virgo (\citealp{lim2020}), Fornax (\citealp{prole2018}), and Hydra (\citealp{iodice}) clusters and in galaxy groups (\citealp{somalwar,muller21}). These observations have highlighted that UDGs tend to host two to three times as many GCs as dwarf galaxies of the same stellar mass (\citealp{lim2018,lim2020}, but see \citealp{amorisco2018,prole2019,francine} for multiple counter-examples). Overall, it seems that UDGs show a wide range of GC abundances; some are GC$-$poor and some are GC$-$rich (\citealp{gannon2021})

In contrast to GC abundances, other GC properties appear to be similar in UDGs and non-UDGs. The ratio of the GC half-number radius ($R_{\rm GC}$) to the UDG effective radius ($R_{\rm e}$) of $<$\,2, the GC colours (dominantly blue and metal-poor), and the shape of GC luminosity function (GCLF) are all similar to what is found for GCs in dwarf galaxies (\citealp{beasley2016b,peng2016,vd17,amorisco2018,prole2019,lim2020,somalwar,saifollahi2020,muller21})

These results suggest that GCs in UDGs are similar to those in other galaxies and that what differs between GCs in UDGs and other galaxies is simply the relative formation efficiency between them and the stars in the host galaxy. As such, GCs may provide an avenue for discriminating among competing UDG formation/evolution models. We briefly outline some competing models and how $N_{\rm GC}$ might be expected to behave:

\begin{itemize}
\item
The initial UDG studies suggested
that UDGs represent a population of failed galaxies that assembled their dark matter haloes and were rapidly quenched at high redshift due to environmental processes in galaxy clusters, mainly ram pressure stripping (\citealp{vd15,koda,yozin,jose2021}). Additionally, as a consequence of the passive evolution of their stellar populations, UDGs surface brightness have decreased with time (\citealp{javier,javier+red,tremmel}). This scenario is consistent with observations of cluster UDGs as gas-poor (\citealp{karunakaran,chen}) quiescent (\citealp{singh}), old, metal-poor, and alpha-enhanced (\citealp{kadowaski,gu,pandaya,ruizlara2018,anna}). This model stemmed from the idea that UDGs are massive objects with $L^*$-like haloes (\citealp{vd15}). However, recently this model is revised to accommodate the dwarf-like haloes of UDGs (\citealp{beasley2016,beasley2016b}). In this "failed dwarf galaxy" model, the low surface brightness is attributed to the lack of subsequent star formation, and so naturally results in a larger $N_{\rm GC}$/$M_*$, as long as GC formation occurred prior to the introduction of the UDG progenitor into the cluster environment. 
\\
\item
Another environment-motivated class of model, the "tidal interaction" scenario, has UDGs suffering dynamical heating within galaxy clusters and galaxy groups that "puffs" them up \citep[e.g.,][]{carleton19,amorisco19,Sales20}. Studies exist that both favour (\citealp{pavel2019,rong2020_2,jones2021,javier2021}) or disfavour (\citealp{venhola2017,kado}) this scenario. In such a scenario, the galaxy's low surface brightness is attributed to the dynamical expansion of the galaxy, not to a lower star formation efficiency, and so $N_{\rm GC}$/$M_*$ should be comparable to that of dwarf galaxies. If only some UDGs form via this channel, we might expect to find large scatter in $N_{\rm GC}$/$M_*$.
\\
\item
The presence of UDGs in the field (\citep{prole2021}) suggests that environment alone is not responsible for UDG formation. In the "high-spin" model (\citealp{amorisco+loeb}), a higher initial halo spin for a dwarf-like halo produces a galaxy with a large effective radius and hence low surface brightness. Again, current observations, in this case H\,{\sc I} observations of gas-rich UDGs, provide
conflicting results with some finding high-spin \citep{spekkens,liao} and others finding low-spin (\citealp{jones2018,sengupta}) field UDGs. In the high-spin UDGs, the baryons are less concentrated than in the low-spin analogs and so, in addition to the stars being more spread out, a lower star formation efficiency may also plausibly be responsible for the galaxy's low surface brightness. It is unclear whether one would also expect GC formation to have been affected and so the expectations for $N_{GC}/M_*$ are unclear. 
\\
\item
Alternatively, the "stellar feedback" model
\citep{cintio2017} invokes episodic and intense star formation activity in gas-rich galaxies to push the gas to larger radii. In a variant of this model, feedback from the GCs themselves is invoked by \citet{gomez}. The redistribution of the gas, which is the dominant baryonic mass component in these galaxies, modifies the total mass distribution within the galaxy, causing the formation of a dark matter core \citep{navarro96, pontzen12} and lowering the surface brightness (\citealp{cintio2017,jiang,chan,martin,freundlich}). As star formation occurs after GC formation, one might expect different radial distributions for GCs and stars. Interestingly, in the model variant where GCs are responsible for the intense feedback \citep{gomez}, we might expect a relation between $N_{\rm GC}$/$M_*$ and $R_{\rm e}$. 
\\
\item
Other models rely on distinctive merger histories to produce a range of surface brightness among otherwise similar galaxies. The "lack of mergers" scenario (\citealp{wright,vannest}) suggests that
UDGs are those galaxies with no late-time ($z$\,<\,1) major mergers. In this case, the spin parameter of the galaxy is systematically higher than in analogs that did have such a merger and in turn star formation occurs at larger radii, which leads to a lower central surface brightness. If mergers play a role in GC formation, then $N_{\rm GC}$/$M_*$ might actually be expected to be lower than in analogs. However, if GCs form only at high redshift, then $N_{\rm GC}$/$M_*$ might be larger in UDGs.

\end{itemize}

\noindent
Each of these proposed mechanisms do not necessarily act at the exclusion of others. For example, \cite{martin} propose that a combination of a distinctive star formation history and environmental effects produces large, low surface brightness galaxies. Even without this additional complication,
it is evident from this discussion that theoretical expectations for the number and properties of GCs in the various UDG formation/evolution models need further development. Fortunately, some studies have begun this exploration in detail for specific models \citep{carleton2021}. Our goal here is to motivate further work by providing higher fidelity GC observations against which to do the comparison.

We address three specific shortcomings of the available $N_{\rm GC}$ measurements. First, and foremost, we provide a consistent analysis of a significant sample (six) of similar UDGs. Second, we focus on a population of large UDGs, $3 < R_e/{\rm kpc}< 5$, with a mean effective radius of $\langle R_e \rangle = 3.6$\,kpc. These are expected to be more massive and, therefore, to each have a significant number of GCs. Third, we use images that are sufficiently deep to allow us to address two key sources of systematic uncertainty. Every estimate of $N_{\rm GC}$ for UDGs includes completeness corrections, both for GCs too faint to detect and for radial incompleteness due to highly uncertain statistical background corrections at large radii. Our photometry reaches the GCLF turnover, minimizing the photometric completeness correction. Furthermore, we detect a sufficient number of GCs per galaxy, which allows us to empirically constrain the GC radial distribution. 

\begin{figure*}
\centering
\includegraphics[width=0.9\linewidth]{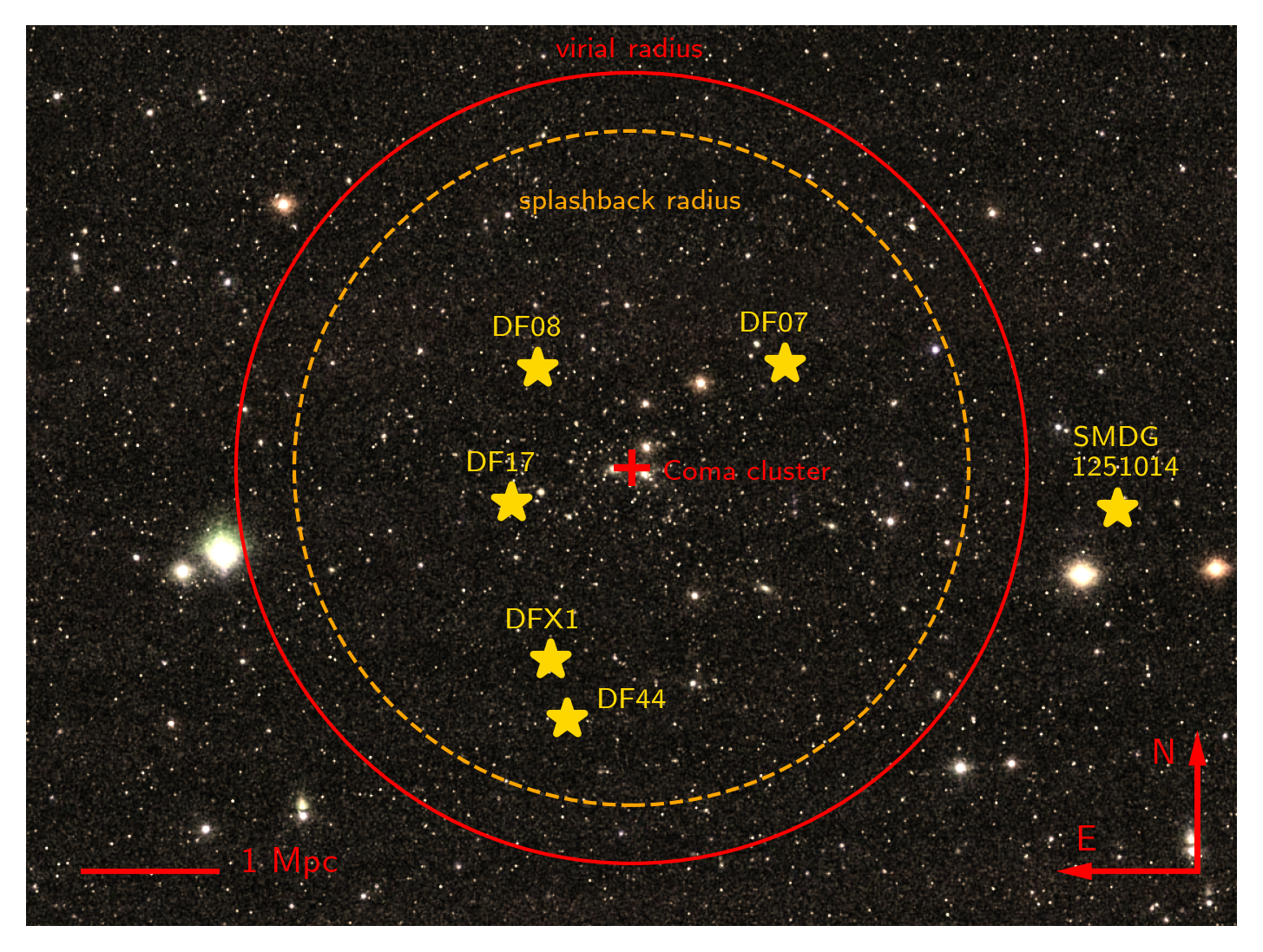}
\caption{Projected distribution of the six Coma cluster UDGs in this study (colour image: Sloan Digital Sky Survey DR9).}
\label{comacluster}
\end{figure*}


The structure of the paper is as follows. In Section~\ref{obs}, the UDG sample and the observational data are described. In Section~\ref{analysis}, after source extraction and photometry, we measure the compactness and colours of sources and identify the GCs around our UDGs. Section~\ref{results} presents the observed properties of the identified GCs, their total number, spatial distribution, colours and luminosity function. Based on these results, in Section~\ref{discussion} we discuss the implications for various UDG formation scenarios. We summarize our findings in Section~\ref{summary}. Throughout this paper, magnitudes and colours are expressed in the AB magnitude system. We adopt the standard cosmological model with $\Omega_{\rm M}=0.3$, $\Omega_{\Lambda}=0.7$ and $H_{0}$=70\,km\,s$^{-1}$\,Mpc$^{-1}$. 

\begin{figure*}
\centering
\includegraphics[width=0.33\linewidth]{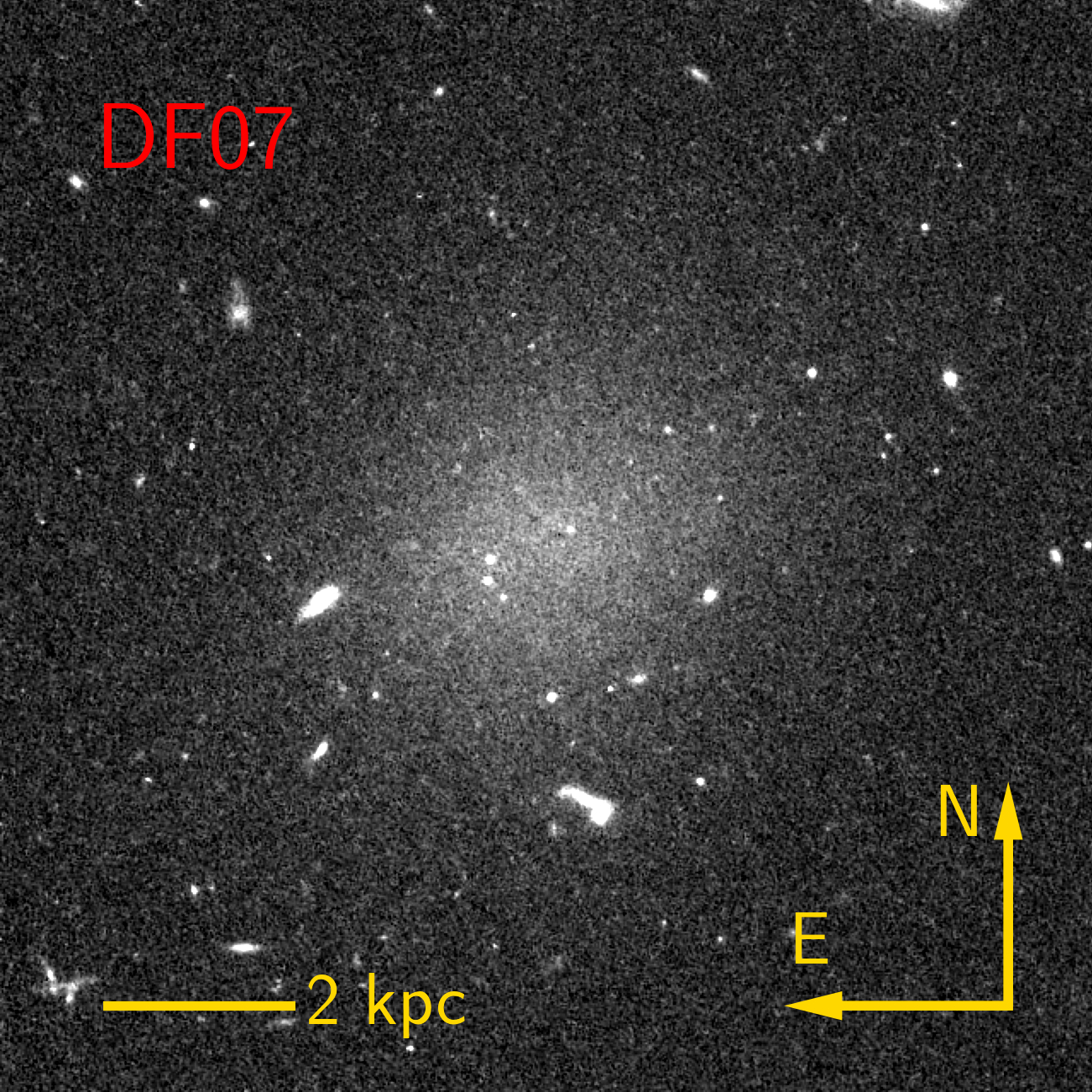}
\includegraphics[width=0.33\linewidth]{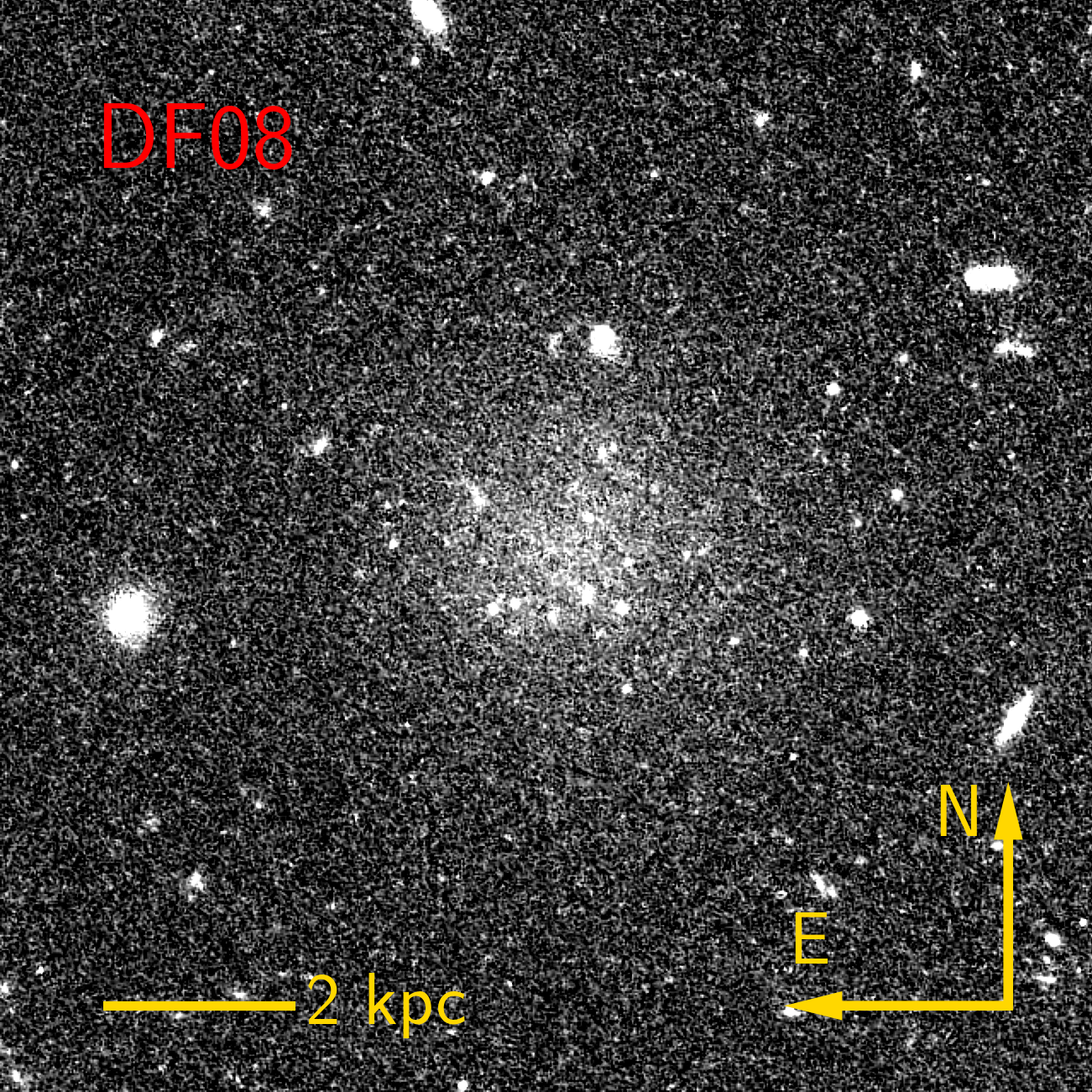}
\includegraphics[width=0.33\linewidth]{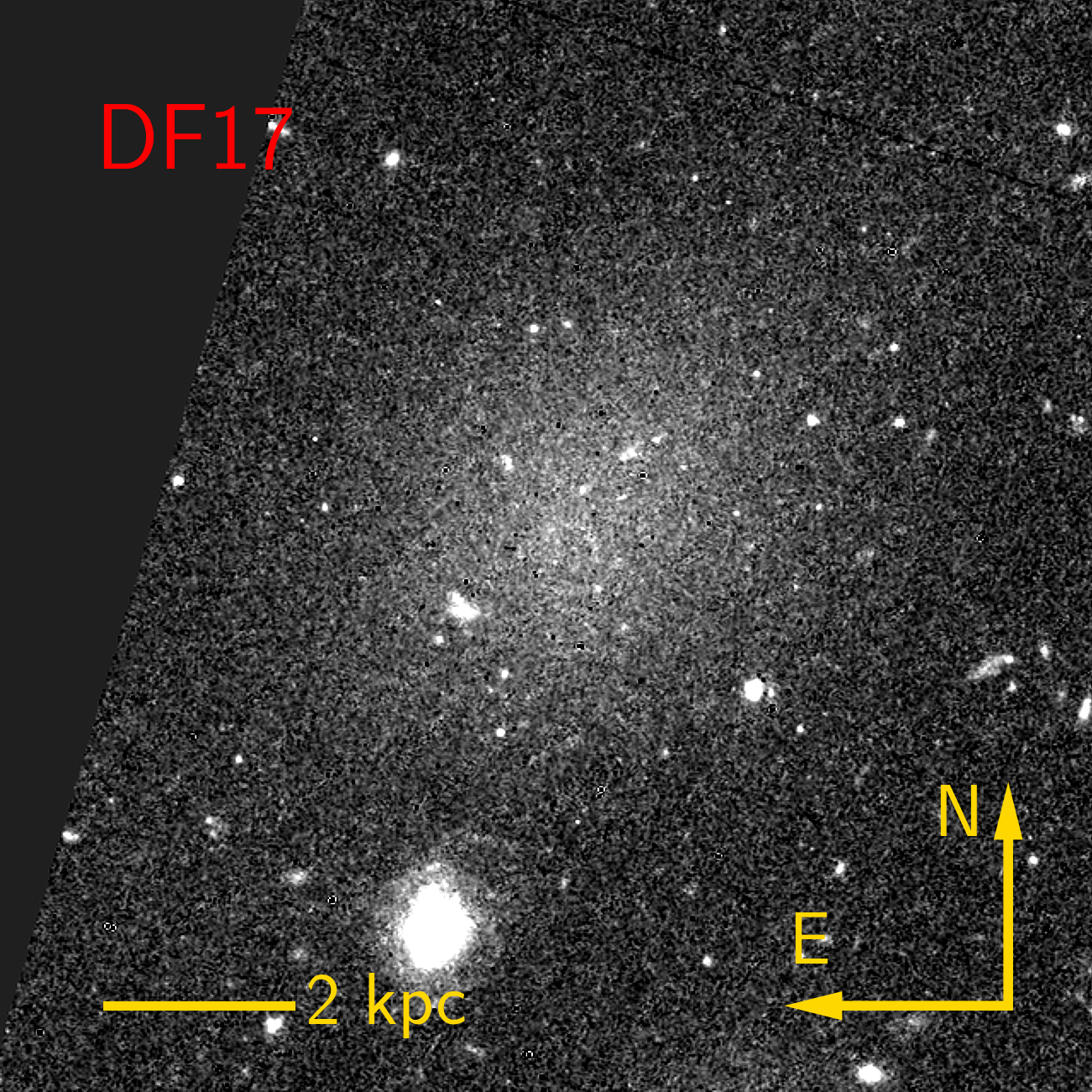}
\includegraphics[width=0.33\linewidth]{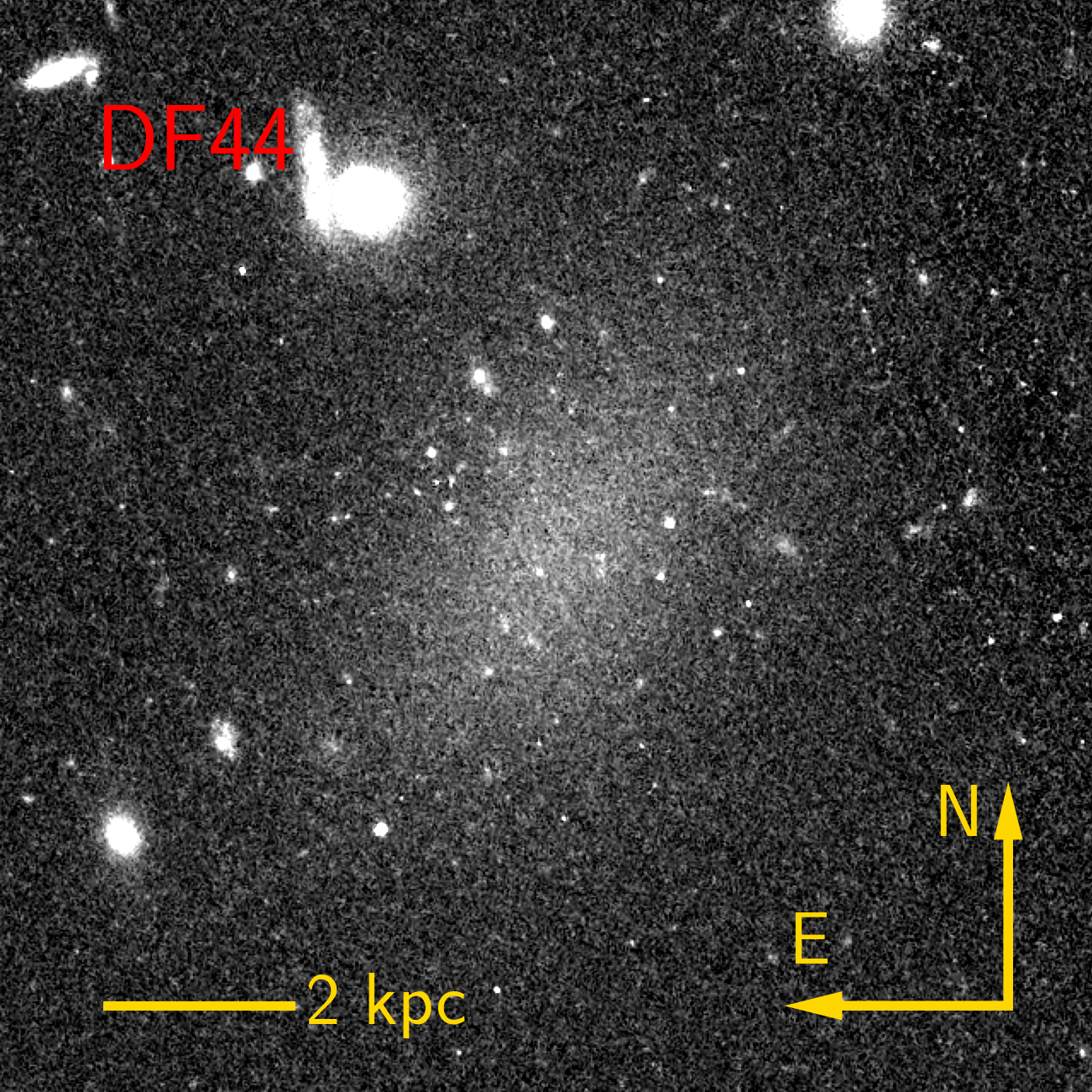}
\includegraphics[width=0.33\linewidth]{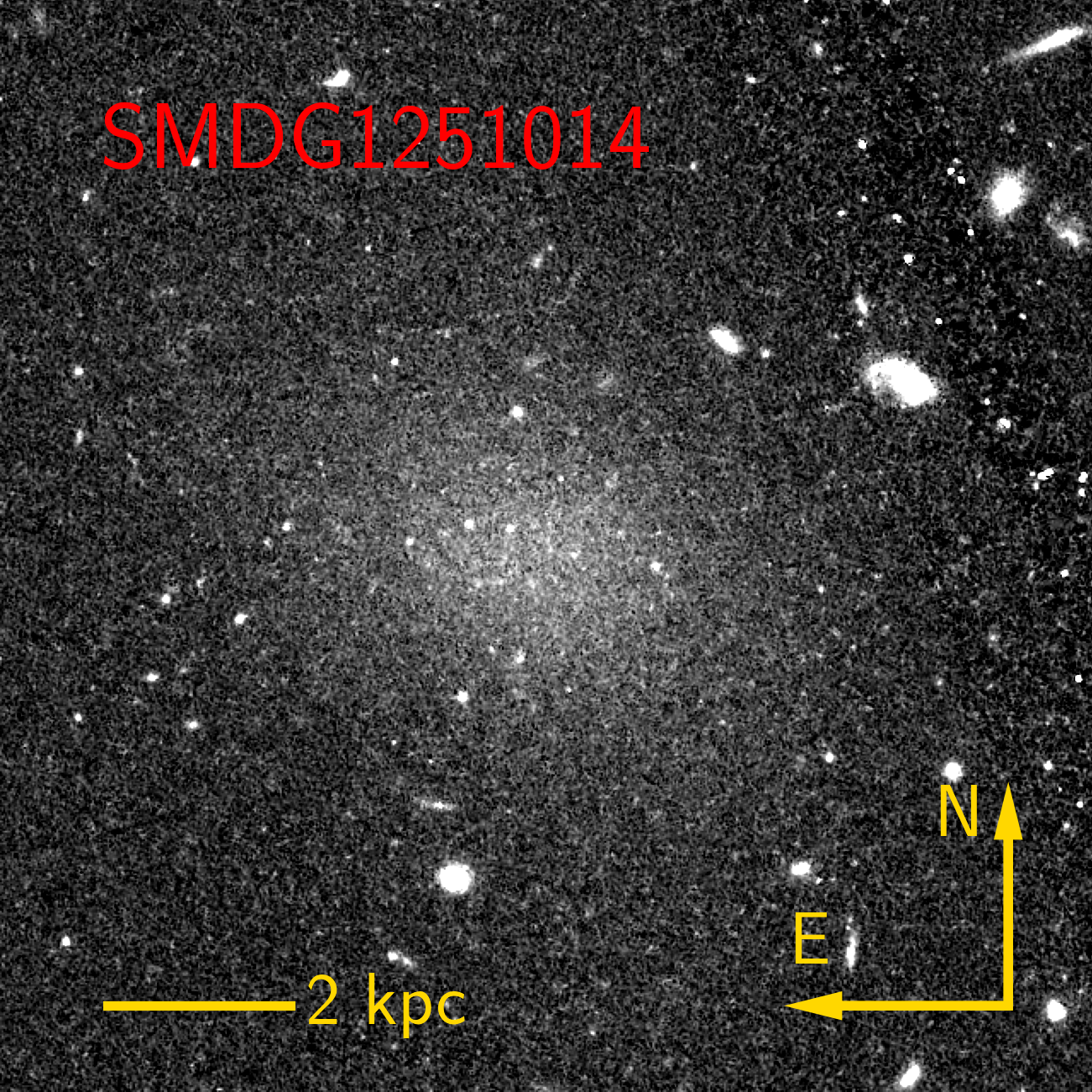}
\includegraphics[width=0.33\linewidth]{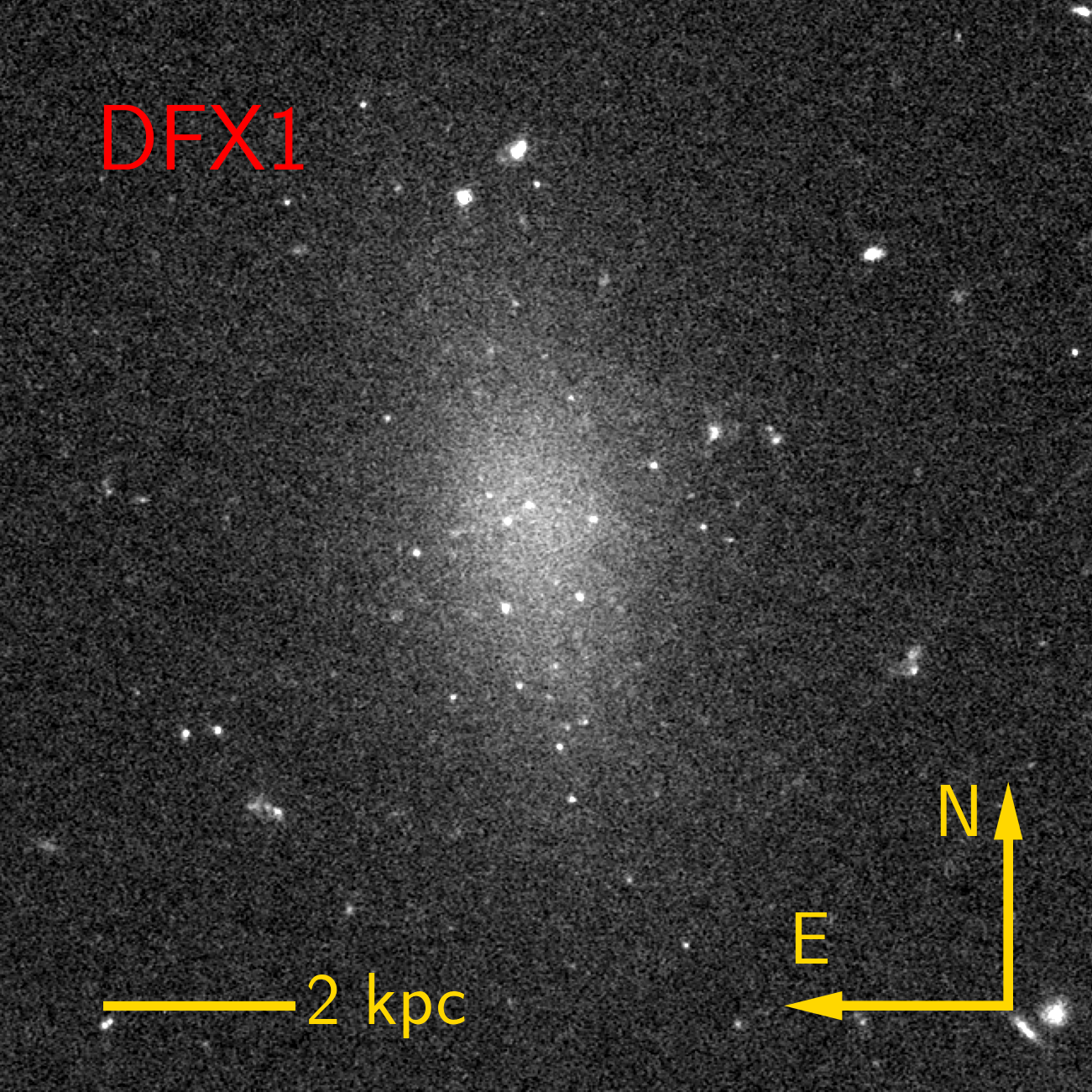}
\caption{The WFC/ACS/{\textit{HST}} view of the UDG sample of this work in $F814W$ for DF07, DF08, DF17, DF44, SMDG1251014 and $F606W$ for DFX1.}
\label{sample-fig}
\end{figure*}

\section{Observations}
\label{obs}

\subsection{UDG sample}

The UDGs in our sample are Coma cluster galaxies with an available spectroscopic redshift, a large effective radius ($R_{\rm e}$\,$\sim$\,4\,kpc), and deep \textit{HST} observations that reach the GCLF turn-over magnitude ($\mu_{\rm peak}$\,$\sim$\,-8.0\,mag in the $I$-band). These criteria narrow down the sample to six UDGs in the Coma cluster with an average effective radius, $R_{\rm e}$, of $\sim$\,3.6\,kpc. The Coma cluster is at a distance of 100\,Mpc (\citealp{distcoma1,distcoma2})\footnote{Radial velocity of 6925\,km/s (\citealp{coma-vel}).} and has a mass of $M_{200}$\,=\,1.3\,$\times$\,10$^{15}$\,$M_{\odot}$, making it one of the most massive galaxy clusters in the local universe (\citealp{coma-virial}).

These six galaxies are drawn from previous Coma cluster UDG surveys (\citealp{vd15,yagi,smudges1}) and, therefore, are known by multiple names. Five of these UDGs have DF names (Dragonfly Coma cluster survey, \citealp{vd15}) namely, DF07, DF08, DF17, DF44 and DFX1\footnote{This galaxy, which was originally named "GMP 2175" (\citealp{dfx1}), is not included in the main DF catalogue (\citealp{vd15}) but was added later, in \citet{vd17}, and renamed DFX1.}. For convenience, throughout this paper, we use the DF names for these five galaxies, although they all appear in both the \cite{yagi} and \cite{smudges1} catalogs as well. The sixth galaxy, known as SMDG1251014+274753 (hereafter we refer to this object as SMDG1251014) is identified by the SMUDGes survey of the Coma cluster (\citealp{smudges1,smudges2,dennis2}). The UDG sample is presented in Table~\ref{sample} and their distribution relative to the Coma cluster is shown in Fig.~\ref{comacluster}. The Coma membership of these objects is spectroscopically confirmed (\citealp{vd15b,vd16,vd17,kadowaski,gu,kadowaski2}). All, except SMDG1251014 are located within the virial radius (2.85\,Mpc, \citealp{coma-virial}) and the splashback radius of the cluster (2.43\,Mpc, \citealp{kadowaski2,diemer}). 

\subsection{Photometric data}

\begin{table*}
\centering
\caption{The UDG sample and the observational data of this work. Columns from left to right represent galaxy names in the SMUDGes (\citealp{smudges1}) and Yagi (\citealp{yagi}) catalogs (a), R.A. (b), Declination (c), heliocentric radial velocity (d) and exposure times in three filters, $F475W$, $F606W$, $F814W$, for available data (e, f and g). The * and ** symbols indicate the primary and secondary datasets, respectively. For convenience, throughout this paper, we use the DF name for five galaxies and the SMDG name for one galaxy. These names are indicated in boldface.}
\begin{tabular}{ llcccccc } \hline  Names (SMDG,DF,Yagi) & J2000 RA (deg) & J2000 DEC (deg) & $V_{\rm rad}$ (km/s) & $F475W$ & $F606W$ & $F814W$ \\ 
(a) & (b) & (c) & (d) & (e) & (f) & (g)  \\
\hline
SMDG1257017+282325, \textbf{DF07}, 680 &194.25716 & 28.390250 &  6,864$\pm$33 & 5,000s** & --- & 5,000s* \\
SMDG1301304+282228, \textbf{DF08}, 194 & 195.37646 & 28.374578 & 7,319$\pm$97 & 5,000s** & --- & 5,000s* \\
SMDG1301582+275011, \textbf{DF17}, 165 & 195.49266 & 27.836445 & 8,583$\pm$43 & 5,100s** &  5,800s & 5,100s* \\
SMDG1300580+265835, \textbf{DF44}, 11 & 195.24162 & 26.976355 & 6,661$\pm$38 & 5,000s** & 7,280s & 5,000s* \\
\textbf{SMDG1251014+274753}, ---, --- & 192.75563 & 27.798033 & 6,404$\pm$45  & 5,000s** & --- & 3,730s* \\
SMDG1301158+271238, \textbf{DFX1}, 13 & 195.31600 & 27.210484 & 8,105$\pm$6 & --- & 7,280 s* & 2,420s** \\
\hline 
\end{tabular}
\label{sample}
\end{table*}

Our data come from three different programs carried out with the {\sl Hubble Space Telescope} (\textit{HST}). The first (\textit{HST} program ID 15121; PI Zaritsky) provides WFC/ACS $F475W$ and $F814W$ observations of DF07, DF08, DF44 and SMDG1251014. These observations were defined to deliver similar data to those available for DF17 from the second program (\textit{HST} program ID 12476; PI Cook). This program provides WFC/ACS observations in $F475W$, $F606W$ and $F814W$, which has been used previously (\citealp{beasley2016b,peng2016}). 
The final program (\textit{HST} program ID 14643; PI van Dokkum) provides WFC3/UVIS $F606W$ and $F814W$ observations of DFX1 and DF44. These data have been also used previously (\citealp{vd17,saifollahi2020}). However, the $F814W$ observations of DF44 are shallower than our $F814W$ so we do not use them here. In Table~\ref{sample} we summarize the observations and basic properties (coordinates and recessional velocity) of our UDGs.

We retrieve the reduced data from the MAST server \footnote{All data are now public and accessible via \url{https://mast.stsci.edu/portal/Mashup/Clients/Mast/Portal.html}}. When more than one frame for a given filter is available, we median stack the frames using \textsc{SWarp} (\citealp{swarp}). We remove cosmic rays from the stacked frames using the L.A.Cosmic\footnote{L.A.Cosmic \textsc{Python} package is available here: \url{https://lacosmic.readthedocs.io/en/latest/}}(\citealp{cosmic}). These processed frames, with cropped versions shown in Fig.~\ref{sample-fig}, are used in the next section to measure the S\'ersic parameters of galaxies, perform photometry and select GC candidates. 
For each galaxy, we select the image that has the highest signal-to-noise ratio as the primary frame that we will use as a reference.
For all except DFX1, the $F814W$ images are defined as primary. For DFX1, the $F606W$ images is defined as the primary. 

\section{Analysis}
\label{analysis}

We use the primary and secondary images to measure the structural properties of UDGs, extract sources and ultimately, identify the globular clusters. In this section, we describe our methodology.

\subsection{Galaxy structural parameters}
To derive the structural properties of the UDGs, we fit a single S\'ersic function using \textsc{GALFIT} (\citealp{galfit1,galfit2}) to the 32\,$\times$\,32\,arcsec (16\,$\times$\,16\,kpc) cropped images. This crop size covers beyond 2$R_{\rm e}$ for all of our UDGs. To estimate the background flux, we make larger cropped frames of galaxies (80\,$\times$\,80\,arcsec or 40\,$\times$\,40\,kpc). Bright contaminating objects in the images are extracted using \textsc{SExtractor} (\citealp{sex}) with the default parameters except that we set
BACK\_TYPE\,=\,GLOBAL,
BACK\_SIZE\,=\,16, and BACK\_FILTERSIZE\,=\,1. We find that this configuration extracts sources located within the galaxy more effectively. At this point, we aim to identify bright sources that may influence the model fitting. Later, for GC identification, we will use a more sophisticated background subtraction (described in Section~\ref{back-sub}).

To determine the background level, 
we place a circular mask with a radius of 10 times the estimated minor axis (\textsc{SExtractor} output $B\_IMAGE$) for each source other than the UDG and a square mask of 32\,arcsec on a side for the UDG. We calculate the sigma-clipped median of the sky values (5 iterations with 3$\sigma$ clipping) for 1,000 randomly placed 5\,$\times$\,5\,arcsec (100\,$\times$\,100\,pixel) boxes. Finally, the median of these median values is used as the background flux input to \textsc{GALFIT}, which floats during fitting. Figures~\ref{sersicmodel-app} and \ref{sersicmodel2-app} in the appendix display the different steps of the S\'ersic modeling. 

\begin{figure}[hbt!]
\centering
\includegraphics[width=\linewidth]{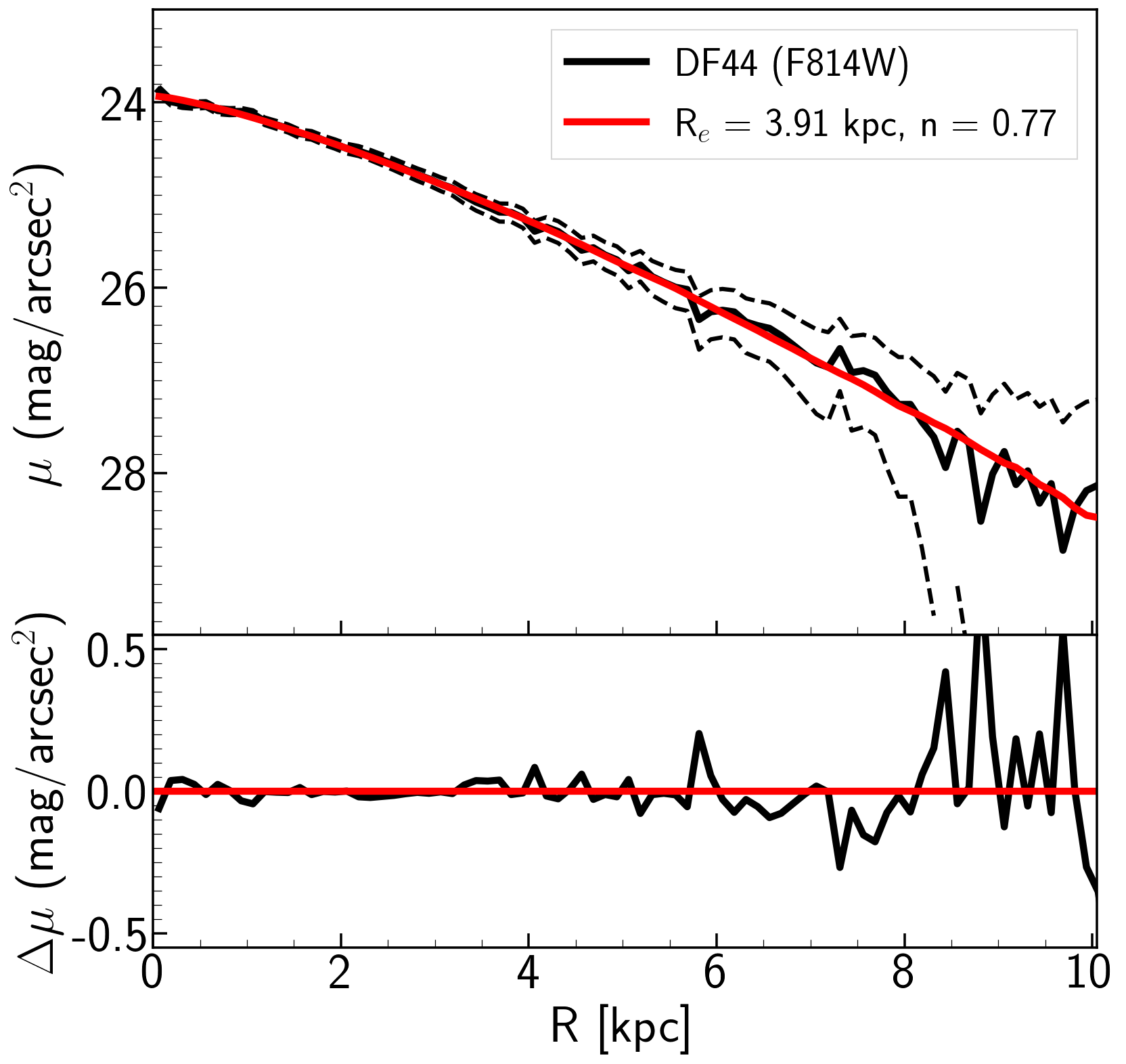}
\caption{The light profile (solid black lines), best-fit S\'ersic function (red curve) for the UDG DF44 (top) and fitting residuals (bottom). The black dashed lines indicate the 1-sigma uncertainty of the light profile.}
\label{sersicprofile}
\end{figure}

We present in Table~\ref{sersicparams} the resulting S\'ersic parameters for the galaxies and show the surface brightness profiles and fits in Fig.~\ref{sersicprofile} for DF44 and in Figure
\ref{sersicprofile-app} for the rest of the sample. In the cases where the same images are used, our derived values for the effective radii, $R_{\rm e}$, are consistent with previous measurements (\citealp{vd15,peng2016,beasley2016b} for DF17; \citealp{vd17} for DFX1; \citealp{saifollahi2020} for DF44). For DF44, using the \textit{HST} data in $F606W$, \citealp{vd17} measured an $R_{\rm e}$ that is on average $\sim$0.8\,kpc larger than our measurements in any of the three available filters. A possible explanation for this difference lies in the estimation of the background level, which can lead to different S\'ersic indices and ultimately, different effective radii. 

\begin{table*}
\centering
\caption{UDG structural parameters of our UDGs. Columns from left to right represent (a) galaxy name as referred to in this work, (b) observed filter, (c) effective radius, (d) S\'ersic index, (e) major axis position angle measured counterclockwise from North to East, (f) ellipticity, (g) total apparent magnitude, (h) total absolute magnitude, (i) effective surface brightness, (j) and the inferred stellar mass. The * and ** symbols denote the primary and secondary datasets. Absolute magnitudes are calculated using an adopted distance modulus $m-M = 35$, assuming that all Coma galaxies are at 100\,Mpc. Uncertainties come from the model fitting using \textsc{GALFIT}. The stellar masses are estimated using the equations in \citet{tom} after converting $M_{814}$, $m_{475}-m_{814}$, and $m_{606}-m_{814}$ to $I$, $B-I$ and $V-I$. The conversion are made using $m_{V,{\rm Vega}} - m_{I,{\rm Vega}}$\,=\,$m_{606,{\rm AB}}-m_{814,{\rm AB}} + 0.58$ and $m_{B,{\rm Vega}}-m_{I,{\rm Vega}}$\,=\,$m_{475,{\rm AB}}- m_{814,{\rm AB}} + 0.82$ (\citealp{blanton,wfc3,harris2018}). Systematic magnitude errors are small in comparison to the uncertainties in model fitting and neglected.}
\begin{tabular}{ llcccccccc } \hline  
Galaxy & Filter & $R_{\rm e}$ & $n$ & PA & $\epsilon$ & $m$  & $M$  & $<\mu_{\rm e}>$ & $M_*$ \\
- & - &  kpc & - & deg & - & mag & mag & mag/arcsec$^2$ & 10$^8$\,$M_{\odot}$\\
(a) & (b) & (c) & (d) & (e) & (f) & (g) & (h) & (i) & (j) \\
\hline
DF07 & $F475W$** & 3.74$\pm$0.37 & 0.85$\pm$0.08 & $-46.5\pm$0.63 & 0.77$\pm$0.01 & 19.38$\pm$0.15 & $-15.61\pm$0.15 & 25.74$\pm$0.36 & \\ 
     & $F814W$* & 3.55$\pm$0.04 & 0.81$\pm$0.01 & $-45.9\pm$0.04 & 0.79$\pm$0.01 & 18.45$\pm$0.02 & $-16.54\pm$0.02 & 24.69$\pm$0.04 & 2.8$\pm$0.7\\
\hline
DF08 & $F475W$** & 3.07$\pm$0.86 & 1.03$\pm$0.22 & $-81.6\pm$1.29 & 0.99$\pm$0.01 & 20.65$\pm$0.35 & $-14.34\pm$0.35 & 26.58$\pm$0.95 & \\  
     & $F814W$* & 2.95$\pm$0.17 & 0.88$\pm$0.05 & 70.13$\pm$0.31 & 0.90$\pm$0.01 & 19.77$\pm$0.09 & $-15.22\pm$0.09 & 25.61$\pm$0.21 & 0.6$\pm$0.2\\   
\hline
DF17 & $F475W$** & 3.75$\pm$0.35 & 0.71$\pm$0.08 & $-48.0\pm$0.77 & 0.75$\pm$0.01 & 20.00$\pm$0.16 & $-14.99\pm$0.16 & 26.36$\pm$0.36 & \\ 
    & $F606W$  & 3.48$\pm$0.20 & 0.65$\pm$0.04 & $-53.5\pm$0.13 & 0.73$\pm$0.01 & 19.66$\pm$0.10 & $-15.33\pm$0.10 & 25.86$\pm$0.22 & \\ 
    & $F814W$* & 3.45$\pm$0.20 & 0.65$\pm$0.04 & $-49.7\pm$0.31 & 0.73$\pm$0.01 & 19.16$\pm$0.10 & $-15.83\pm$0.10 & 25.34$\pm$0.22 & 1.4$\pm$0.4\\ 
\hline
DF44 & $F475W$** & 4.21$\pm$0.61 & 0.77$\pm$0.11 & $-20.1\pm$1.25 & 0.67$\pm$0.01 & 19.39$\pm$0.24 & $-15.60\pm$0.24 & 26.01$\pm$0.55 & \\ 
    & $F606W$  & 3.83$\pm$0.40 & 0.74$\pm$0.09 & $-29.2\pm$0.46 & 0.64$\pm$0.01 & 19.19$\pm$0.17 & $-15.80\pm$0.17 & 25.60$\pm$0.40 & \\ 
    & $F814W$* & 3.92$\pm$0.01 & 0.77$\pm$0.01 & $-24.4\pm$0.03 & 0.68$\pm$0.01 & 18.62$\pm$0.01 & $-16.37\pm$0.01 & 25.08$\pm$0.01 & 2.1$\pm$0.5\\ 
\hline
SMDG1251014 & $F475W$** & 5.06$\pm$0.80 & 0.94$\pm$0.11 & 49.25$\pm$0.88 & 0.71$\pm$0.01 & 19.16$\pm$0.24 & $-15.83\pm$0.24 & 26.17$\pm$0.58 & \\   
          & $F814W$* & 4.99$\pm$0.29 & 1.01$\pm$0.04 & 50.03$\pm$0.40 & 0.74$\pm$0.01 & 18.22$\pm$0.09 & $-16.77\pm$0.09 & 25.20$\pm$0.20 & 4.9$\pm$1.2\\  
\hline
DFX1 & $F606W$* & 3.73$\pm$0.17 & 0.93$\pm$0.04 & 9.85$\pm$0.05 & 0.60$\pm$0.01 & 19.18$\pm$0.07 & $-15.81\pm$0.07 & 25.53$\pm$0.16 & \\ 
     & $F814W$** & 3.69$\pm$0.22 & 0.92$\pm$0.05 & 10.86$\pm$0.22 & 0.59$\pm$0.01 & 18.77$\pm$0.09 & $-16.22\pm$0.09 & 25.10$\pm$0.21 & 1.5$\pm$0.4\\ 
\hline 
\end{tabular}
\label{sersicparams}
\end{table*}

\begin{figure}
\centering
\includegraphics[width=\linewidth]{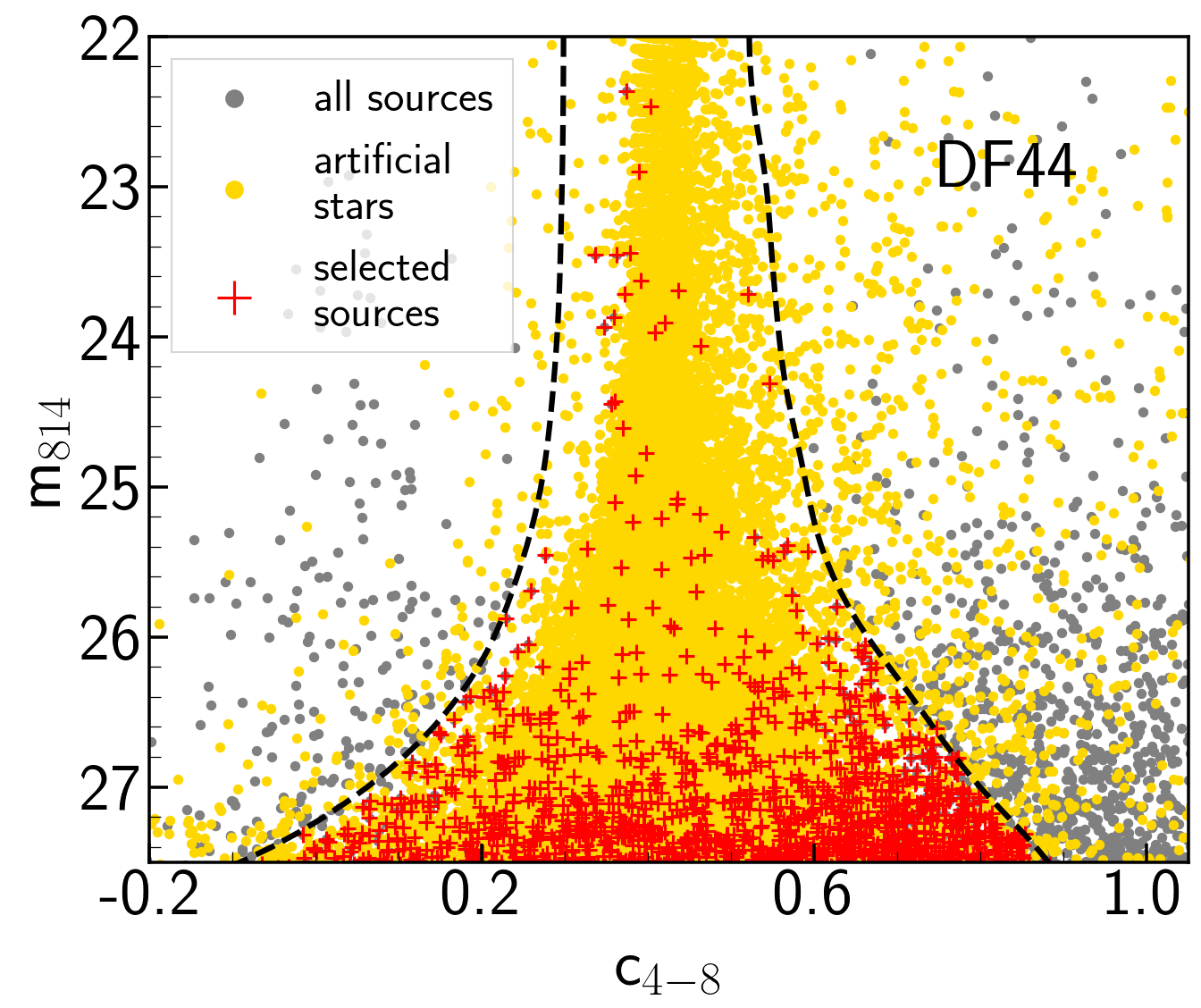}
\includegraphics[width=\linewidth]{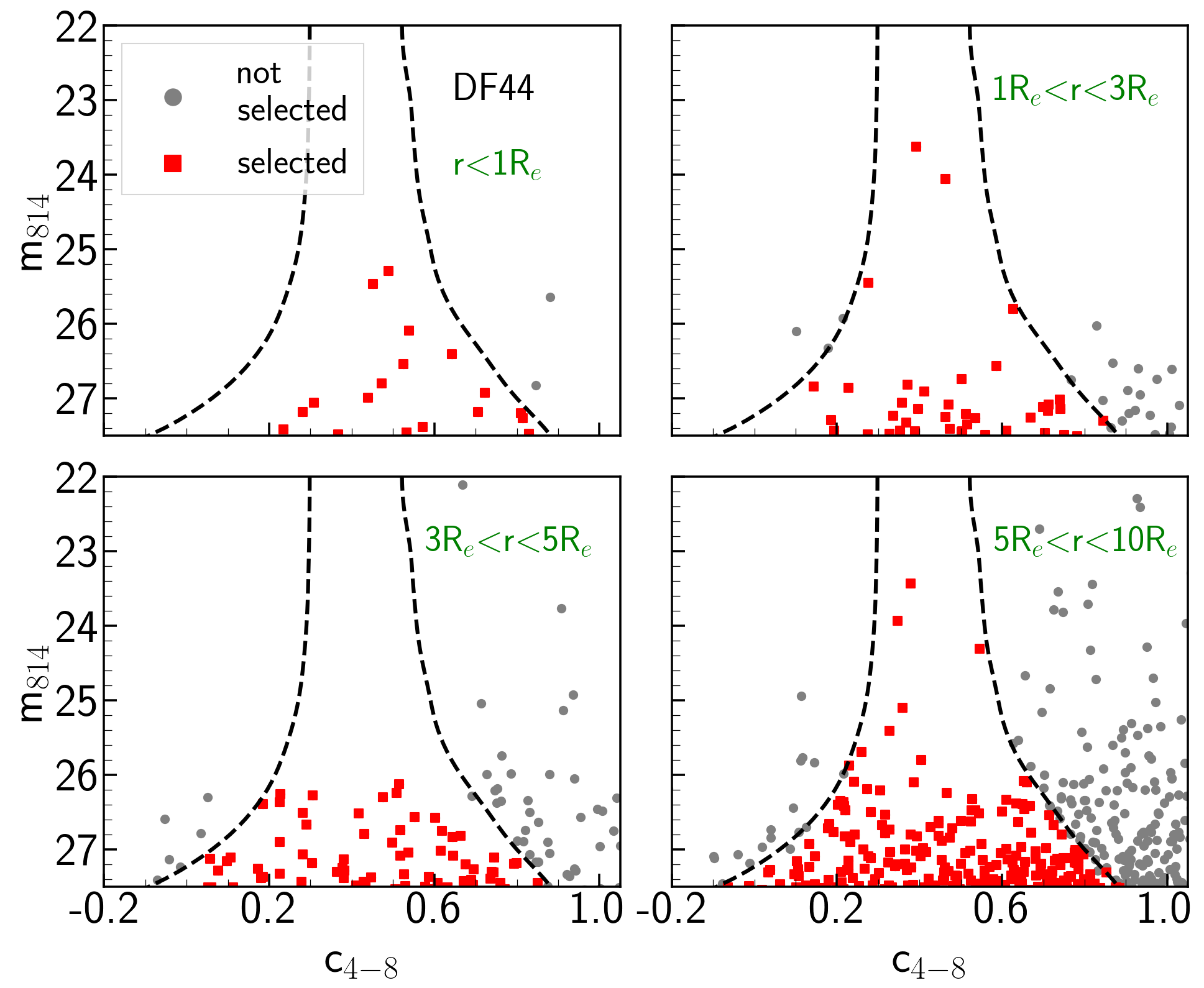}
\caption{Top: Compactness index ($c_{4-8}$)\,-\,magnitude diagram for DF44 in its primary frames ($F814W$). The boundaries of point source selection are indicated with black dashed curves. All the sources in the primary frames, simulated stars and selected point sources are shown in grey dots, yellow dots and red crosses respectively. Bottom: Similar diagram as the diagram on top for sources in different radial distances from the host UDG. Selected sources and not-selected (discarded) sources are shown in red and grey dots.}
\label{compactness}
\end{figure}

\begin{figure}
\centering
\includegraphics[width=\linewidth]{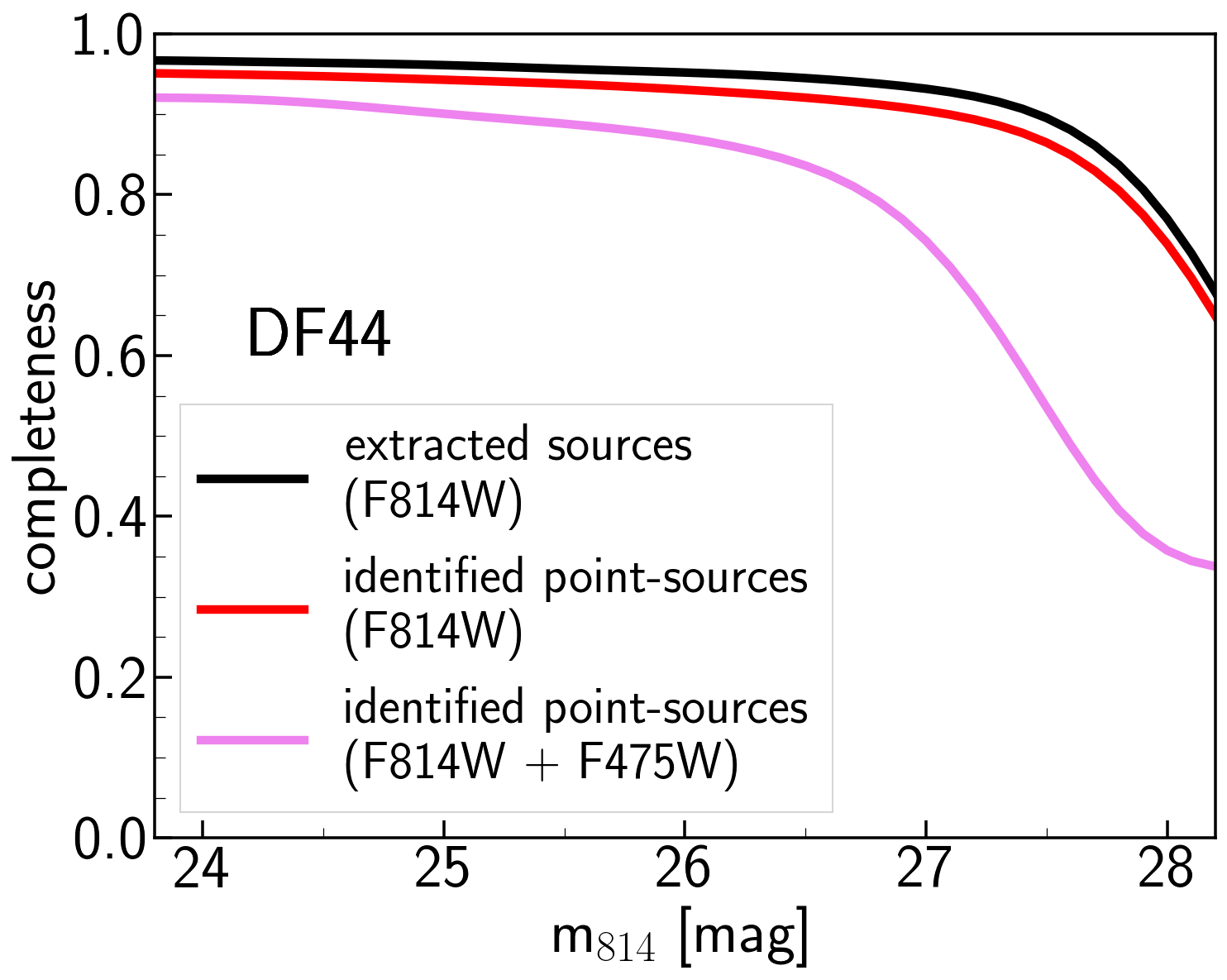}
\caption{The completeness of source extraction (black), point source selection (red for $F814W$ primary filter) and source selection after including the secondary filter (shallower data, purple), for the galaxy DF44.}
\label{comp}
\end{figure}

\begin{figure*}
\centering
\includegraphics[width=\linewidth]{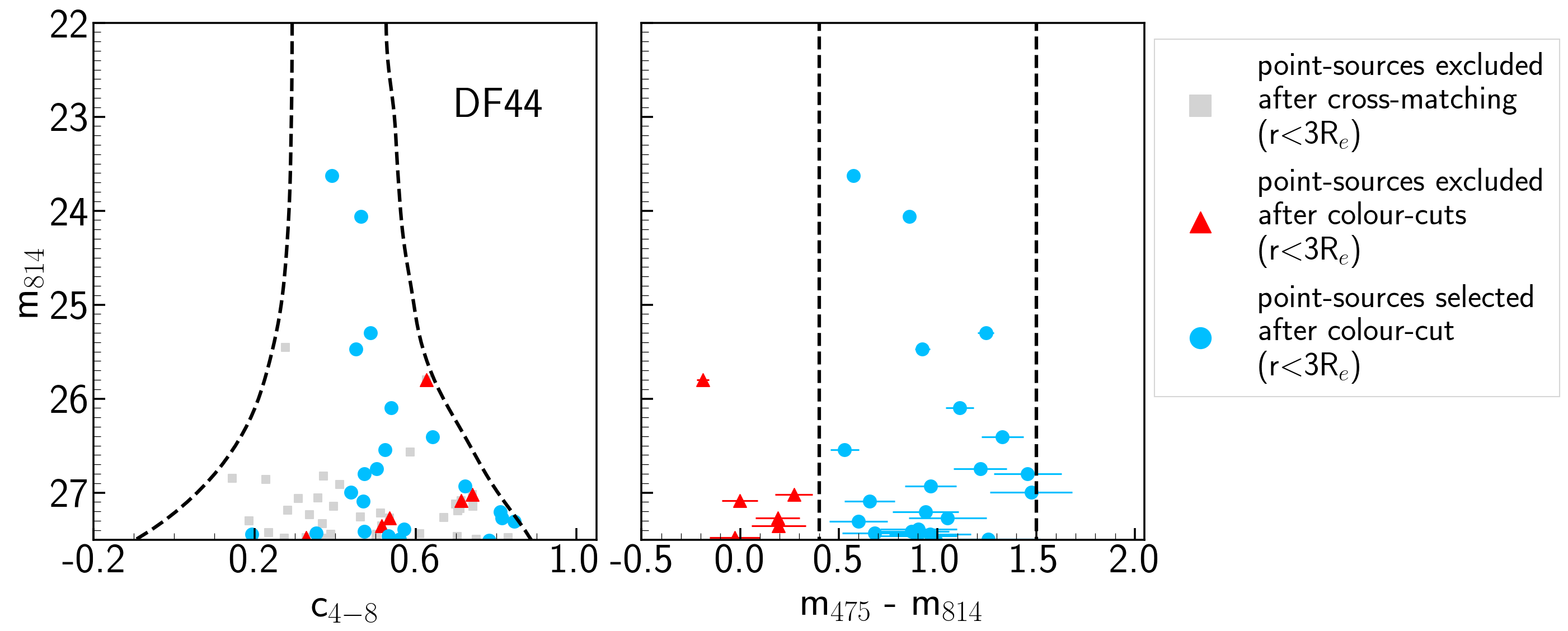}
\caption{Compactness index ($c_{4-8}$) - magnitude diagram (left) and colour-magnitude diagram (right) of point sources within 3$R_{\rm e}$ of DF44. After cross-matching sources in the primary data with the extracted sources in the secondary data, those without corresponding detection are likely remaining cosmic rays and rejected (grey symbols). With the colour limits as indicated in the right panel (vertical dashed lines), the remaining sources are either selected (blue) or rejected (red) as GC candidates.}
\label{colour-select}
\end{figure*}

\subsection{Source extraction and photometry}
\label{back-sub}

We perform source extraction and source photometry using \textsc{SExtractor} to construct the source catalogue that is needed for the next steps in our analysis. To improve the source extraction, we subtract an 8\,$\times$\,8\,pixel (spatial) median-filter from each frame. We identify and extract sources using \textsc{SExtractor} and its default parameters. Detected sources are masked and the procedure (filtering, source extraction, masking) is repeated three more times. The final median-subtracted frames are used to construct the source catalogues. For this search we use the default \textsc{SExtractor} values with a few adjustments: DETECT\_MINAREA\,=\,4.0, DETECT\_THRESH\,=\,1.5, ANALYSIS\_THRESH\,=\,1.5, DEBLEND\_NTHRESH\,=\,4, DEBLEND\_MINCONT\,=\,0.005, BACK\_TYPE\,=\,GLOBAL, BACK\_SIZE\,=\,32, and BACK\_FILTERSIZE\,=\,3.
We calculate aperture magnitudes within apertures of 4 and 8\,pixel diameter. We use the 4\,pixels aperture magnitudes ($\sim$\,2\,$\times$\,FWHM) to measure the source aperture magnitude and, in combination with the
8\,pixel aperture magnitudes, to measure a source compactness index.
The aperture corrections we apply are as presented in the instrument handbook and are 0.44, 0.45 and 0.54\,mag for $F475W$, $F606W$ and $F814W$ respectively. We adopt zeropoints from the ACS Zeropoints calculator\footnote{\url{https://acszeropoints.stsci.edu/}}. These values are slightly different between observations that have been carried out at different times.

\subsection{Identifying GC candidates}

Here we describe how we identify GC candidates. Because ACS has a pixel size of 0.05\,arcsec and the point spread function full width half maximum (FWHM) is 0.1\,arcsec (2\,pixels), this instrument is the best available for GC searches in UDGs at distances between that of the Virgo and Coma clusters. Although GCs are unresolved in the \textit{HST} observations of systems at the distance of the Coma cluster (\citealp{harris2009}), the high-resolution imaging of the ACS can distinguish them from high-redshift background galaxies, which are resolved. Furthermore, the 202\,$\times$\,202\,arcsec
field of view compares well to the average effective radius of UDGs in the sample (7.2\,arcsec on average), providing enough field coverage to allow us to statistically estimate any remaining contamination after the source selection procedure described below.

As already indicated,
GCs at the Coma distance are unresolved. To help distinguish them from background galaxies, we define a compactness index, $c_{4-8}$, as the difference between the 4 and 8\,pixels aperture magnitudes. Because the primary frames were selected to have a higher signal-to-noise ratio, we adopt the $c_{4-8}$ measured from those.
In the following, we select point sources as GC candidates based on this compactness criteria. We then use the measured colours to further clean the sample of possible contamination (foreground/background). Next, we describe each step in detail.

\subsubsection{Artificial stars and point source selection}

GC candidate selection based on the compactness index is a well-established approach (e.g. \citealp{jordan2007,jordan2015,amorisco2018,lim2018}). The compactness index threshold is established for each different data set through comparisons of the resulting compactness indices of implanted artificial stars. The point spread function (PSF) used to create those stars is typically determined using bright and unsaturated stars in the data. 

To implant artificial stars, we first discard all objects brighter than 22nd magnitude because these are $\sim$\,5\,mag brighter than the expected GCLF peak. Assuming a GCLF width of $\sigma$\,=\,1.0\,mag (\citealp{brodie}), this limit is $\sim$\,5$\sigma$ from the GCLF peak. We conclude that compact objects brighter than 22\,mag are foreground stars. From the remaining sources, we select between 10 and 20 of the brightest unsaturated stars in each primary frame. These are brighter than 25\,mag in $F814W$ and with a compactness index between 0.25 and 0.45 mag. This range of compactness index is chosen by visual inspection of the compactness index\,-\,magnitude diagram of sources. After selecting stars within primary frames, we re-sample their PSFs at a pixel size 10 times smaller, combine all of the PSFs, and reconstruct the PSF of each primary frame. The reconstructed PSFs have a pixel size of 0.005\,arcsec. Given the small number of bright unsaturated stars in each frame, we choose to adopt a constant PSF across the frame.

We simulate 1,000 stars with magnitudes between 22 and 29\,mag per 0.1\,mag bin. We randomly locate artificial stars to sub-pixel precision (along X-axis and Y-axis, within 10 pixels from the PSF centroid), re-sample to the instrumental pixel size (0.05\,arcsec), add Poisson noise, and randomly distribute them throughout the original images.

For each of the new images produced, we perform cosmic ray rejection, source extraction and photometry as we did before. In Fig.~\ref{compactness} we show the $c_{4-8}$ - magnitude diagram of the simulated stars for DF44 (the diagrams for the other galaxies are in Fig.~\ref{compactness-app}). Based on the median and standard deviation of $c_{4-8}$ at each magnitude, we define a range of $c_{4-8}$ as characteristic of point sources (indicated by black dashed curves) and select objects within these curves as GC candidates (red crosses). The boundaries correspond to 3$\sigma$ deviations from the median compactness index of the artificial stars plus 0.1\,mag to account for compactness variations across a frame. 

Using the artificial star results we estimate that we are 90\,per\,cent complete in identifying point sources near the expected GCLF turnover (27.8 and 27.3\,mag in $F606W$ and $F814W$). The full completeness function is presented in Fig.~\ref{comp} for DF44 and Fig.~\ref{comp-app} for the other UDGs.

\begin{figure*}
\centering
\includegraphics[width=0.49\linewidth]{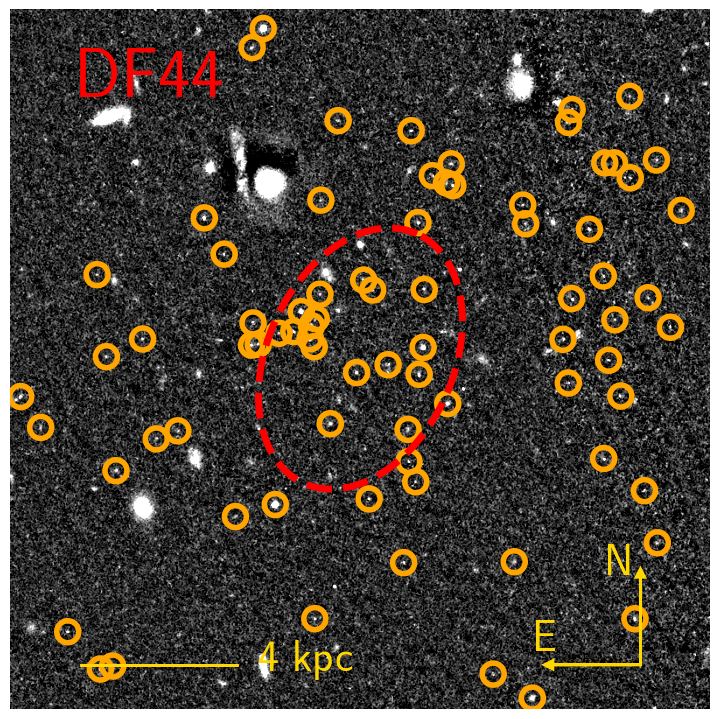}
\includegraphics[width=0.49\linewidth]{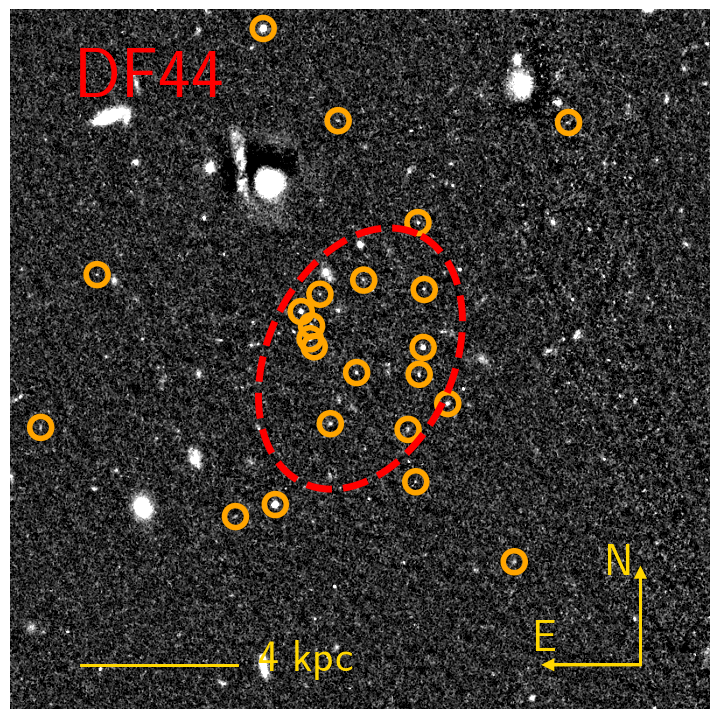}
\caption{Selected point sources (left panel) and remaining sources after cross-matching and colour selection (right panel) for DF44. Sources brighter than $F814W$\,=\,27.5\,mag are shown. The dashed red ellipse corresponds to the effective radius of the galaxy.}
\label{gc-cands}
\end{figure*}

\begin{figure*}
\centering
\includegraphics[width=0.32\linewidth]{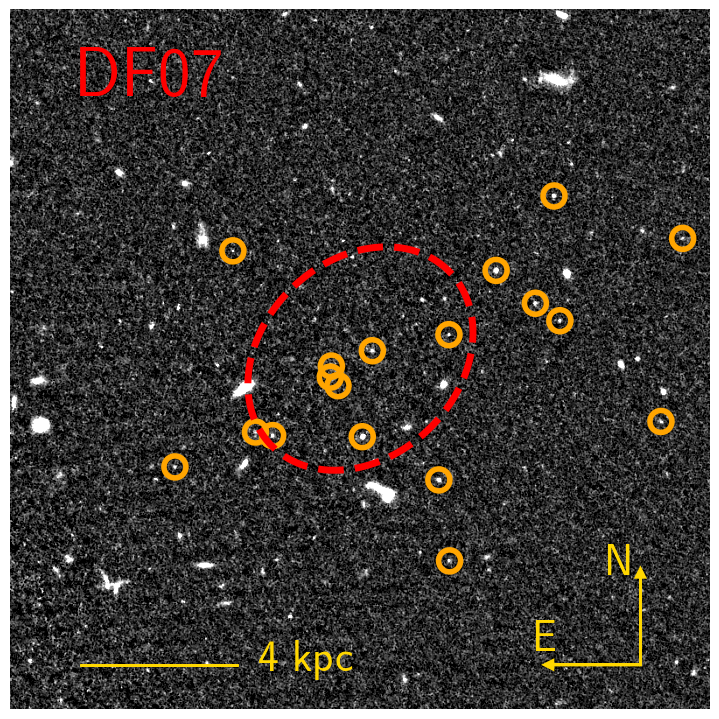}
\includegraphics[width=0.32\linewidth]{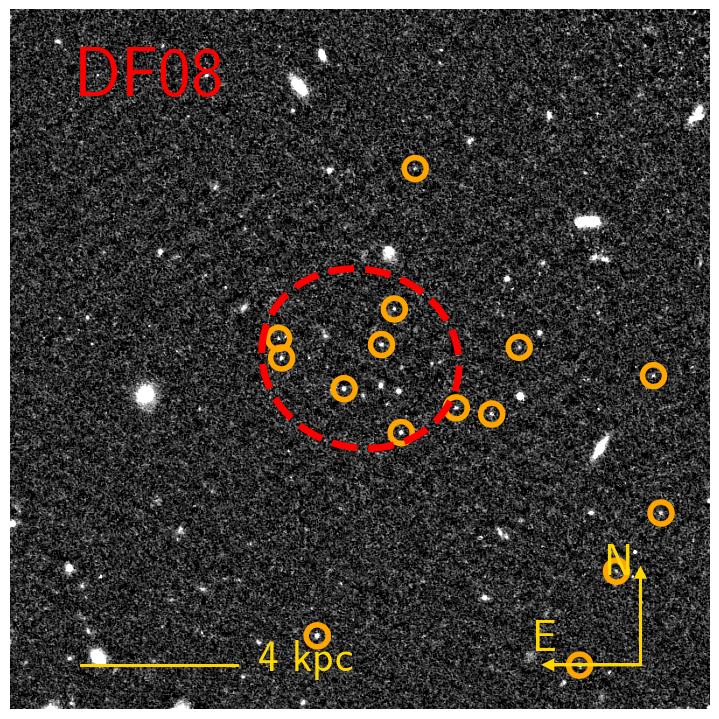}
\includegraphics[width=0.32\linewidth]{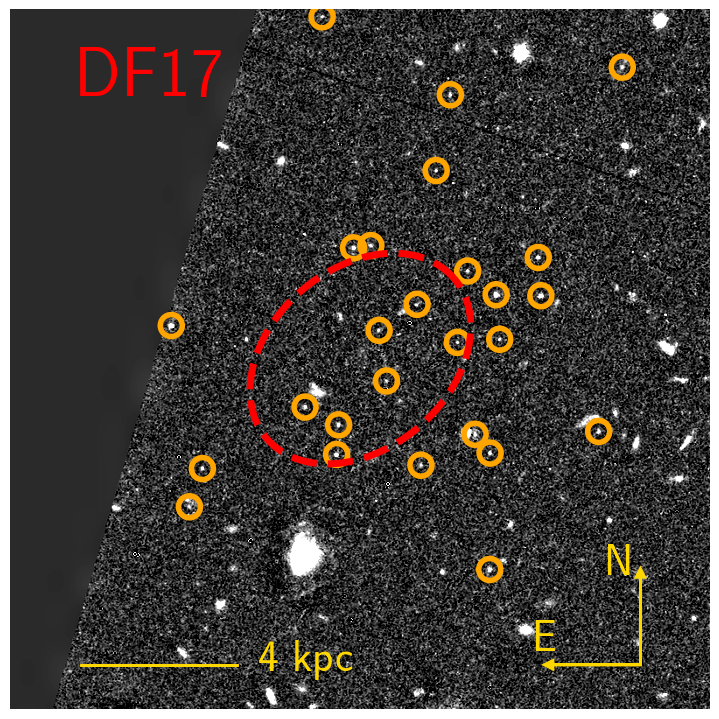}
\includegraphics[width=0.32\linewidth]{selected_sources_around_galaxy_DF44_814_27.5.png}
\includegraphics[width=0.32\linewidth]{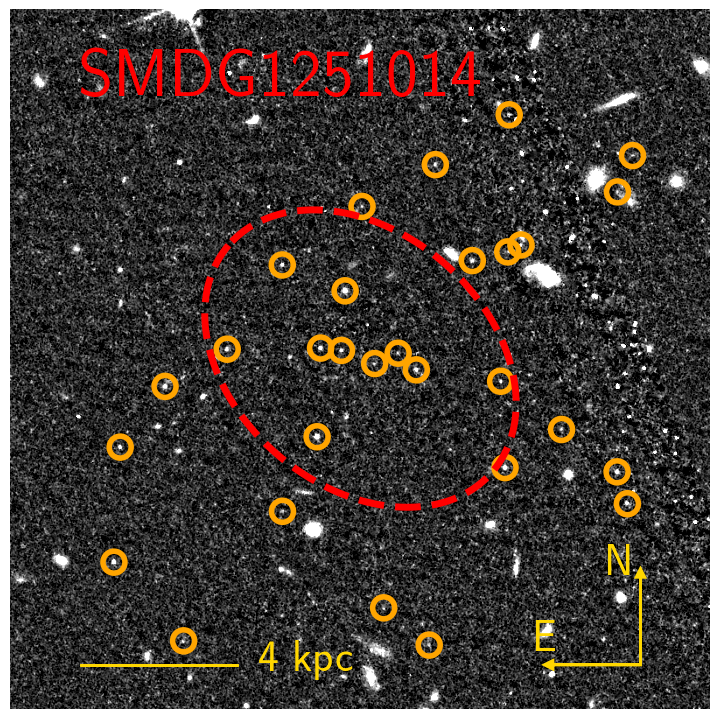}
\includegraphics[width=0.32\linewidth]{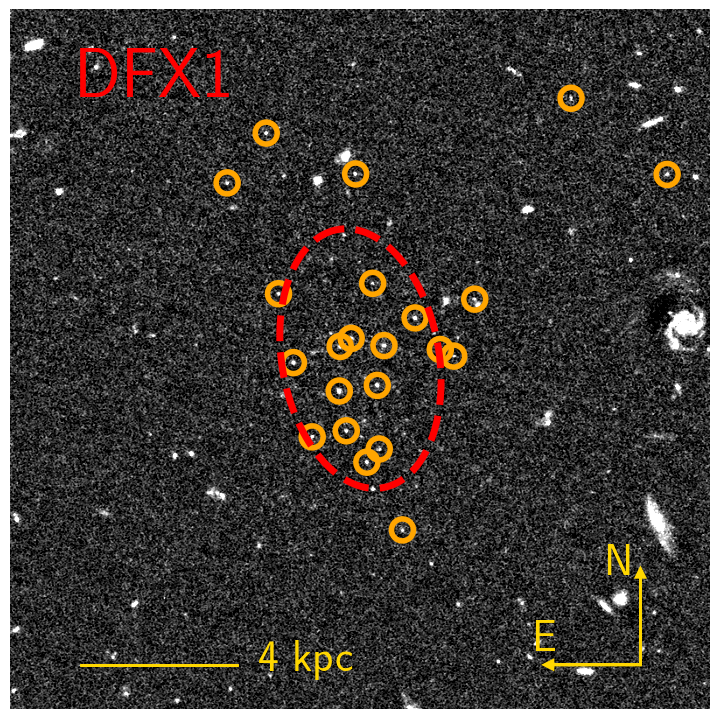}
\caption{GC candidates selected based on compactness index and colours. Sources brighter than $F814W$\,=\,27.5\,mag are shown (for DFX1, brighter than $F606W$\,=\,28.0\,mag).}
\label{gc-cands2}
\end{figure*}

\begin{figure*}
\centering
\includegraphics[width=0.32\linewidth]{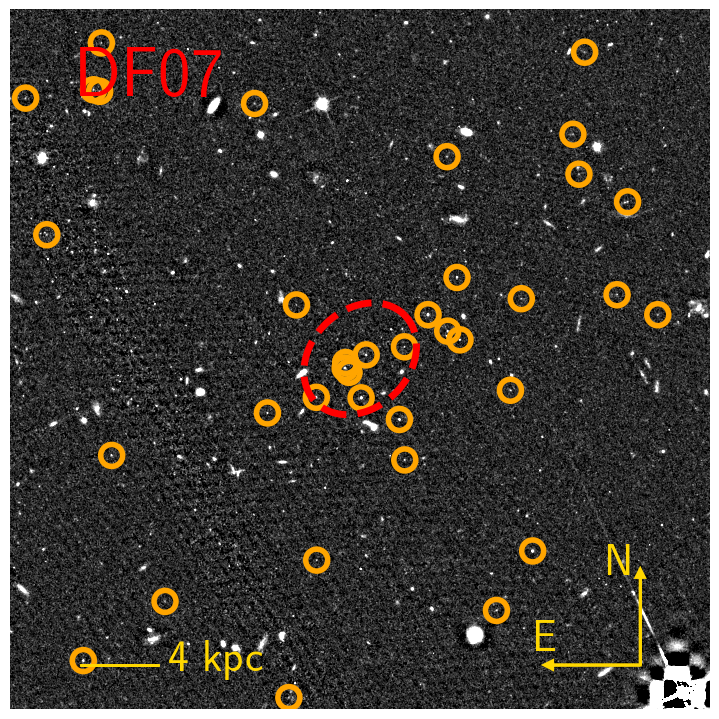}
\includegraphics[width=0.32\linewidth]{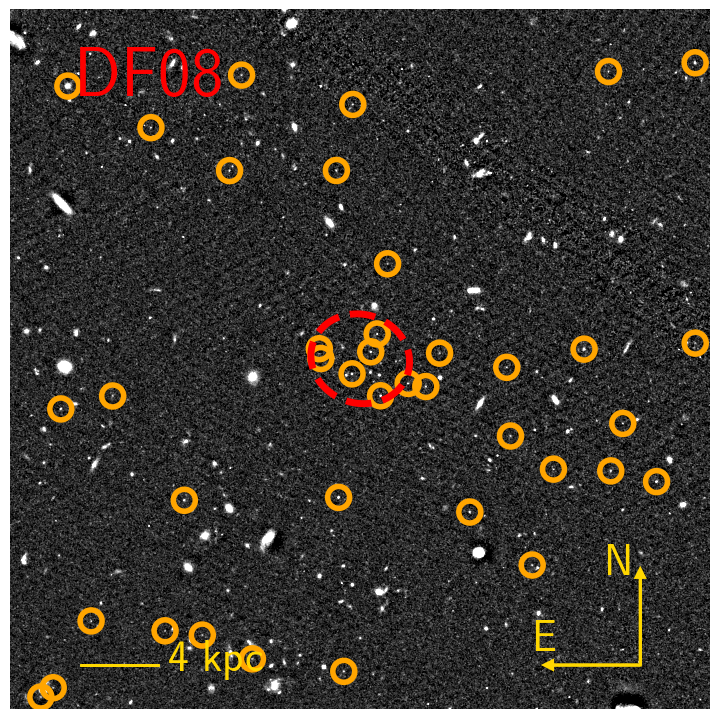}
\includegraphics[width=0.32\linewidth]{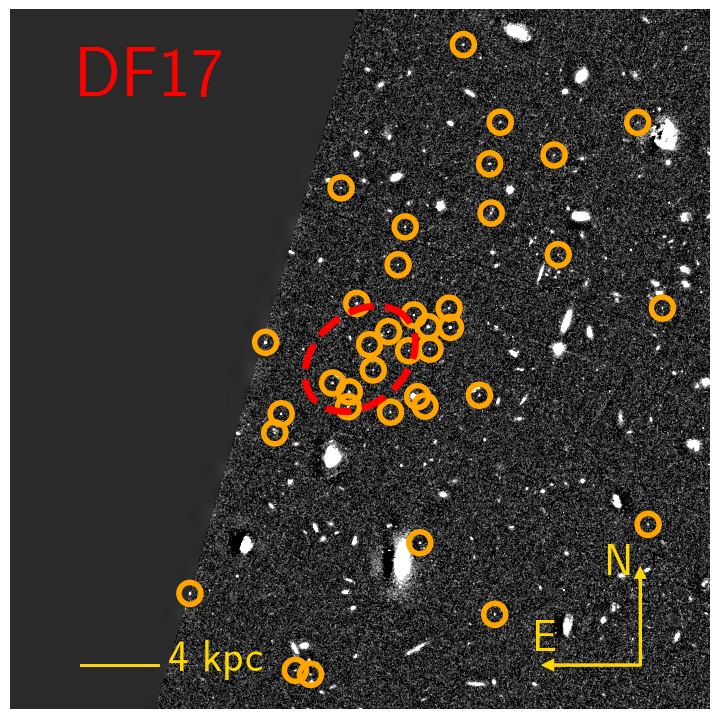}
\includegraphics[width=0.32\linewidth]{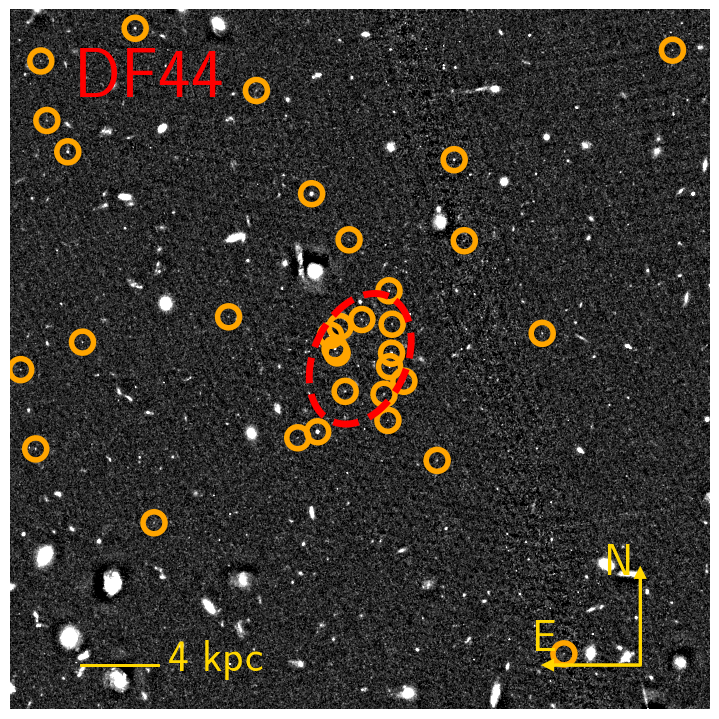}
\includegraphics[width=0.32\linewidth]{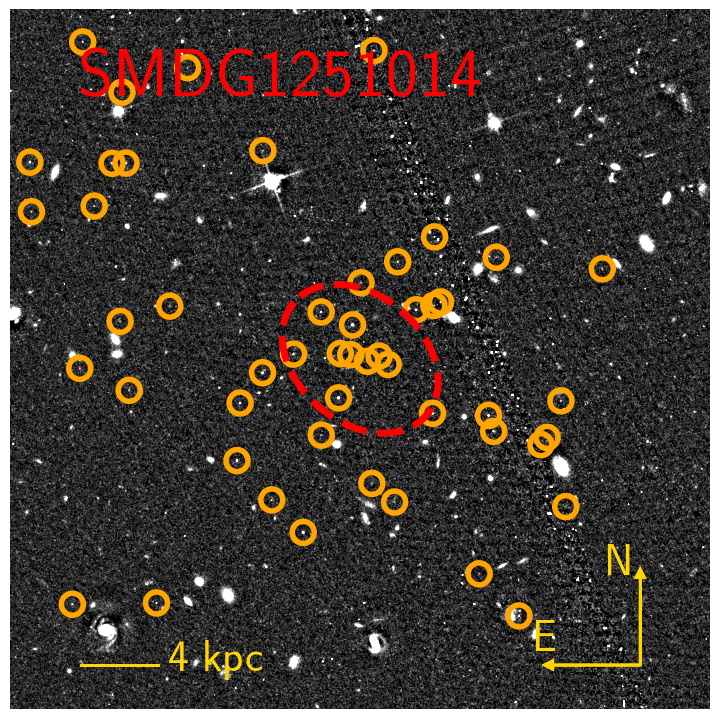}
\includegraphics[width=0.32\linewidth]{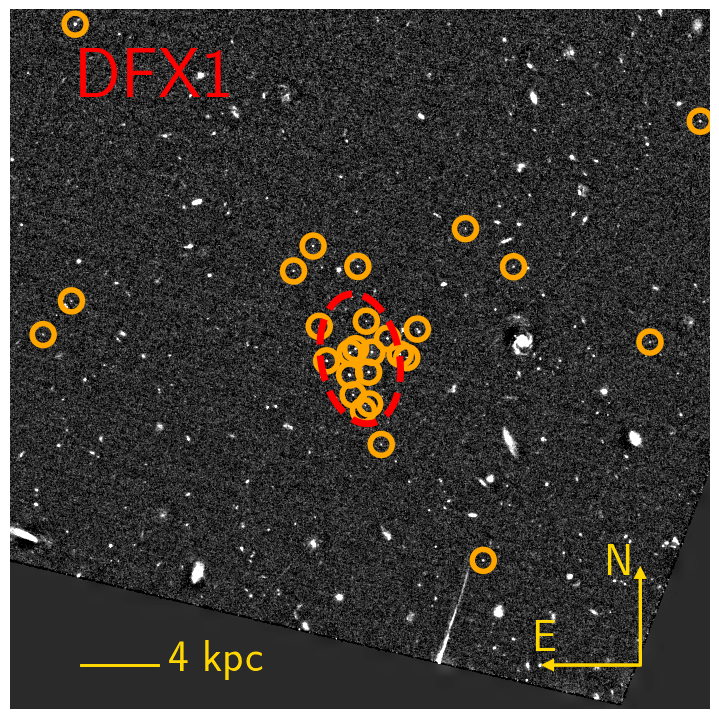}
\caption{Zoomed-out view of GC candidates selected based on compactness index and colours. Sources brighter than $F814W$\,=\,27.5\,mag are shown (for DFX1, sources brighter than $F606W$\,=\,28.0\,mag are shown).}
\label{gc-cands3}
\end{figure*}
\subsubsection{Colour-based selection}

Following the point source selection using the primary image, we use the secondary data to apply a colour cut. Requiring a positional match in the secondary image (within 0.1\,arcsec or 2\,pixels) rejects most remaining cosmic rays. Unfortunately, some cosmic rays, regardless of the cosmic ray cleaning and compactness criteria, remain and make their way into the sample of selected point sources. We will account for these and other remaining contaminants statistically in the next steps of the analysis.

Because the secondary data are shallower than the primary, completeness will be a function of colour. To account for this, we repeat the artificial star simulations for the secondary filter and measure its completeness. To derive the combined completeness, we assume that GCs have average colours of $m_{606}-m_{814}$\,=\,0.35\,mag and $m_{475}-m_{814}$\,=\,0.80\,mag (based on the photometric prediction of the E-MILES stellar library for a 10-14 Gyr single stellar population ith metallicities between $-2 < Z < 0.5$; \citep{vazdekis,vazdekis2016}. The combined completeness function is shown in Fig.~\ref{comp} in purple. The multi-wavelength catalogues are 75\,per\,cent complete around the expected GCLF turn-over, around $m_{814}$\,$\sim$\,27.0\,mag (\citealp{miller,kundu}).

We apply colour cuts to the selected point sources. We use the colour 0.1\,$<$\,$m_{606}-m_{814}$\,$<$\,0.75\,mag for DFX1 and 0.4\,$<$\,$m_{475}-m_{814}$\,$<$\,1.5\,mag for rest of the sample. This colour range is estimated based on the photometric prediction of the E-MILES stellar library as described earlier. We retain candidates that exceed these limits by less than 1$\sigma$.

In Fig.~\ref{colour-select} we demonstrate the effects of cross-matching and colour-selection in removing sources in the DF44 field (for the other UDGs in Fig.~\ref{colour-select-app}) for candidates within 3$R_{\rm e}$ of their host galaxies.
We highlight in grey the (suspected) cosmic rays that are not cleaned by L.A.Cosmic but are removed after cross-matching between primary and secondary images. The red triangles highlight those sources rejected due to their colour. In Fig.~\ref{gc-cands} we show the GC candidates in DF44 before and after applying the colour-cuts (left and right panels). 

As already mentioned, even the best set of criteria cannot exclude the possibility of contamination. As such, the surviving objects are only GC candidates. To correct for contamination, we estimate and take into account the average properties of objects that satisfy all of our criteria but lie sufficiently far from the UDG that they are unlikely to be associated with the UDG. These contaminants may include surviving cosmic rays, Galactic stars, or actual GCs, perhaps free-floating Coma GCs, that are not associated with the targeted UDG. In Fig.~\ref{gc-cands2} we show the selected GC candidates after compactness and colour selection for the six UDGs in our sample. We provide a wider view in Fig.~\ref{gc-cands3}.

\begin{figure*}
\centering
\includegraphics[width=0.8\linewidth]{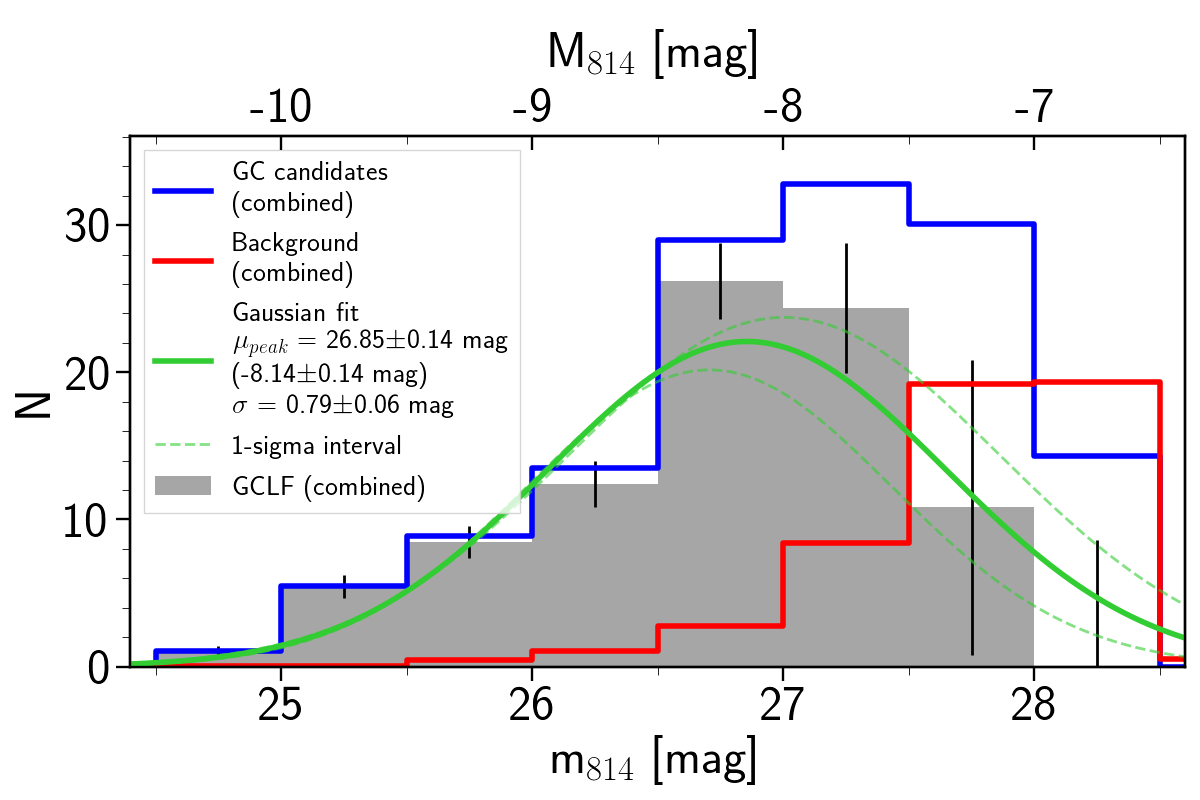}
\caption{The combined globular clusters luminosity function (GCLF) in $F814W$ (grey histogram), the best-fit Gaussian function (green curve), the combined GCLF before background subtraction (blue histogram) and the background luminosity function (red histogram). The errorbar at a given magnitude bin is the sum of uncertainties (Poisson noise) in background subtraction and incompleteness correction in that bin.}
\label{gclf-combined}
\end{figure*}

\begin{figure*}
\centering
\includegraphics[width=0.46\linewidth]{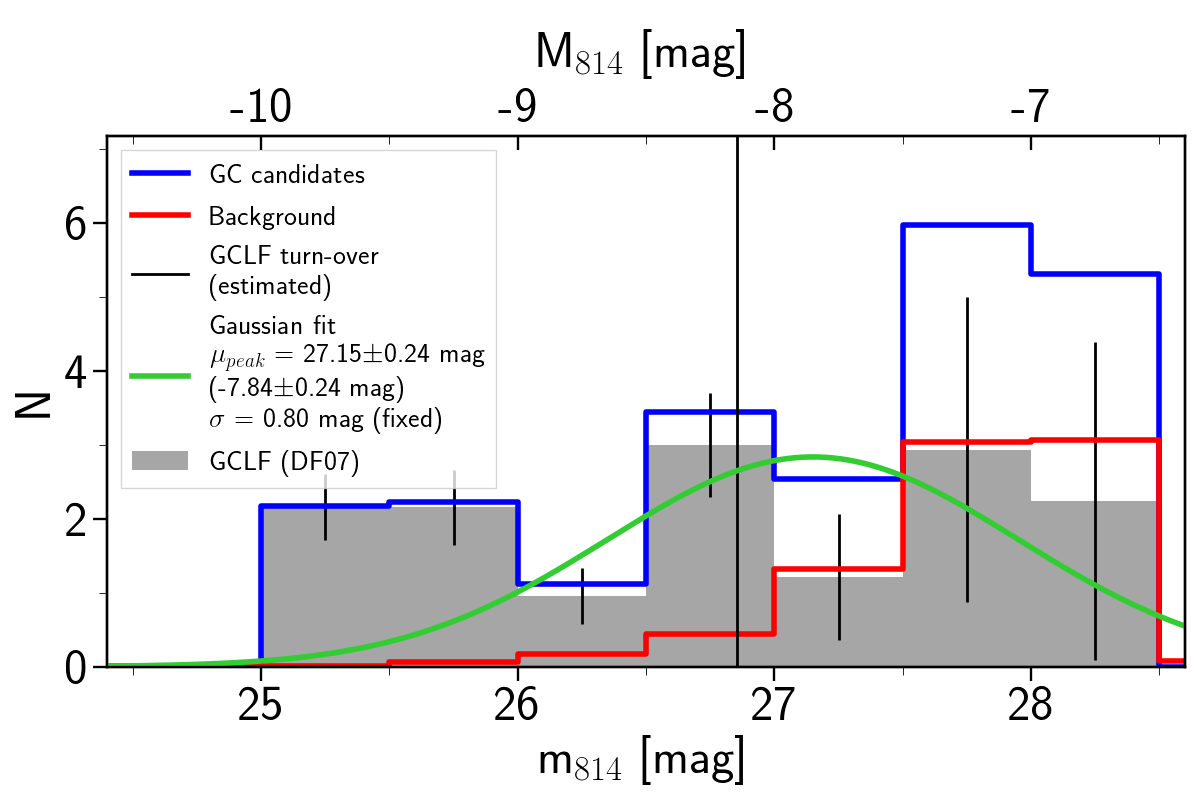}
\includegraphics[width=0.46\linewidth]{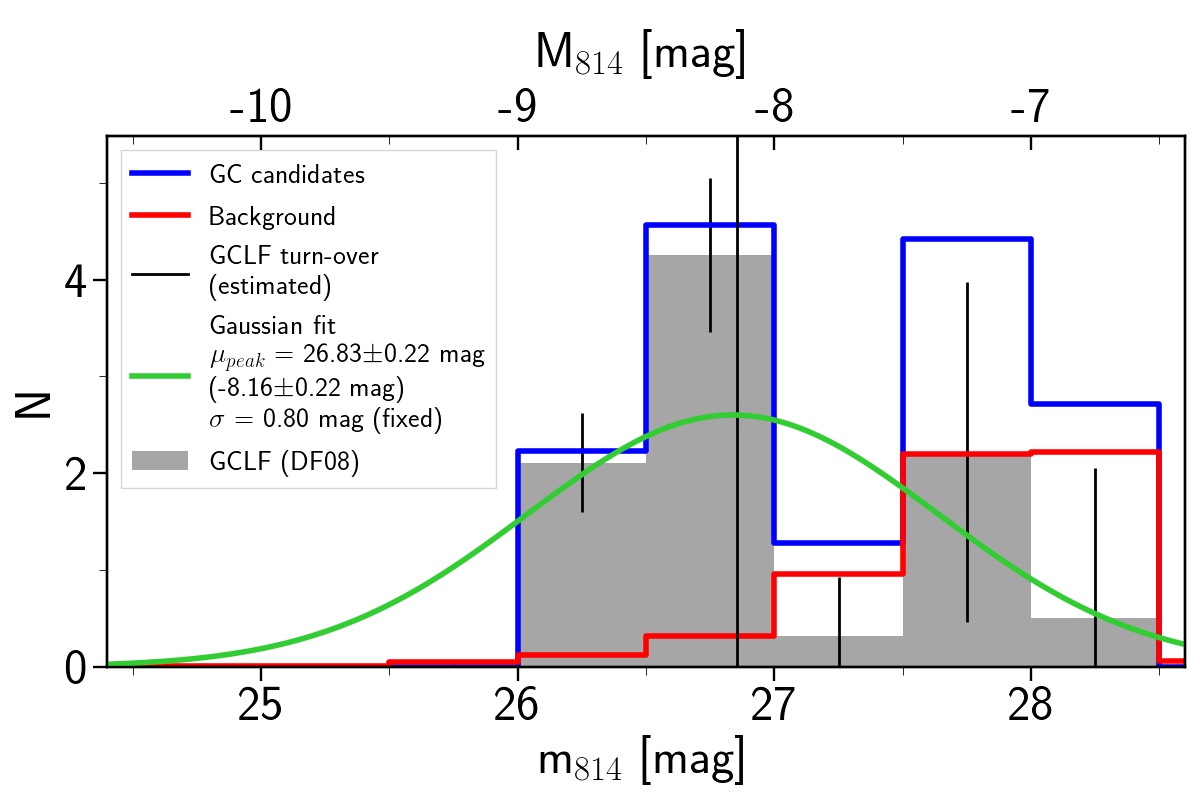}
\includegraphics[width=0.46\linewidth]{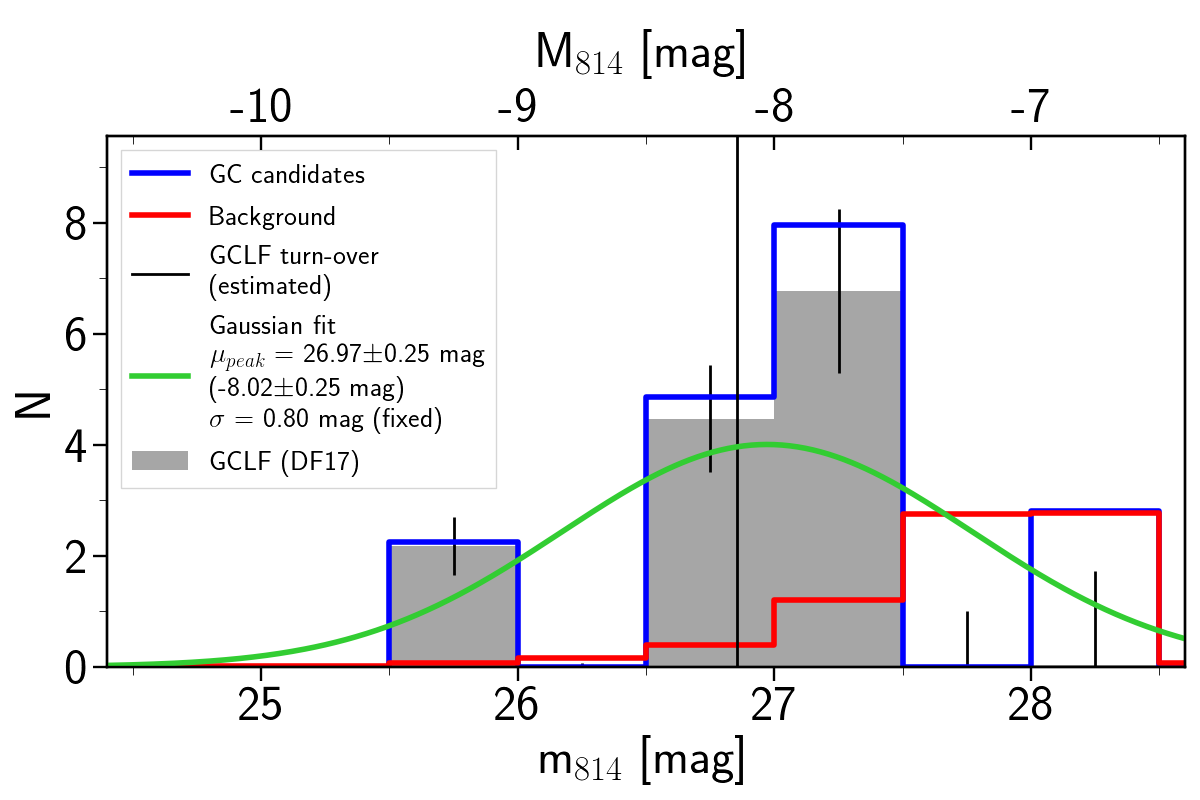}
\includegraphics[width=0.46\linewidth]{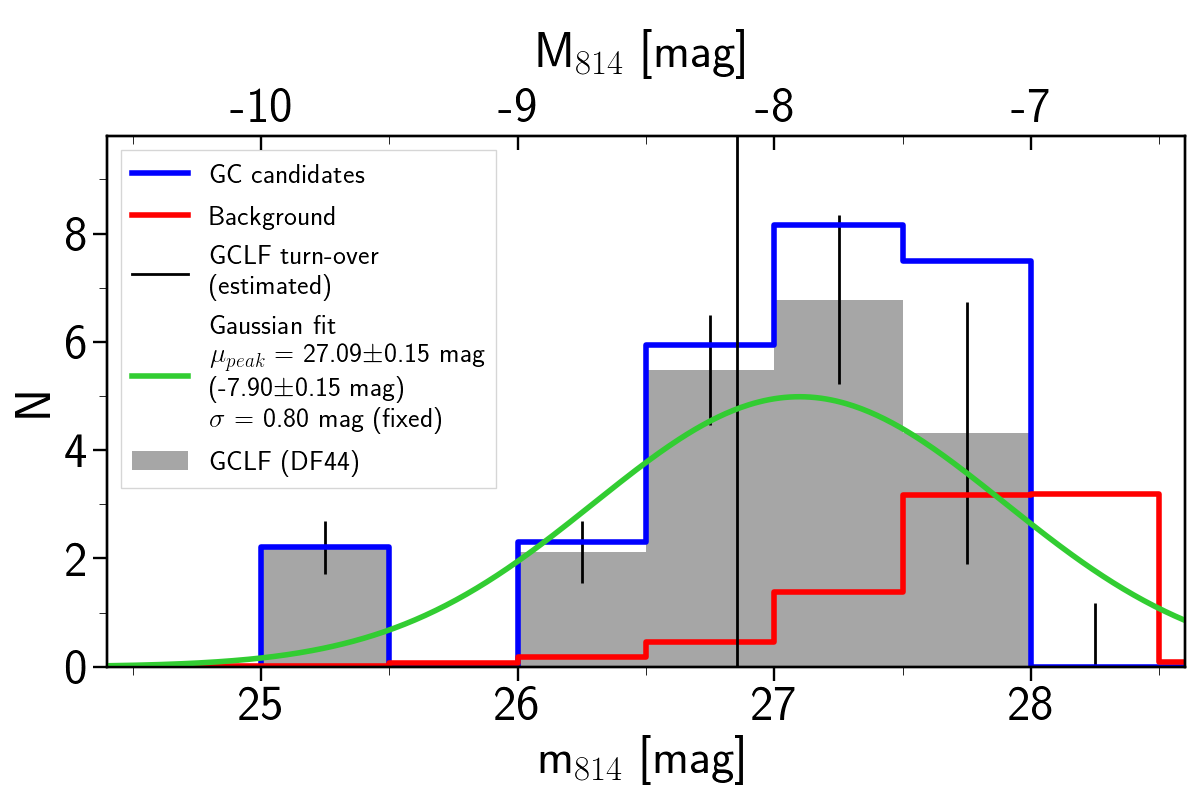}
\includegraphics[width=0.46\linewidth]{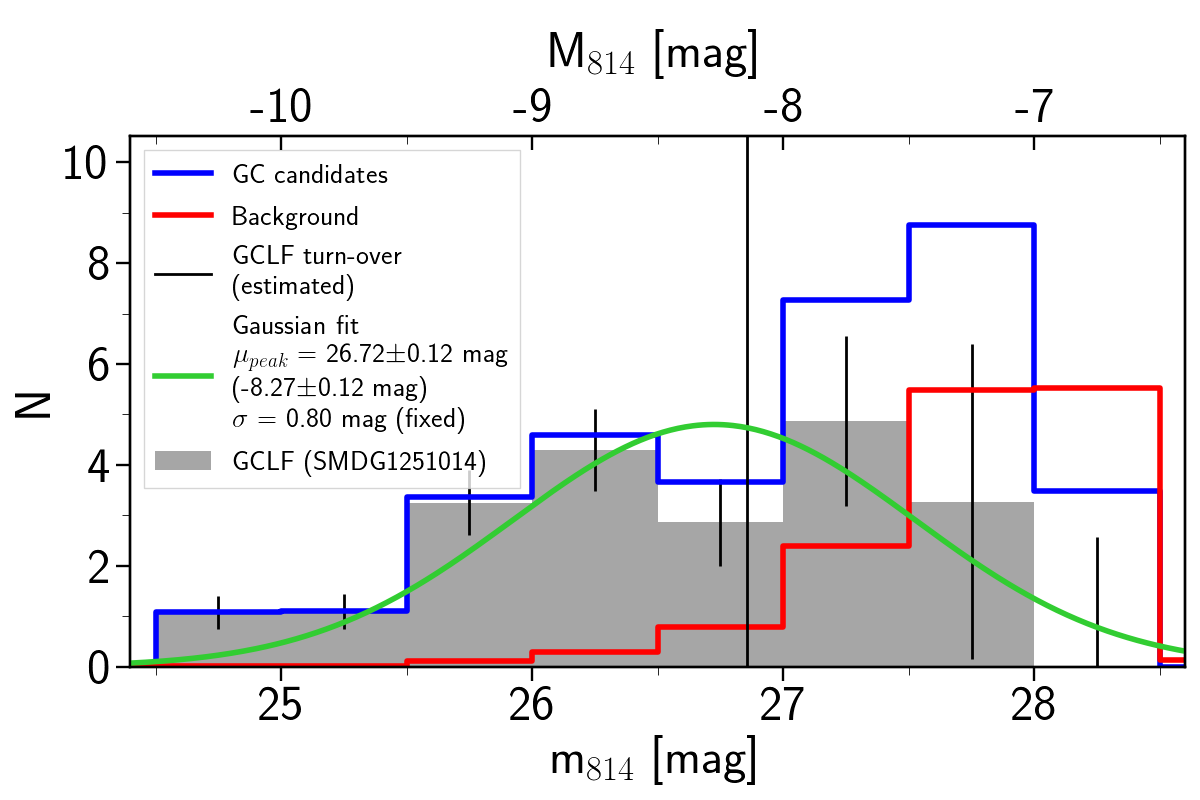}
\includegraphics[width=0.46\linewidth]{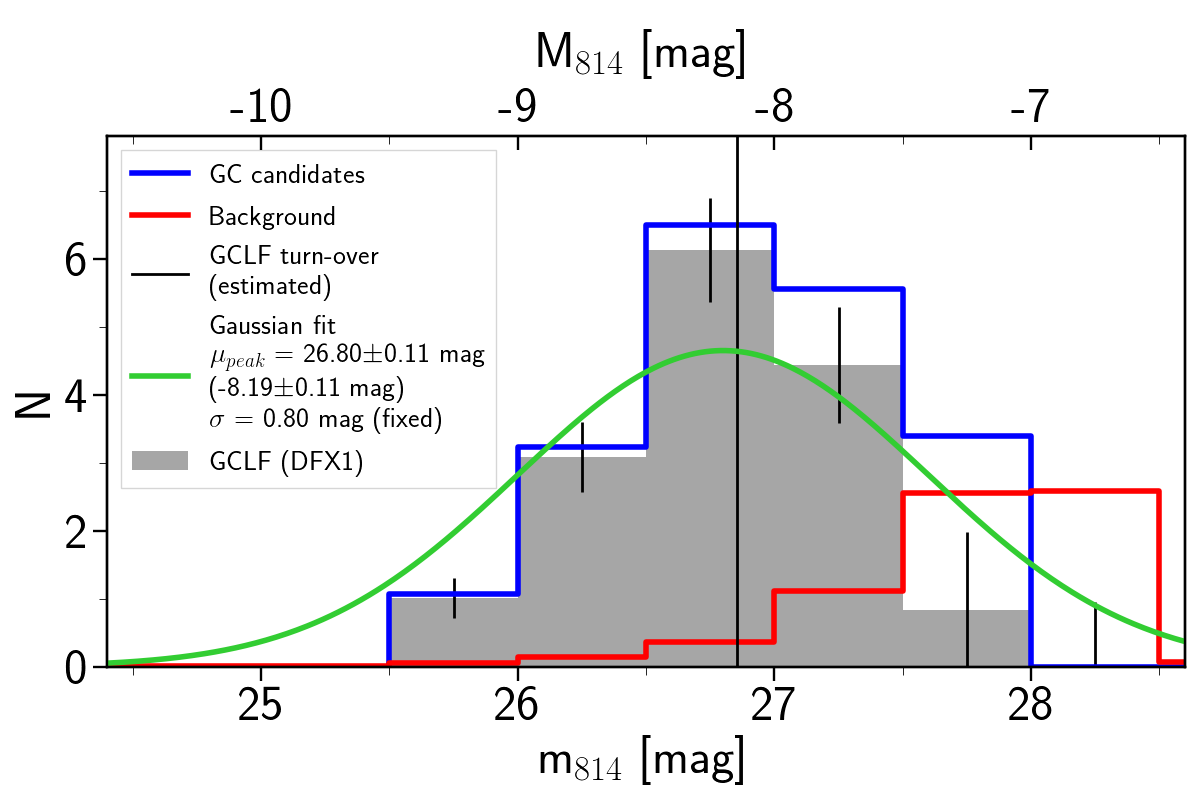}
\caption{As Fig.~\ref{gclf-combined}, but now showing the GCLF in $F814W$ for individual UDGs and the best-fit Gaussian function with a fixed $\sigma$\,=\,0.80\,mag.}
\label{gclfs}
\end{figure*}

\begin{figure*}
\centering
\includegraphics[width=0.405\linewidth]{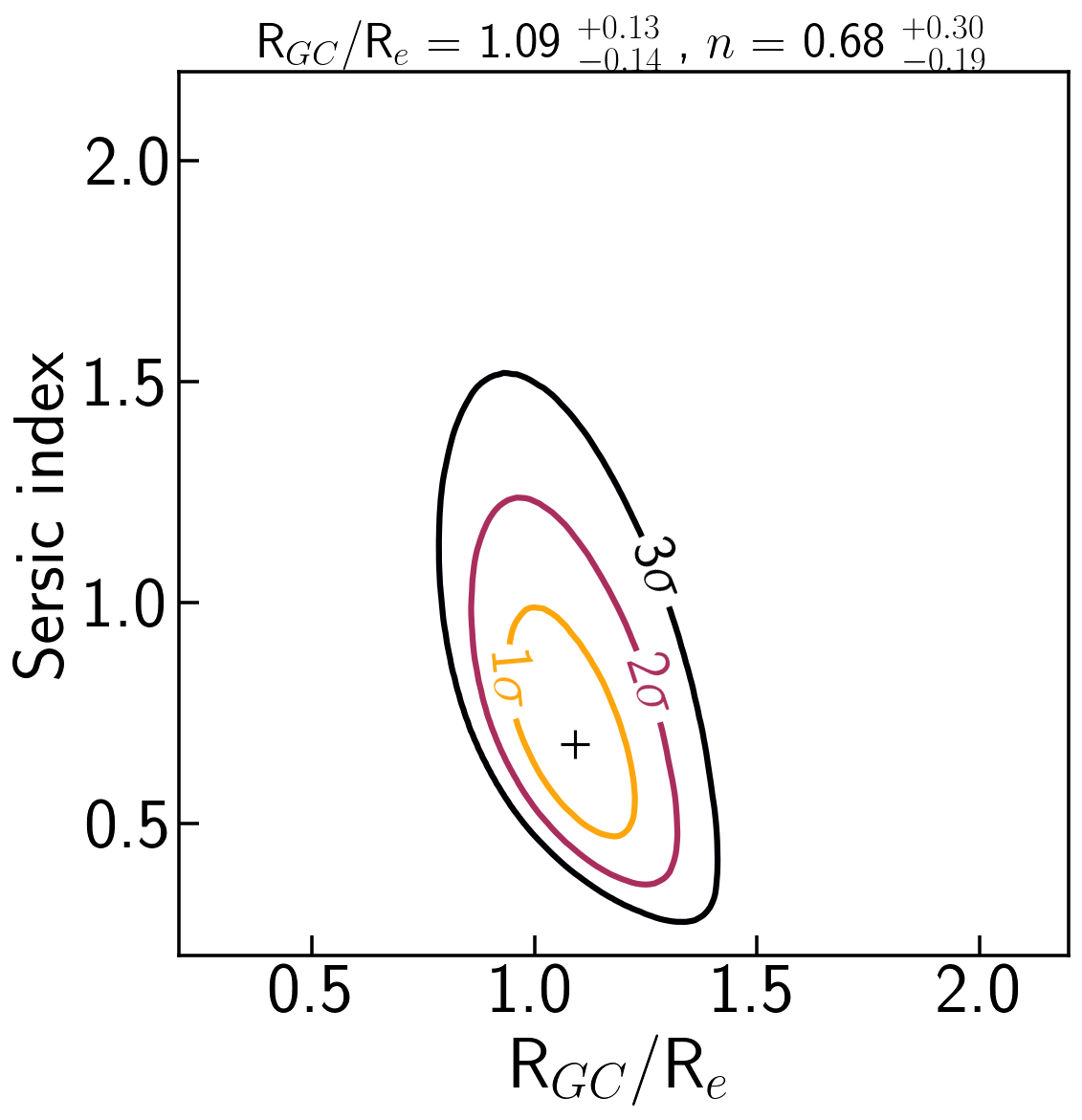}
\includegraphics[width=0.585\linewidth]{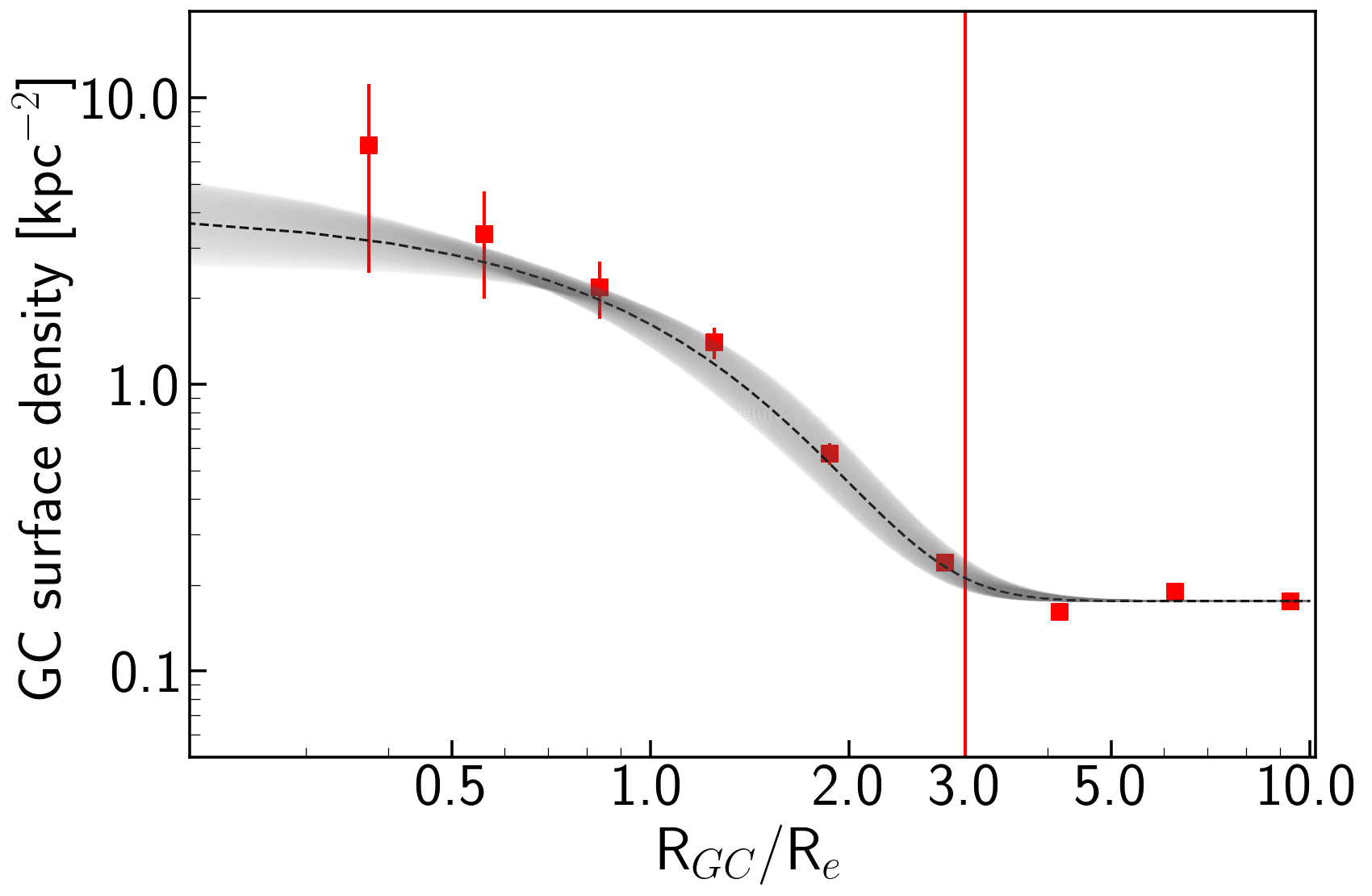}
\caption{Left: Likelihood maps for the derived S\'ersic parameters of the combined GC distribution. Right: The combined GC radial profile and the derived S\'ersic profile from MLE. The estimated profile (black dashed line) is not a fit to the binned data. The grey region indicates the 1$\sigma$ uncertainties in the estimated S\'ersic profile (based on our maximum likelihood estimator). The red vertical line indicates $R_{\rm GC}$\,=\,3$R_{\rm e}$, which is the radius-cut that is used for selecting GCs for estimating their S\'ersic parameters.}
\label{gc-radial}
\end{figure*}

\begin{figure}
\centering
\includegraphics[width=\linewidth]{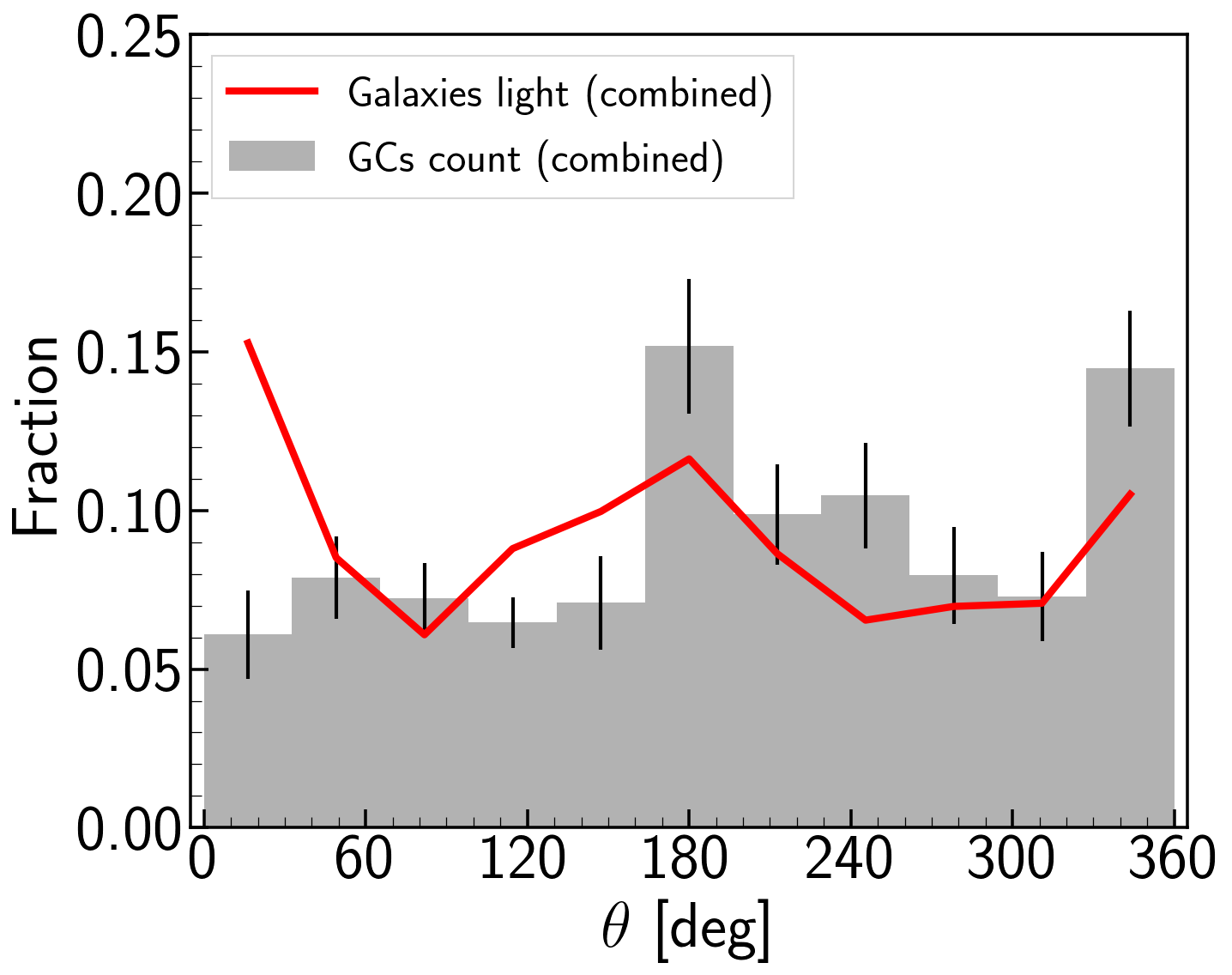}
\caption{The stacked azimuthal distribution of the globular clusters (grey) and galaxies field stars (red) relative to the galaxy's major axis. The peak in GC number around 0/360$^{\circ}$ and 180$^{\circ}$ corresponds closely to a peak in the galaxies light. GCs are thus well aligned with the stars.}
\label{gc-gal-az-dist}
\end{figure}

\begin{figure*}
\centering
\includegraphics[width=0.95\linewidth]{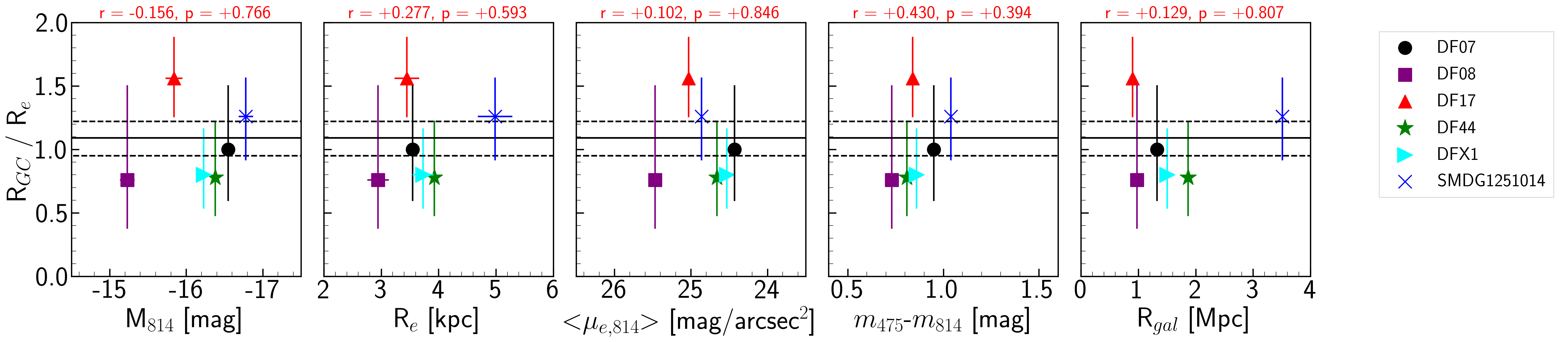}
\caption{The ratio between the GC half number radius $R_{\rm GC}$ and galaxy effective radius $R_{\rm e}$ for the UDG sample. The horizontal lines indicated the estimated value from combined GC distribution and its standard deviation. The $R_{\rm GC}$/$R_{\rm e}$ values do not show any strong correlation with UDG properties: (from left to right) absolute magnitude in $F814W$, effective radius. effective surface brightness in $F814W$, colour and clustercentric distance).}
\label{gc-params2}
\end{figure*}

\begin{figure}
\centering
\includegraphics[width=0.95\linewidth]{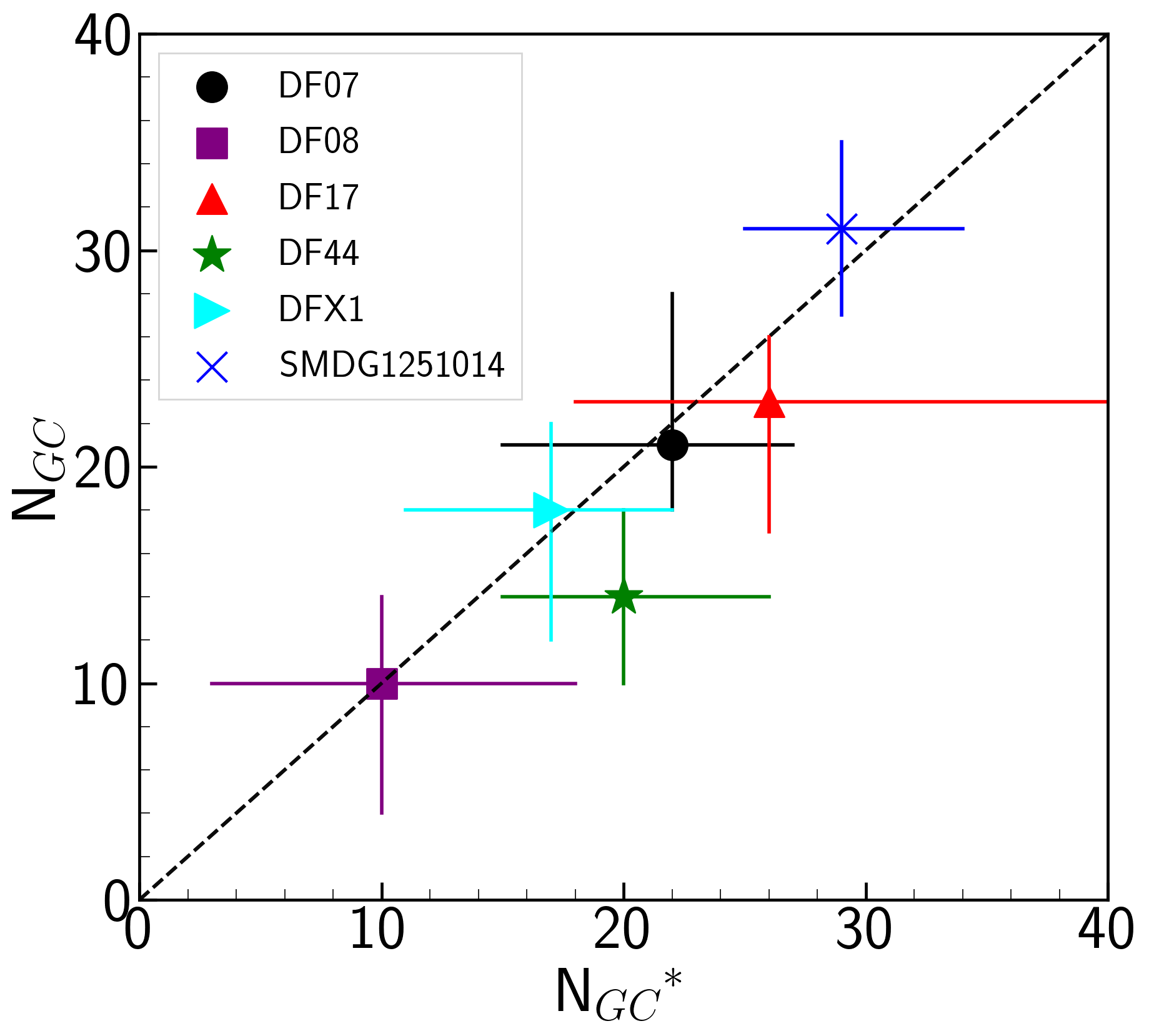}
\caption{Comparison between the number of GCs derived using individual $R_{\rm GC}$/$R_{\rm e}$ values ($N_{\rm GC}$) and combined $R_{\rm GC}$/$R_{\rm e}$ value ($N_{\rm GC}^*$). The dashed line indicates the one-to-one values of $N_{\rm GC}$ and $N_{\rm GC}^*$.}
\label{ngc-compare}
\end{figure}

\section{Results}
\label{results}

\begin{figure*}
\centering
\includegraphics[width=0.95\linewidth]{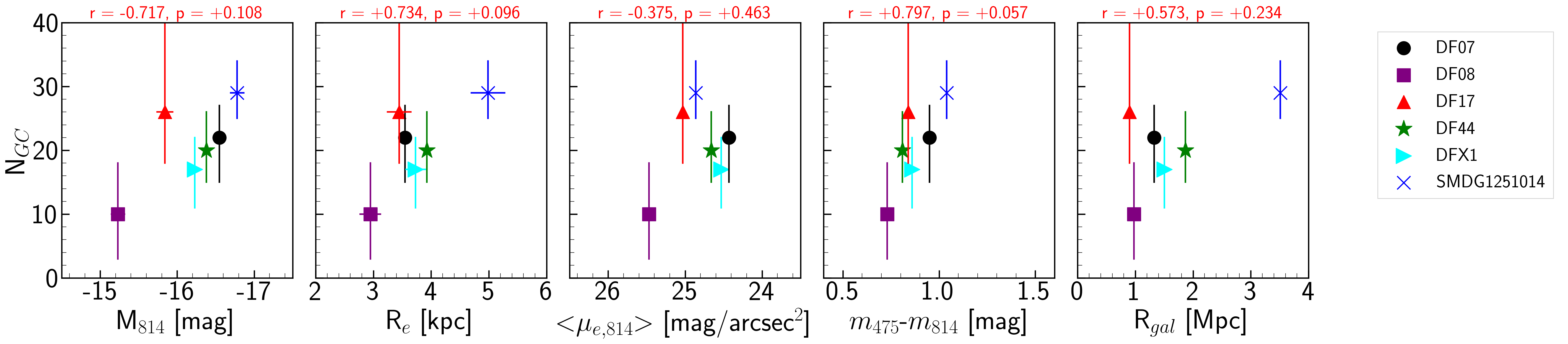}
\caption{$N_{\rm GC}$ versus galaxy property for our UDG sample (from left to right: absolute magnitude in $F814W$, effective radius. effective surface brightness in $F814W$, colour and clustercentric distance). The Pearson correlation coefficient, $r$, and $p$-values are included above each panel.}
\label{gc-params}
\end{figure*}

\begin{figure*}
\centering
\includegraphics[width=0.95\linewidth]{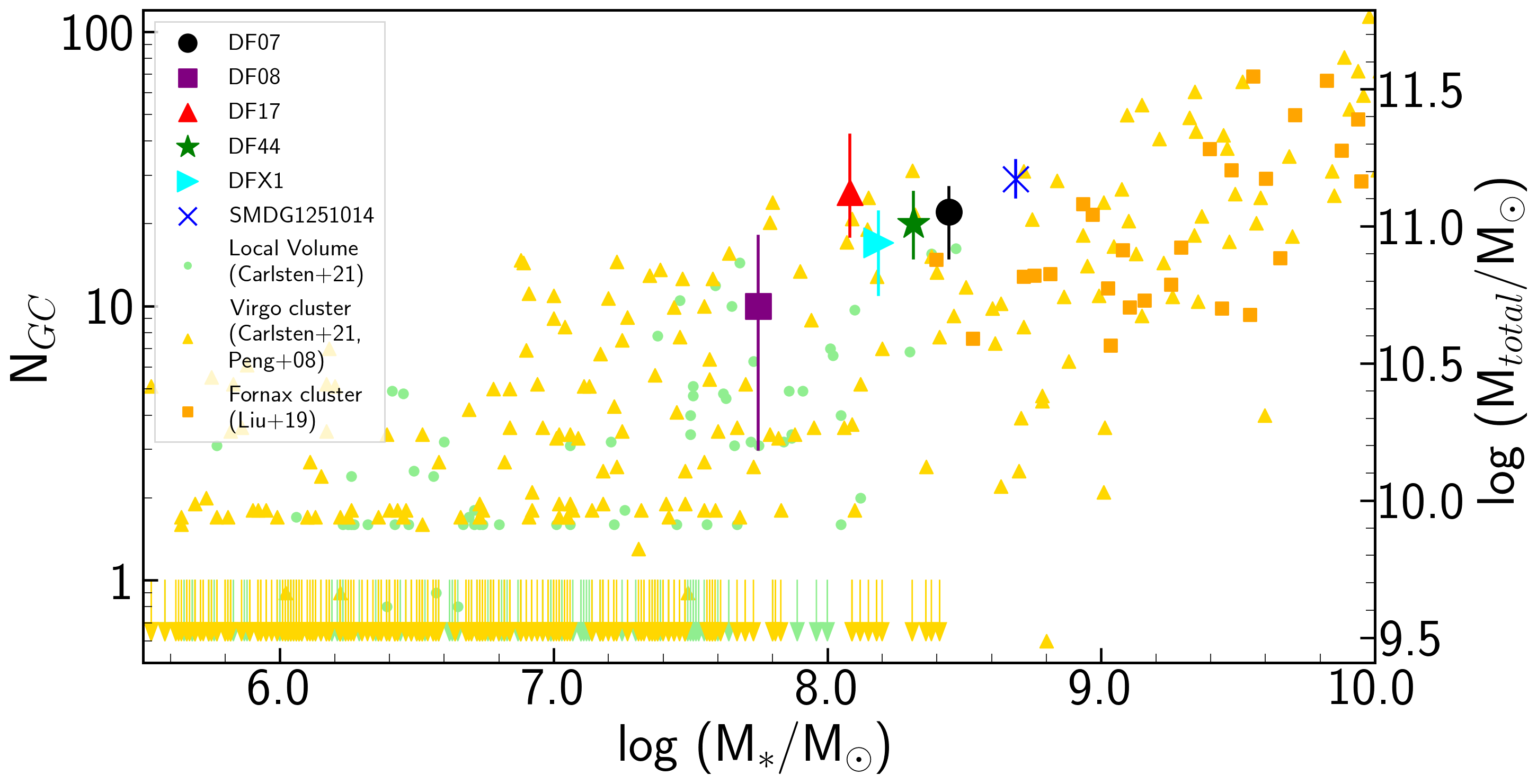}
\caption{$N_{\rm GC}$ versus the stellar mass, $M_*$, for our UDGs and for galaxies in the Virgo cluster (yellow triangles), Fornax cluster (orange squares) and galaxy groups in the local universe (green dots). Galaxies where $N_{\rm GC}$ is consistent with zero are shown with the downward arrows. On the right side of the diagram we present the corresponding total galaxy mass calculated using $M_{\rm total} = 5.12 \times 10^9 \times N_{\rm GC}$\,$M_{\odot}$. Most galaxies from the literature with masses log ($M_*$/$M_{\odot}$)\,<\,8.5 are dwarf early type galaxies and only five are considered to be UDGs. Stellar masses for the UDGs are calculated based on equations provided in \citealp{tom}. The Pearson correlation coefficient, r, and $p$-values are included above each panel.}
\label{ngc-m-star}
\end{figure*}

\begin{figure*}
\centering
\includegraphics[width=0.475\linewidth]{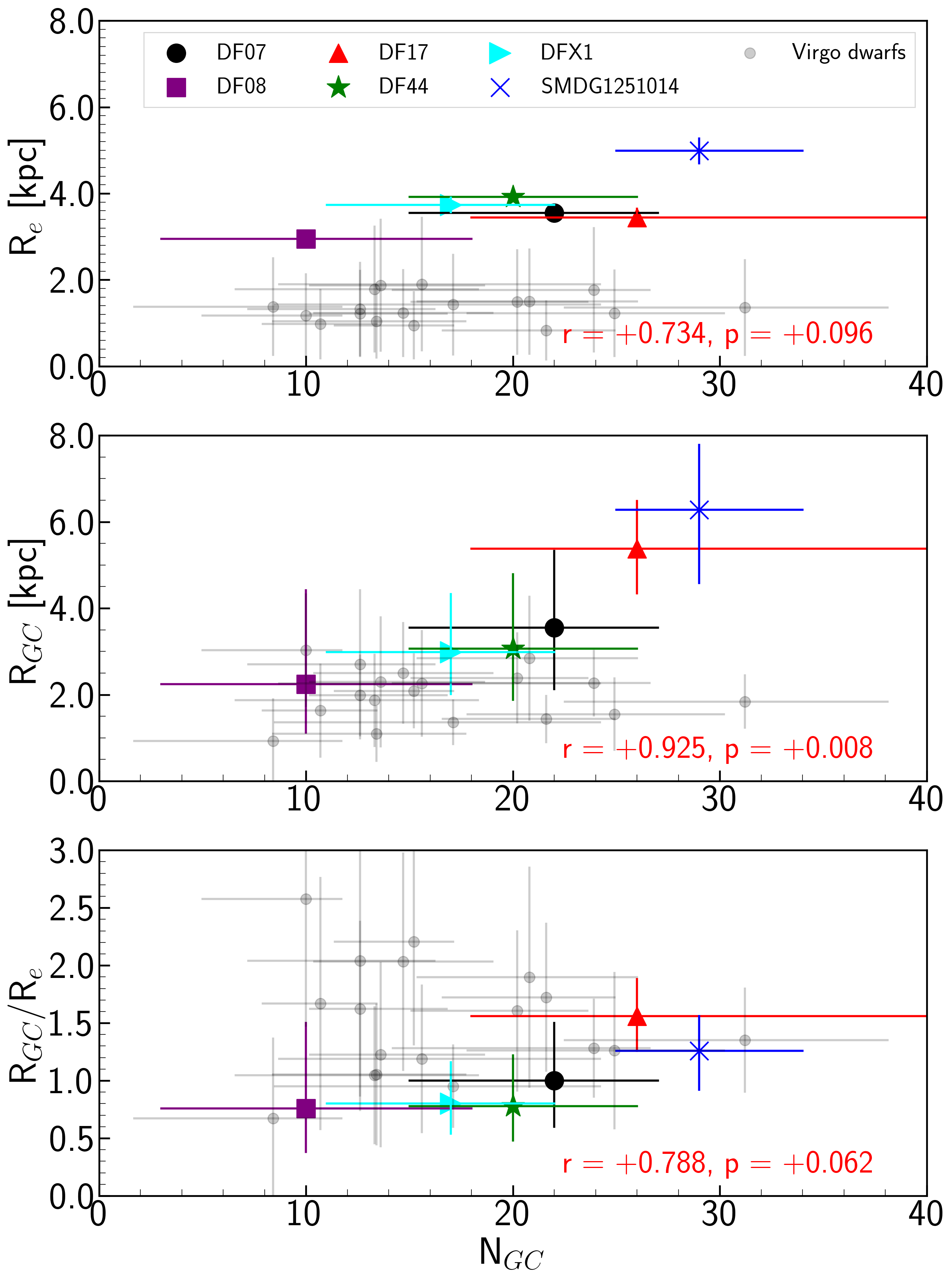}
\includegraphics[width=0.475\linewidth]{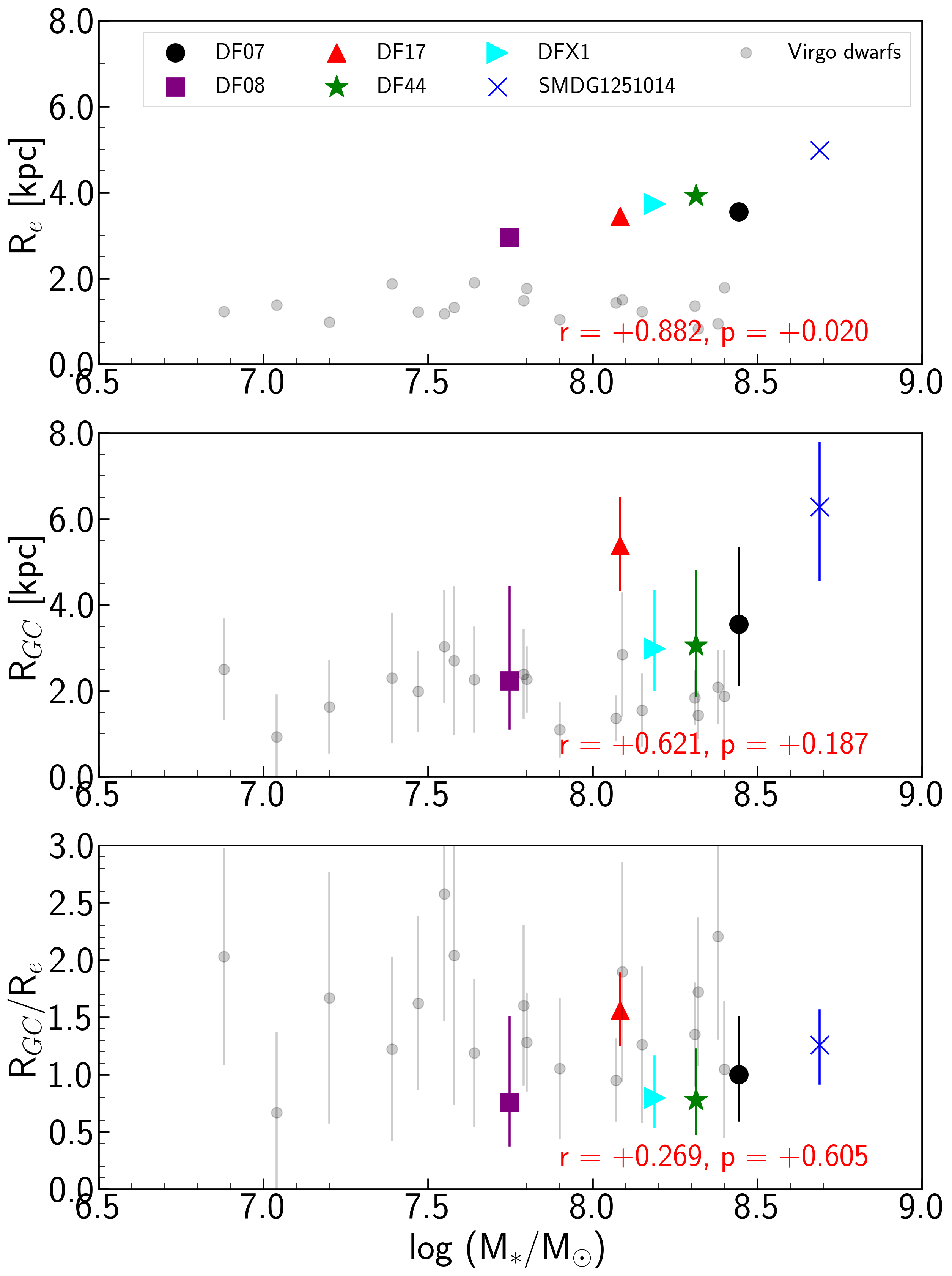}
\caption{The galaxies effective radius ($R_{\rm e}$, upper frames), GC half-number radius ($R_{\rm GC}$, middle frames) and their ratio ($R_{\rm GC}$/$R_{\rm e}$, lower panel) as a function of the GC total number (left) and the host galaxy stellar mass (right) for the UDGs in this work and the dwarf sample (\citealp{carlsten2021}). The Pearson correlation coefficient, $r$, and $p$-values are included within each panel.}
\label{ngc-rgc}
\end{figure*}

\begin{figure*}
\centering
\includegraphics[width=0.95\linewidth]{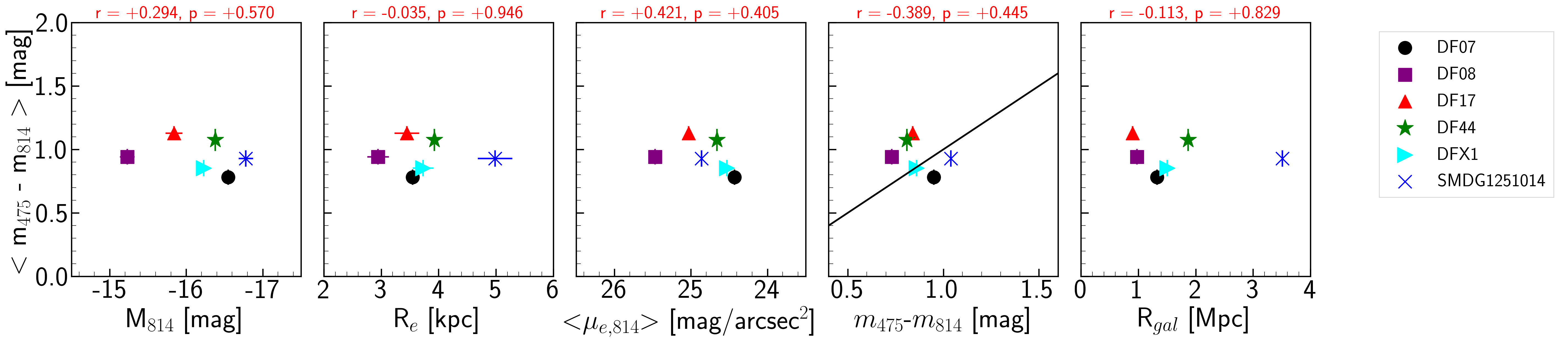}
\includegraphics[width=0.95\linewidth]{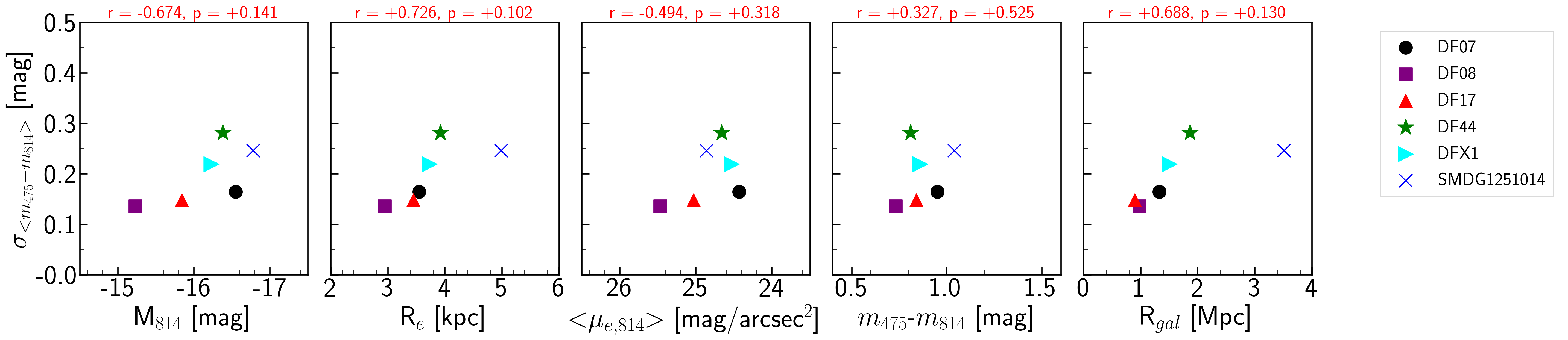}
\caption{GC average colour and colour spread versus the properties of host UDGs: (from left to right) absolute magnitude in $F814W$, effective radius. effective surface brightness in $F814W$, colour and clustercentric distance. For DFX1, the values are converted from $m_{606}-m_{814}$ to $m_{475}-m_{814}$ using the transformations in \citet{blak2010}. Galaxies below/above the diagonal line (second diagram from right) have bluer/redder GCs than their field stars. The Pearson correlation coefficient ($r$) and $p$-values are indicated on top of the diagram.}
\label{gc-params3}
\end{figure*}

We now use our GC candidate catalogues to explore the GC luminosity function, radial profile, total number, and colours. 

\subsection{GC Luminosity Function}
\label{gclf-cal}

The GCLF describes the number of GCs as a function of luminosity. Observations of GCs in other classes of galaxies show that this function, when expressed in terms of the GC magnitude, can be approximated by a Gaussian function (\citealp{secker}) described by the value of the mean, or turn-over, magnitude ($\mu_{\rm peak}$) and its width ($\sigma$). These parameters carry information on the formation history and dynamical evolution of GCs in galaxies, in this case for large UDGs. 
One puzzling aspect of GC populations is the near universality of the peak of the GCLF (\citealp{rejkuba2012}). Over the range of environments in which GCs are found one might expect both different formation conditions and significantly different rates of tidal disruption and dissolution.

Our large UDGs
are expected to host a few tens of GCs each of which about half of that many are brighter than the GCLF turn-over magnitude \citep{peng2016,beasley2016b,saifollahi2020}. Additionally, we are 75\,per\,cent complete near the peak of the GCLF, and so expect a fraction (more than 75\,per\,cent) of that half in our catalogues. Such low numbers render the resulting GCLFs and their parameters ($\mu_{\rm peak}$ and $\sigma$) highly uncertain for individual galaxies. Instead, we construct the combined GCLF and adopt its parameters for all our galaxies. To do this, we select GC candidates around each galaxy within an ellipse with a semi-major axis of 1.5$R_{\rm e}$ and the axis ratio and position angle of the host UDG. 
The value of 1.5$R_{\rm e}$ for the semi-major axis is a compromise between encompassing the entire GC system and minimizing contamination. The choice of axis ratio and position angle is supported by the expectation that (in-situ) GCs follow the azimuthal distribution of the field stars in galaxies (\citealp{kissler,kave,gomez1}) and we will return to this choice later (Section~\ref{gc-dist}). 

To correct statistically for remaining contamination, we estimate a combined background using selected GC candidates farther than 5$R_{\rm e}$ from each galaxy. We normalize by the area and subtract from our observed LF within 1.5$R_{\rm e}$. We then correct for incompleteness (source extraction in both bands and point source selection) to obtain our final estimate of the GCLF. In Fig.~\ref{gclf-combined} we present this combined GCLF for the six UDGs in our sample and the Gaussian fit. We find for $F814W$ that $\mu_{{\rm peak},{\rm 814}} = -8.14\pm$0.14\,mag 
and $\sigma_{814}$\,=\,0.79$\pm$0.06\,mag, which corresponds to $\mu_{{\rm peak},I,AB} = -8.10\pm0.14$\,mag\footnote{using $m_{\rm F814W, AB}$ = $m_{I, {\rm AB}} - 0.04$\,mag  (\citealp{wfc3})}.

Within the uncertainties, the turnover we measure is the same as that
observed for dwarf ellipticals ($\mu_{I,AB} = -8.1$ ;\citealp{miller}) and less than $3\sigma$ different from that found for more massive ellipticals ($\mu_{{\rm peak},I,AB} = -8.46$; \citealp{kundu}). Similarly, our measured value of $\sigma_{814} = -0.79\pm$0.06 is consistent with the typical value measured for dwarf elliptical galaxies ($\sigma$\,$<$\,1\,mag ;\cite{miller,harris2014}). In turn, this is different from what found in more massive galaxies ($\sigma$ $>$ 1.2 mag; \cite{harris2014}).

The stacked GCLF described here is representative of the GCs around UDGs with large effective radii. 
Because the completeness-corrected
estimates of the total number of GCs in each galaxy, $N_{\rm GC}$, that we present later (in Section~\ref{gc-number}) are sensitive to the turn-over magnitude, we now search for possible galaxy-to-galaxy variations by repeating the above exercise for each individual UDG, but adopting $\sigma$ from the composite GCLF. 
In Fig.~\ref{gclfs} we present the individual galaxy GCLFs and the estimated turn-over magnitude. We find no evidence for significant ($> 2\sigma$) variations in $\mu_{{\rm peak},{\rm 814}}$.

\subsection{GC Distribution}
\label{gc-dist}
Given the relatively small number of GCs per UDG, it is understandable that there have been only a few studies of the GC distribution in UDGs. One way to describe the radial distribution is with the GC half-number radius, $R_{\rm GC}$, which is the radius containing half of all of the GCs in that particular system. 
\citet{peng2016} estimated $R_{\rm GC}$\,=\,1.5\,$R_{\rm e}$ for DF17, which was later adopted in several studies to help determine the completeness corrections needed to estimate $N_{\rm GC}$ (\citealp{vd17,lim2018}). \citet{amorisco2018} found that, even in UDGs 
with abundant GC systems, $R_{\rm GC}/R_{\rm e}<2$ with high probability,
with several systems better described by $R_{\rm GC}/R_{\rm e}\lesssim1$. More recent estimates of $R_{\rm GC}$ for DF44\footnote{\citet{vd17} also estimated $R_{\rm GC}$ from a stacked GC distribution for DF44 and DFX1 and found $R_{\rm GC} = 2.2^{+1.3}_{-0.7}$\,$R_{\rm e}$, however given the large uncertainties in their estimation, they adopted the $R_{\rm GC}$ value from \citet{peng2016}.} \citep{saifollahi2020}, for MATLAS-2019 \citep{muller21} and for NGC1052-DF2 \citep{montes2} find $R_{\rm GC} \le$\,$R_{\rm e}$. 

Again, due to statistical uncertainties, we opt first to work with the composite sample. Here, we describe the radial profile of the combined GC sample using a S\'ersic model and measure both the GC half number radius $R_{\rm GC}$ and the S\'ersic index $n$. To combine the candidates from the different galaxies, we normalize the radial distances relative to the corresponding host effective radius. We select only the GC candidates within $R_{\rm GC}$/$R_{\rm e}$\,=\,3. We doubled the radial distance cut from that used earlier to have better leverage on the model fitting. 

To obtain the parameters, 
we apply a maximum likelihood estimator (MLE), as in \citet{saifollahi2020}. The MLE approach uses a large number of simulations (here 1,000) and in each run, it accounts for the background contamination by randomly excluding a number of GCs corresponding to the number of expected background sources from the background luminosity function. In Fig.~\ref{gc-radial} and
Table~\ref{gc-results-table} we present the derived S\'ersic parameters and radial profile of the combined GC sample. We find $R_{\rm GC}$/$R_{\rm e}$\,=\,1.09$^{+0.13}_{-0.14}$ and $n$\,=\,0.68$^{+0.30}_{-0.19}$. 
This value of $R_{\rm GC}$ is consistent with that found for local dwarf galaxies in groups and in the Virgo cluster ($R_{\rm GC}$\,=\,1.06$^{+0.21}_{-0.18}$\,$R_{\rm e}$ and $R_{\rm GC}$\,=\,1.25$^{+0.24}_{-0.18}$\,$R_{\rm e}$, respectively), as well as for galaxies from that sample selected to span the same stellar mass range \citep[7.8\,$<$\,log ($M_*$/$M_{\odot}$)\,$<$\,8.4; $R_{\rm GC}$\,=\,1.40$\pm$0.38 $R_{\rm e}$;][]{carlsten2021}. In Fig.~\ref{gc-params2} we compare the individual $R_{\rm GC}/R_{\rm e}$ values with a set of UDG properties and find no significant trends.

We also estimate $R_{\rm GC}$ for each UDG. For DF17, we find $R_{\rm GC}$\,=\,1.54$^{+0.28}_{-0.30}$\,$R_{\rm e}$, consistent with the value from \citet{peng2016}. For DF44, using a new dataset, we measure $R_{\rm GC}$\,=\,0.78$^{+0.44}_{-0.30}$\,$R_{\rm e}$, consistent with the reported value in \citet{saifollahi2020} ($R_{\rm GC}$\,=\,0.8$^{+0.3}_{-0.2}$\,$R_{\rm e}$).

In Fig.~\ref{gc-gal-az-dist} we inspect the azimuthal distribution of the composite GC candidate sample out to 1.5\,$R_{\rm e}$ relative to the distribution of stellar light. Measuring the azimuthal distribution relative to the position of the stellar major axis, the GC numbers clearly peak around 0$^{\circ}$ and 180$^{\circ}$, confirming the alignment of the two distributions that we had assumed in \S\ref{gclf-cal}.

\subsection{GC Number}
\label{gc-number}

The number of GCs in a galaxy ($N_{\rm GC}$) is known to scale more strongly with the total mass of the host galaxy ($M_{\rm total}$) (\citealp{harris2013}) than with the stellar mass. This somewhat surprising result can be recovered for massive galaxies with the total masses typically larger than 10$^{11}$\,$M_{\odot}$, at least, as the natural outcome of hierarchical formation \citep{BK17,elbadry,lucas}. Recent simulations incorporating GC formation in cosmological hydrodynamical simulations (E-MOSAICS, \citealp{kuij,Pfeffer-2018}) also reproduce the sense of the relation between $N_{\rm GC}$ and $M_{\rm total}$ (\citealp{bastian}) for galaxies more massive than 5\,$\times$\,10$^{11}$\,$M_{\odot}$. These total masses are similar or slightly larger the total mass of UDGs in our sample. However, among lower-mass galaxies, where the stellar mass is observed to drop precipitously, there is no corresponding observed drop in $N_{\rm GC}$ and the origin of this relation remains in question \citep{forbes2018}. As such, more measurements of $N_{\rm GC}$ in low stellar mass systems are needed to confirm the trend and explore possible physical sources of scatter. The relation between $N_{GC}$ and $M_{\rm total}$ has been assumed to hold for UDGs, although that is worth confirming, and has been utilized as an alternative to direct, but more observationally expensive, ways to estimate masses. We will also estimate the total masses using $N_{GC}$.

We measure $N_{GC}$ as follows.
First, we select the GC candidates within 3$R_{\rm GC}$ of each galaxy. We estimate that within this radius, for the mean S\'ersic index $n$\,=\,0.68 that describes the radial distribution of GCs in our UDGs, we capture more than 95\,per\,cent of all the GCs (\citealp{trujilo2001}). Then, because our data does not cover the full GCLF magnitude range, we assume that the GCLF is symmetric around $\mu_{\rm peak}$ and correct the measured $N_{\rm GC}$ up to $\mu_{\rm peak}$ for fainter GCs by doubling the number. We make two different estimates of $N_{\rm GC}$. In the first, which we indicate with $N_{\rm GC}$, we use our estimate of $R_{\rm GC}$/$R_{\rm e}$ for each individual galaxy, and in the other, which we refer to with $N_{\rm GC}^*$, we use the mean value of $R_{\rm GC}$/$R_{\rm e}$ estimated from the composite. We
find no difference between these two approaches within the uncertainties (Fig.~\ref{ngc-compare}). Therefore, we adopt our values $N_{\rm GC}$ in the following. In Table~\ref{gc-results-table} we present the final values of $R_{\rm GC}$/$R_{\rm e}$, $n$ and $N_{\rm GC}$ for each galaxy. The uncertainties in $N_{\rm GC}$ take into account the uncertainties in $R_{\rm GC}$, $\mu_{\rm peak}$ and the background number density.

Adopting the $N_{\rm GC}$-$M_{\rm total}$ relation, $M_{\rm total}$\,=\,5.12\,$\times$\,10$^9$\,$\times$\,$N_{\rm GC}$\,$M_{\odot}$ (drawn from \citealp{harris2017} relation between total GC population mass and halo mass, and adopting an average GC mass of 2\,$\times$\,10$^5$\,$M_{\odot}$), we find total masses for our UDGs between 0.51$^{+0.26}_{-0.41}$\,$\times$\,10$^{11}$ and 1.54$^{+0.26}_{-0.20}$\,$\times$\,10$^{11}$\,$M_{\odot}$ and a corresponding range of total to stellar mass ratios between 300 to 1,000. 

In Fig.~\ref{gc-params} we show the relations between $N_{\rm GC}$ and other UDG properties. As expected, the number of GCs increases with the luminosity of the galaxy, as a proxy for its stellar mass. The same increasing trend is observed between $N_{\rm GC}$ and other galaxy parameters, which can be the outcome of the existing scaling relations between the properties of UDG (such as that between $R_{\rm e}$ and total mass (\citealp{dennis}). 

When the UDG $N_{\rm GC}$ values are compared to those of other galaxies, the UDGs show systematically twice larger $N_{\rm GC}$ at a given stellar mass (Fig.~\ref{ngc-m-star}). The galaxy sample in Fig.~\ref{ngc-m-star} includes galaxies in the Virgo cluster (\citealp{peng2008,carlsten2021}), the Fornax cluster (ACSFCS, \citealp{liu2019}) and galaxy groups in the local volume (\citealp{carlsten2021}). Our UDGs are in the Coma cluster, which is more massive than the other two clusters, and the relative increase in $N_{\rm GC}$ at a given stellar mass may be an indication of environmental effects in GC formation or UDG evolution \citep{peng2008,carleton2021,carlsten2021}. It is noteworthy that among the best-studied UDGs in term of their GCs, the UDG known as MATLAS-2019 (\citealp{matlas}) with $R_{\rm e}$\,=\,2.2\,kpc (smaller effective radius than our UDGs) shows a large number of GCs for its stellar mass (\citealp{muller21,matlas-danieli}). Although it constitutes a single example, it lies in a low-density environment and is therefore a counterexample to the possible trend with environment.

In the middle panel of Fig.~\ref{ngc-rgc} we show a statistically significant positive correlation between $R_{\rm GC}$ and $N_{\rm GC}$ for the six UDGs in our sample (coloured points), suggesting that more massive haloes (with more GCs) have larger $R_{\rm GC}$. The trend is still there, but with less statistical significance after normalizing $R_{\rm GC}$ by $R_{\rm e}$, suggesting that some of the effect in the top panel is driven simply by the overall size of the system. Nevertheless, some indication remains that even after accounting for the differences in $R_{\rm e}$, the distribution of globular clusters is more extended in the more massive galaxies. Furthermore, Fig.~\ref{ngc-rgc} compares these UDGs to a sample of dwarf galaxies in the Virgo cluster (\citealp{carlsten2021}\footnote{The sample that we used here consists of those 18 dwarf galaxies for which $R_{\rm GC}$ has been estimated by the authors. Therefore this sample does not include all of the Virgo cluster dwarfs galaxies in \citet{carlsten2021}, but the ones with enough GCs that $R_{\rm GC}$ could be estimated.}). We exclude SMDG1251014 and DF17 from this comparison because the dwarf sample does not provide enough points for their total mass (based on $N_{\rm GC}$). The four remaining UDGs (DF07, DF08, DF44 and DFX1) have $R_{\rm e}$ on average 3 times larger than the average value for dwarfs of the same (total) mass while their $R_{\rm GC}$ is only 1.5 times higher which leads to a lower $R_{\rm GC}$/$R_{\rm e}$ for these UDGs. The implications of this result are discussed in Section~\ref{discussion}.


\subsection{GC Colour}

GCs around massive galaxies show complex colour distributions that are mostly attributed to metallicity differences (\citealp{brodie}). It is hypothesized that massive galaxies host both in-situ metal-rich/red GCs and ex-situ metal-poor/blue GCs. The latter formed in lower-mass galaxies, where the bluer colour is attributed to lower metallicity and accreted GCs. One of the implications of this scenario is a correlation between GC average colour and galaxy luminosity (\citealp{peng2006}). Finally, the GC colour distributions and a comparison to the properties of the UDG main body may provide clues on star/GC formation episodes within UDGs.

We measure the GC average colour ($m_{475}-m_{814}$) and colour spread ($\sigma_{m_{475}-m_{814}}$) using GCs within 1.5\,$R_{\rm e}$ (Table~\ref{gc-results-table}) and compare those quantities to the UDG properties in Fig.~\ref{gc-params3}. The average GC colours cover a range between 0.78\,mag and 1.13\,mag with an average value of 0.95\,mag. This value is consistent with the average colour of GCs around lower luminosity galaxies (\citealp{peng2006}). As evident from Fig.~\ref{gc-params3}, these colours do not show any evident correlation with UDG properties. Furthermore, in one panel of this figure (second panel from right) we compare GC colour to galaxy host colour (top, second plot from right), where the diagonal line separates the UDGs with redder and bluer GC population than their corresponding field stars. We do not see any obvious pattern and some are slightly redder than their host UDG while some are slighty bluer. GC colour spreads range between 0.1\,mag to 0.3\,mag and increase for the brighter UDGs.

\begin{table*}
\centering
\caption{The derived properties of the GCs around the UDGs in the sample. Columns from left to right represent galaxy name (a), the ratio between GC half-number radius and galaxy's half light radius (b), the S\'ersic index of the GC radial profile (c), GC number count (d), average colour and colour spread of GCs (e, f) and the derived total mass using $N_{\rm GC}$ (g). For DFX1, the average colour is converted from $m_{606}-m_{814}$ to $m_{475}-m_{814}$ using transformations in \citet{blak2010}.}
\begin{tabular}{ lcccccc } \hline  
Galaxy & $R_{\rm GC}$/$R_{\rm e}$ & $n$ & $N_{\rm GC}$ & $<$ $m_{475}-m_{814}$ $>$ & $\sigma_{<m_{475}-m_{814}>}$ & $M_{\rm total}$ \\ 
- & - & - & - & mag & mag &\,$M_{\odot}$\\ 
(a) & (b) & (c) & (d) & (e) & (f) & (g) \\
\hline
DF07 & 0.98$^{+0.50}_{-0.38}$ & 1.12$^{+1.86}_{-0.88}$ & 22$^{+5}_{-7}$ &  0.78$\pm$0.05 & 0.16 & 1.12$^{+0.26}_{-0.36}$ x 10$^{11}$\\\\

DF08 & 0.68$^{+0.52}_{-0.38}$ &	0.36$^{+3.50}_{-0.16}$ & 10$^{+5}_{-8}$ &  0.94$\pm$0.05 & 0.14 & 0.51$^{+0.26}_{-0.41}$ x 10$^{11}$\\\\

DF17 & 1.54$^{0.28}_{-0.30}$ & 0.20$^{+0.34}_{0.00}$ & 26$^{+17}_{-7}$ &  1.13$\pm$0.05 & 0.15 & 1.33$^{+0.87}_{-0.36}$ x 10$^{11}$\\\\

DF44 & 0.78$^{+0.44}_{-0.30}$ & 0.94$^{+1.74}_{-0.54}$ & 20$^{+6}_{-5}$ & 1.07$\pm$ 0.08 & 0.28 & 1.02$^{+0.31}_{-0.26}$ x 10$^{11}$\\\\

SMDG1251014 & 1.28$^{+0.32}_{-0.32}$ & 0.38$^{+0.72}_{-0.18}$ & 30$^{+5}_{-4}$ &  0.93$\pm$0.06 & 0.25 & 1.54$^{+0.26}_{-0.20}$ x 10$^{11}$\\\\

DFX1 & 0.8$^{+0.34}_{-0.26}$ & 1.06$^{+1.20}_{-0.54}$ & 17$^{+5}_{-6}$ & 0.86$\pm$0.06 & 0.22 &0.87$^{+0.26}_{-0.41}$ x 10$^{11}$\\
\hline 
\end{tabular}
\begin{flushleft}
\end{flushleft}
\label{gc-results-table}
\end{table*}

\section{Discussion}
\label{discussion}

We aim to answer three questions regarding these large effective radius UDGs : \\\\

\noindent
\textit{(i) Do UDGs with {$L^*$-like} haloes exist?} \\\textit{(ii) Are GCs in UDGs similar to those in other galaxies?} \\\textit{(iii) Can GC properties constrain UDG formation models?}\\

\subsection{Do UDGs with {$L^*$-like} haloes exist?}

The first studies on UDGs speculated that UDGs might be hosted by $L^*$-like (e.g. like that of the Milky Way) dark matter halos of mass $\sim$\,10$^{12}$\,$M_{\odot}$ (\citealp{vd15,koda,yozin,vd16,vd17}). This idea has lost appeal as dynamical mass measurements demonstrated that the most massive UDGs have masses of at most a few $\times 10^{11}$\,$M_{\odot}$ (\citealp{beasley2016,toloba,amorisco2018,vd19}), placing them among massive dwarf galaxies like the LMC. 

Among known UDGs, DF44 is an exceptional case and the original one thought to possibly have an $L^*$-like halo, in part because the initial estimates found $N_{\rm GC}$\,=\,94$^{+25}_{-20}$ (\citealp{vd16}) and $N_{\rm GC}$\,=\,74$\pm$18 (\citealp{vd17}). Subsequent kinematic mass estimates based on spatially resolved kinematics \citep[$M_{\rm halo}$\,=\,1.6$^{+5.0}_{-1.2}$\,$\times$\,10$^{11}$\,$M_{\odot}$;][] {vd19}, $N_{\rm GC}$ measurements \citep[$N_{\rm GC}=21^{+7}_{-9}$;][]{saifollahi2020} and constraints from X-ray data (\citealp{bogdan}) have helped push the mass estimates downward. We now extend the $N_{GC}$ measurements to an additional 5 UDGs with properties similar to those of DF44 and confirm that these are consistent with being in halos of $\sim 10^{11}$ M$_\odot$ rather than $\sim 10^{12}$ M$_\odot$. In agreement with the recent consensus, we conclude that we do not find evidence for physically large UDGs inhabiting $L_*$-like halos. 

\subsection{Are GCs in UDGs similar to those in other galaxies?}

We find that the GC numbers, the GCLF ($\mu_{\rm peak}$ and width $\sigma$), GC radial distribution, and GC average colour in our UDGs are consistent with the values derived for GCs in dwarf galaxies. 
Among these, 
$N_{\rm GC}$ has received the most attention because some studies claim that UDGs host more GCs than other galaxies of the same stellar mass (\citealp{vd17,lim2018,lim2020}, but see also \citealp{amorisco2018,prole2018}). Given the heterogeneity of the UDG population, all the previous conclusions can be correct. 

We have focused on UDGs with large $R_{\rm e}$. These are thought to be the most massive \citep{dennis} and have, at least in a couple of cases, been shown to host significant numbers of GCs. With a sample of six such UDGs, we can investigate whether any differences are systematic among UDGs or susceptible to object-to-object variation. Our images are deeper than the above-mentioned work, which allows us to achieve lower measurement uncertainties. As we show in Fig.~\ref{ngc-m-star}, these UDGs are at the upper end of the $N_{\rm GC}$ range for galaxies of the same stellar mass. We find that $N_{\rm GC}$ for these UDGs is roughly twice that of the average value for galaxies of the same stellar mass. Nonetheless, they are well within the global distribution considering the scatter of the relation. As noted by \citet{carleton2021}, whose model attributes such an offset not to an overabundance of GCs but rather to a paucity of stars in UDGs, there are different ways to view the result. Because we do not have independent measurements of $M_{\rm total}$ for the full sample, we cannot yet determine if these systems are overabundant in GCs or deficient in stars. Indications from DF44 suggest the latter.

\subsection{Can GC properties constrain UDG formation models?}

\begin{table*}
\centering
\caption{The implications of the observed GC properties for the proposed formation models. Here we discuss implications of two observed GC properties for the UDG formation models in the literature: i. the excessive GC number of UDGs compared to dwarf galaxies, ii. the similarity between the radial profile of GCs and the fields stars of UDGs ($R_{\rm GC}$/$R_{\rm e}$) (\textcolor{green}{\checkmark}: observable favours this model, \textcolor{red}{x}: observable disfavours this model).}
\begin{tabular}{lcccccc} \hline  
Observable & Failed dwarf galaxy & Tidal interactions & Stellar feedback & High-spin dwarfs & Lack of mergers \\
\hline

i. Excessive GC number &  \Large \textcolor{green}{\checkmark}& \Large \textcolor{red}{x} & \Large \textcolor{red}{x} & \Large \textcolor{red}{x} & \Large \textcolor{green}{\checkmark}\\\\

ii. GC radial profile & \Large \textcolor{green}{\checkmark} & \Large \textcolor{red}{x}  & \Large \textcolor{red}{x}  & \Large \textcolor{red}{x}  & \Large \textcolor{green}{\checkmark}\\

\hline 
\end{tabular}
\label{formation}
\end{table*}

Several UDG formation models are proposed to explain the current appearance of these galaxies. These models commonly describe UDGs as dwarf galaxies that undergo various processes that shape the current properties of UDGs. Until now, all the proposed models agree that cluster UDGs are quenched due to environmental effects taking place in galaxy groups and galaxy clusters. These models provide different explanations for the low surface brightness and large effective radius. The properties of GCs, as studied here, provide another set of observables to constrain the proposed models.

The elevated $N_{\rm GC}$/$M_*$ of UDGs argues against models that seek to explain the low surface brightness nature of UDGs primarily by redistributing the stars to larger radii ("high-spin", "tidal interaction", "stellar feedback" models\footnote{Here and in Table~\ref{formation}, the "stellar feedback" model only refers to the model proposed in \citet{cintio2017}.}). \citet{alexa} made a similar conclusion based on the stellar population parameters of DF44. These results are independent of the total mass of the galaxy.

To the degree that such models also affect the star formation efficiency, for example, the "high-spin" model also presumably leads to lower gas densities and lower star formation rates, the models may be salvageable. The $N_{\rm GC}$ values suggest that unless a model specifically predicts a greater GC formation efficiency, then it must account for a lower than average star formation efficiency. 
However, such modifications to the mean star formation efficiency must occur without affecting the GC formation efficiency. This suggests that whatever is responsible acts after the bulk of GC formation. Additionally, the astrophysical processes involved in these models (i.e. dynamical heating in the "tidal interaction" model) have a similar impact on the distribution of stars and GCs. In this case, it is expected that the star and the GC distribution within galaxies expand ($R_{\rm e}$ and $R_{\rm GC}$ increase) similarly and therefore, the ratio between them ($R_{\rm GC}$/$R_{\rm e}$) remain the same. Therefore, the smaller $R_{\rm GC}$/$R_{\rm e}$ of our UDGs compared to the $R_{\rm GC}$/$R_{\rm e}$ in dwarf galaxies of similar total mass (based on their $N_{\rm GC}$) disfavours these models. 

Furthermore, models that allow for star formation after some critical event, such as a merger or strong feedback effect, might face challenges in maintaining the close correspondence of $R_{\rm GC}$ and $R_{\rm e}$, as well as $R_{\rm GC}$/$R_{\rm e}$ and the azimuthal alignment. These considerations place additional constraints on these models. Moreover, a slightly larger $R_{\rm GC}$ (about 1.5 times) of UDGs compared to the dwarf sample can be the indication of a secondary process or a more complicated model for UDG formation.

In Table~\ref{formation} we have catalogued our understanding of how the models fare against these constraints, but caution the reader that specific predictions from the models are not yet available and we have taken our best guess. In addition, in some cases the models can probably be easily modified to address perceived challenges.

\section{Conclusions}
\label{summary}

We present an analysis of six ultra-diffuse galaxies (UDG) and their globular clusters (GCs) to study the nature of GCs in UDGs and provide additional constraints on formation models of UDGs. The galaxy sample is comprised of UDGs with large effective radii ($\langle$$R_e\rangle$\,$\sim$\,3.6\,kpc) in the Coma galaxy cluster. We combine deep new and archival \textit{HST} observations to identify GCs up to and fainter than the turnover of the GC luminosity function (M$_I$ $\sim -$8.1\,mag). We find that:

\begin{itemize}
    \item The GC luminosity function (GCLF) is consistent with that of similar galaxies, with a turnover magnitude, $\mu_{{\rm peak},{\rm 814}}$, of $-8.14\pm 0.14$\,mag and a width, $\sigma_{814}$, of $0.79\pm$0.06. The nature of the luminosity function is important to establish because the total GC count includes a correction for those GCs that are too faint to detect. 
    \\
    \item The GC distribution, in both radius and azimuth, is consistent with that of the underlying stars. We find that the half number radius for GCs relative to the half light radius, $R_{\rm GC}$/$R_{\rm e}$, is 1.09$^{+0.13}_{-0.14}$ and that the GC systems are also extended along the galaxy's major axis. Understanding the distribution of GCs is critical because the total GC count includes a correction for those GCs that are at large radii and difficult to distinguish from contaminants. 
    \\
    \item The number of GCs in each galaxy, $N_{\rm GC}$, spans 10 to 30, with an average of\,$\sim$\,21\,GCs for each of our galaxies. We find no statistically convincing trend between UDG properties and $N_{\rm GC}$, although this is not surprising because our sample was specifically selected to be composed of similar galaxies. We do find that all six galaxies lie at the upper end of the $N_{\rm GC}$ range for galaxies of similar stellar mass (about a factor of two larger than the mean) and well within the scatter of galaxies with the same stellar mass. 
    \\
    \item Adopting a relation between $N_{\rm GC}$ and total mass, $M_{\rm total}$, we find that our UDGs have $M_{\rm total} \sim 10^{11}$\,$M_{\odot}$ and that they have large $M_{\rm total}$/$M_*$. The latter implies that these galaxies are relatively inefficient at forming field stars, while forming GCs at the standard rate. Confirming the $N_{\rm GC}$-$M_{\rm total}$ relationship with additional kinematic mass measurements is a priority. 
    \\
    \item The lower ratio $R_{\rm GC}$/$R_{\rm e}$ that we find for these UDGs ($\sim$\,1) compared to dwarf galaxies ($\sim$\,1.5) suggests that the process that is responsible for the larger $R_{\rm e}$ of UDGs does not have the same influence on the GC distribution: either it has a weaker effect or no effect. In the case of the latter, a secondary process might play a role in increasing the $R_{\rm GC}$ of UDGs.
    \\
    \item These findings disfavour UDG models that appeal primarily to a redistribution of the stars to larger radii to reduce a galaxy's surface brightness and suggest that a decline in integrated star formation efficiency is needed. However, that decline should occur after GC formation, because the GC formation efficiency appears to have been normal. This would disfavour 'intrinsic' property models, such as that appealing to high-spin. Complex histories, with multiple star formation or dynamical events may be difficult to reconcile with the close spatial alignment of GCs and field stars in UDGs without invoking some fine tuning.
\end{itemize}

We look forward both to enlarging the sample of UDGs with high-fidelity GC measurements with which to test and extend the results presented here, and to comparisons of GC properties with UDG models specifically tailored to make predictions regarding GCs. Such comparisons have the potential to help us reach an accurate understanding of both UDG and GC formation.In the upcoming years, several wide-field surveys, including the ESA mission \textit{Euclid}, observe hundreds of thousand of LSB galaxies, UDGs and their GCs in the Local Universe within 100\,Mpc (\citealp{ariane}), which provide a much deeper picture of LSBs/UDGs, their formation and evolution.

\section*{Acknowledgements}
We thank the referee for a number of excellent suggestions to improve the quality of the paper. We would like to thank Scott Carlsten for sharing the data of the GCs around dwarf galaxies and his kind response to our queries. We thank Arianna Di Cintio for discussion on formation of UDGs, and Edwin Valentijn for his comments on this work. We thank Andrea Afruni for his help and assistance in developing the MLE code used in this work. T.S., R.P., J.H.K., I.T. acknowledge financial support from the European Union's Horizon 2020 research and innovation programme under Marie Sk\l odowska-Curie grant agreement No 721463 to the SUNDIAL ITN network. D.Z. acknowledges support from NASA/STScI through NAS5-26555(HST-GO-15121.001-A), which also provided many of the observations used here. J.H.K. acknowledges financial support from the State Research Agency (AEI-MCINN) of the Spanish Ministry of Science and Innovation under the grant "The structure and evolution of galaxies and their central regions" with reference PID2019-105602GB-I00/10.13039/501100011033, from the ACIISI, Consejer\'{i}a de Econom\'{i}a, Conocimiento y Empleo del Gobierno de Canarias and the European Regional Development Fund (ERDF) under grant with reference PROID2021010044, and from IAC project P/300724, financed by the Ministry of Science and Innovation, through the State Budget and by the Canary Islands Department of Economy, Knowledge and Employment, through the Regional Budget of the Autonomous Community. N.C.A. is supported by an STFC/UKRI Ernest Rutherford Fellowship, Project Reference: ST/S004998/1. I.T. acknowledges financial support from the State Research Agency (AEI-MCINN) of the Spanish Ministry of Science and Innovation under the grant with reference PID2019-107427GB-C32, and from IAC project P/300624, financed by the Ministry of Science and Innovation, through the State Budget and by the Canary Islands Department of Economy, Knowledge and Employment, through the Regional Budget of the Autonomous Community. M.A.B acknowledges support from the grant PID2019-107427GB-C32 from the Spanish Ministry of Science, Innovation and Universities (MCIU) and from the Severo Ochoa Excellence scheme (SEV-2015-0548). This work has been backed through the IAC project TRACES which is partially supported through the state budget and the regional budget of the Consejería de Economía, Industria, Comercio y Conocimiento of the Canary Islands Autonomous Community. This research has made use of the SIMBAD database (\citealp{simbad}), operated at CDS, Strasbourg, France, the VizieR catalogue access tool (\citealp{vizier}), CDS, Strasbourg, France (DOI : 10.26093/cds/vizier), and the Aladin sky atlas (\citealp{aladin1,aladin2}) developed at CDS, Strasbourg Observatory, France and SAOImageDS9 (\citealp{ds9}). This work has been done using the following software, packages and \textsc{python} libraries: \textsc{Numpy} (\citealp{numpy}), \textsc{Scipy} (\citealp{scipy}), \textsc{Astropy} (\citealp{astropy}), \textsc{Scikit-learn} (\citealp{scikit-learn}).

\section*{Data Availability}
The data underlying this article were retrieved from the Hubble Space Telescope archive and it provides the reduced/drizzled frames by the standard \textit{HST} pipeline. The software and packages that are used in this work are publicly available. The catalogue of the Globular Cluster candidates around the UDG sample generated in this research are available in the article and its online supplementary material. 



\bibliographystyle{mnras}
\bibliography{mnras} 

\begin{thebibliography}{}
\makeatletter
\relax
\def\mn@urlcharsother{\let\do\@makeother \do\$\do\&\do\#\do\^\do\_\do\%\do\~}
\def\mn@doi{\begingroup\mn@urlcharsother \@ifnextchar [ {\mn@doi@}
  {\mn@doi@[]}}
\def\mn@doi@[#1]#2{\def\@tempa{#1}\ifx\@tempa\@empty \href
  {http://dx.doi.org/#2} {doi:#2}\else \href {http://dx.doi.org/#2} {#1}\fi
  \endgroup}
\def\mn@eprint#1#2{\mn@eprint@#1:#2::\@nil}
\def\mn@eprint@arXiv#1{\href {http://arxiv.org/abs/#1} {{\tt arXiv:#1}}}
\def\mn@eprint@dblp#1{\href {http://dblp.uni-trier.de/rec/bibtex/#1.xml}
  {dblp:#1}}
\def\mn@eprint@#1:#2:#3:#4\@nil{\def\@tempa {#1}\def\@tempb {#2}\def\@tempc
  {#3}\ifx \@tempc \@empty \let \@tempc \@tempb \let \@tempb \@tempa \fi \ifx
  \@tempb \@empty \def\@tempb {arXiv}\fi \@ifundefined
  {mn@eprint@\@tempb}{\@tempb:\@tempc}{\expandafter \expandafter \csname
  mn@eprint@\@tempb\endcsname \expandafter{\@tempc}}}

\bibitem[\protect\citeauthoryear{{Amorisco}}{{Amorisco}}{2018}]{amorisco2016+}
{Amorisco} N.~C.,  2018, \mn@doi [\mnras] {10.1093/mnrasl/sly012}, \href
  {https://ui.adsabs.harvard.edu/abs/2018MNRAS.475L.116A} {475, L116}

\bibitem[\protect\citeauthoryear{{Amorisco}}{{Amorisco}}{2019}]{amorisco19}
{Amorisco} N.~C.,  2019, \mn@doi [\mnras] {10.1093/mnrasl/slz121}, \href
  {https://ui.adsabs.harvard.edu/abs/2019MNRAS.489L..22A} {489, L22}

\bibitem[\protect\citeauthoryear{{Amorisco} \& {Loeb}}{{Amorisco} \&
  {Loeb}}{2016}]{amorisco+loeb}
{Amorisco} N.~C.,  {Loeb} A.,  2016, \mn@doi [\mnras] {10.1093/mnrasl/slw055},
  \href {https://ui.adsabs.harvard.edu/abs/2016MNRAS.459L..51A} {459, L51}

\bibitem[\protect\citeauthoryear{{Amorisco}, {Monachesi}, {Agnello}  \&
  {White}}{{Amorisco} et~al.}{2018}]{amorisco2018}
{Amorisco} N.~C.,  {Monachesi} A.,  {Agnello} A.,   {White} S.~D.~M.,  2018,
  \mn@doi [\mnras] {10.1093/mnras/sty116}, \href
  {https://ui.adsabs.harvard.edu/abs/2018MNRAS.475.4235A} {475, 4235}

\bibitem[\protect\citeauthoryear{{Astropy Collaboration} et~al.,}{{Astropy
  Collaboration} et~al.}{2018}]{astropy}
{Astropy Collaboration} et~al., 2018, \mn@doi [\aj] {10.3847/1538-3881/aabc4f},
  \href {https://ui.adsabs.harvard.edu/abs/2018AJ....156..123A} {156, 123}

\bibitem[\protect\citeauthoryear{{Barbosa} et~al.,}{{Barbosa}
  et~al.}{2020}]{smudges2}
{Barbosa} C.~E.,  et~al., 2020, \mn@doi [\apjs] {10.3847/1538-4365/ab7660},
  \href {https://ui.adsabs.harvard.edu/abs/2020ApJS..247...46B} {247, 46}

\bibitem[\protect\citeauthoryear{{Bastian}, {Pfeffer}, {Kruijssen}, {Crain},
  {Trujillo-Gomez}  \& {Reina-Campos}}{{Bastian} et~al.}{2020}]{bastian}
{Bastian} N.,  {Pfeffer} J.,  {Kruijssen} J.~M.~D.,  {Crain} R.~A.,
  {Trujillo-Gomez} S.,   {Reina-Campos} M.,  2020, \mn@doi [\mnras]
  {10.1093/mnras/staa2453}, \href
  {https://ui.adsabs.harvard.edu/abs/2020MNRAS.498.1050B} {498, 1050}

\bibitem[\protect\citeauthoryear{{Beasley} \& {Trujillo}}{{Beasley} \&
  {Trujillo}}{2016}]{beasley2016b}
{Beasley} M.~A.,  {Trujillo} I.,  2016, \mn@doi [\apj]
  {10.3847/0004-637X/830/1/23}, \href
  {https://ui.adsabs.harvard.edu/abs/2016ApJ...830...23B} {830, 23}

\bibitem[\protect\citeauthoryear{{Beasley}, {Romanowsky}, {Pota}, {Navarro},
  {Martinez Delgado}, {Neyer}  \& {Deich}}{{Beasley}
  et~al.}{2016}]{beasley2016}
{Beasley} M.~A.,  {Romanowsky} A.~J.,  {Pota} V.,  {Navarro} I.~M.,  {Martinez
  Delgado} D.,  {Neyer} F.,   {Deich} A.~L.,  2016, \mn@doi [\apjl]
  {10.3847/2041-8205/819/2/L20}, \href
  {https://ui.adsabs.harvard.edu/abs/2016ApJ...819L..20B} {819, L20}

\bibitem[\protect\citeauthoryear{{Benavides} et~al.,}{{Benavides}
  et~al.}{2021}]{jose2021}
{Benavides} J.~A.,  et~al., 2021, \mn@doi [Nature Astronomy]
  {10.1038/s41550-021-01458-1}, \href
  {https://ui.adsabs.harvard.edu/abs/2021NatAs...5.1255B} {5, 1255}

\bibitem[\protect\citeauthoryear{{Bertin} \& {Arnouts}}{{Bertin} \&
  {Arnouts}}{1996}]{sex}
{Bertin} E.,  {Arnouts} S.,  1996, \mn@doi [\aaps] {10.1051/aas:1996164}, \href
  {https://ui.adsabs.harvard.edu/abs/1996A&AS..117..393B} {117, 393}

\bibitem[\protect\citeauthoryear{{Bertin}, {Mellier}, {Radovich}, {Missonnier},
  {Didelon}  \& {Morin}}{{Bertin} et~al.}{2002}]{swarp}
{Bertin} E.,  {Mellier} Y.,  {Radovich} M.,  {Missonnier} G.,  {Didelon} P.,
  {Morin} B.,  2002, {The TERAPIX Pipeline}.
p.~228

\bibitem[\protect\citeauthoryear{{Binggeli}, {Sandage}  \&
  {Tammann}}{{Binggeli} et~al.}{1985}]{Binggeli}
{Binggeli} B.,  {Sandage} A.,   {Tammann} G.~A.,  1985, \mn@doi [\aj]
  {10.1086/113874}, \href
  {https://ui.adsabs.harvard.edu/abs/1985AJ.....90.1681B} {90, 1681}

\bibitem[\protect\citeauthoryear{{Blakeslee} et~al.,}{{Blakeslee}
  et~al.}{2010}]{blak2010}
{Blakeslee} J.~P.,  et~al., 2010, \mn@doi [\apj] {10.1088/0004-637X/724/1/657},
  \href {https://ui.adsabs.harvard.edu/abs/2010ApJ...724..657B} {724, 657}

\bibitem[\protect\citeauthoryear{{Blanton} \& {Roweis}}{{Blanton} \&
  {Roweis}}{2007}]{blanton}
{Blanton} M.~R.,  {Roweis} S.,  2007, \mn@doi [\aj] {10.1086/510127}, \href
  {https://ui.adsabs.harvard.edu/abs/2007AJ....133..734B} {133, 734}

\bibitem[\protect\citeauthoryear{{Boch} \& {Fernique}}{{Boch} \&
  {Fernique}}{2014}]{aladin2}
{Boch} T.,  {Fernique} P.,  2014, in {Manset} N.,  {Forshay} P.,  eds,
  Astronomical Society of the Pacific Conference Series Vol. 485, Astronomical
  Data Analysis Software and Systems XXIII. p.~277

\bibitem[\protect\citeauthoryear{{Bogd{\'a}n}}{{Bogd{\'a}n}}{2020}]{bogdan}
{Bogd{\'a}n} {\'A}.,  2020, \mn@doi [\apjl] {10.3847/2041-8213/abb886}, \href
  {https://ui.adsabs.harvard.edu/abs/2020ApJ...901L..30B} {901, L30}

\bibitem[\protect\citeauthoryear{{Bonnarel} et~al.,}{{Bonnarel}
  et~al.}{2000}]{aladin1}
{Bonnarel} F.,  et~al., 2000, \mn@doi [\aaps] {10.1051/aas:2000331}, \href
  {https://ui.adsabs.harvard.edu/abs/2000A&AS..143...33B} {143, 33}

\bibitem[\protect\citeauthoryear{{Bothun}, {Impey}  \& {Malin}}{{Bothun}
  et~al.}{1991}]{bothun91}
{Bothun} G.~D.,  {Impey} C.~D.,   {Malin} D.~F.,  1991, \mn@doi [\apj]
  {10.1086/170290}, \href
  {https://ui.adsabs.harvard.edu/abs/1991ApJ...376..404B} {376, 404}

\bibitem[\protect\citeauthoryear{{Boylan-Kolchin}}{{Boylan-Kolchin}}{2017}]{BK17}
{Boylan-Kolchin} M.,  2017, \mn@doi [\mnras] {10.1093/mnras/stx2164}, \href
  {https://ui.adsabs.harvard.edu/abs/2017MNRAS.472.3120B} {472, 3120}

\bibitem[\protect\citeauthoryear{{Brodie} \& {Strader}}{{Brodie} \&
  {Strader}}{2006}]{brodie}
{Brodie} J.~P.,  {Strader} J.,  2006, \mn@doi [\araa]
  {10.1146/annurev.astro.44.051905.092441}, \href
  {https://ui.adsabs.harvard.edu/abs/2006ARA&A..44..193B} {44, 193}

\bibitem[\protect\citeauthoryear{{Carleton}, {Errani}, {Cooper}, {Kaplinghat},
  {Pe{\~n}arrubia}  \& {Guo}}{{Carleton} et~al.}{2019}]{carleton19}
{Carleton} T.,  {Errani} R.,  {Cooper} M.,  {Kaplinghat} M.,  {Pe{\~n}arrubia}
  J.,   {Guo} Y.,  2019, \mn@doi [\mnras] {10.1093/mnras/stz383}, \href
  {https://ui.adsabs.harvard.edu/abs/2019MNRAS.485..382C} {485, 382}

\bibitem[\protect\citeauthoryear{{Carleton}, {Guo}, {Munshi}, {Tremmel}  \&
  {Wright}}{{Carleton} et~al.}{2021}]{carleton2021}
{Carleton} T.,  {Guo} Y.,  {Munshi} F.,  {Tremmel} M.,   {Wright} A.,  2021,
  \mn@doi [\mnras] {10.1093/mnras/stab031}, \href
  {https://ui.adsabs.harvard.edu/abs/2021MNRAS.502..398C} {502, 398}

\bibitem[\protect\citeauthoryear{{Carlsten}, {Greene}, {Beaton}  \&
  {Greco}}{{Carlsten} et~al.}{2021}]{carlsten2021}
{Carlsten} S.~G.,  {Greene} J.~E.,  {Beaton} R.~L.,   {Greco} J.~P.,  2021,
  arXiv e-prints, \href {https://ui.adsabs.harvard.edu/abs/2021arXiv210503440C}
  {p. arXiv:2105.03440}

\bibitem[\protect\citeauthoryear{{Chan}, {Kere{\v{s}}}, {Wetzel}, {Hopkins},
  {Faucher-Gigu{\`e}re}, {El-Badry}, {Garrison-Kimmel}  \&
  {Boylan-Kolchin}}{{Chan} et~al.}{2018}]{chan}
{Chan} T.~K.,  {Kere{\v{s}}} D.,  {Wetzel} A.,  {Hopkins} P.~F.,
  {Faucher-Gigu{\`e}re} C.~A.,  {El-Badry} K.,  {Garrison-Kimmel} S.,
  {Boylan-Kolchin} M.,  2018, \mn@doi [\mnras] {10.1093/mnras/sty1153}, \href
  {https://ui.adsabs.harvard.edu/abs/2018MNRAS.478..906C} {478, 906}

\bibitem[\protect\citeauthoryear{{Chen} et~al.,}{{Chen} et~al.}{2020}]{chen}
{Chen} H.,  et~al., 2020, \mn@doi [\mnras] {10.1093/mnras/staa1868}, \href
  {https://ui.adsabs.harvard.edu/abs/2020MNRAS.496.4654C} {496, 4654}

\bibitem[\protect\citeauthoryear{{Chilingarian}, {Afanasiev}, {Grishin},
  {Fabricant}  \& {Moran}}{{Chilingarian} et~al.}{2019}]{chil}
{Chilingarian} I.~V.,  {Afanasiev} A.~V.,  {Grishin} K.~A.,  {Fabricant} D.,
  {Moran} S.,  2019, \mn@doi [\apj] {10.3847/1538-4357/ab4205}, \href
  {https://ui.adsabs.harvard.edu/abs/2019ApJ...884...79C} {884, 79}

\bibitem[\protect\citeauthoryear{{Danieli} et~al.,}{{Danieli}
  et~al.}{2021}]{matlas-danieli}
{Danieli} S.,  et~al., 2021, arXiv e-prints, \href
  {https://ui.adsabs.harvard.edu/abs/2021arXiv211114851D} {p. arXiv:2111.14851}

\bibitem[\protect\citeauthoryear{{Di Cintio}, {Brook}, {Dutton}, {Macci{\`o}},
  {Obreja}  \& {Dekel}}{{Di Cintio} et~al.}{2017}]{cintio2017}
{Di Cintio} A.,  {Brook} C.~B.,  {Dutton} A.~A.,  {Macci{\`o}} A.~V.,  {Obreja}
  A.,   {Dekel} A.,  2017, \mn@doi [\mnras] {10.1093/mnrasl/slw210}, \href
  {https://ui.adsabs.harvard.edu/abs/2017MNRAS.466L...1D} {466, L1}

\bibitem[\protect\citeauthoryear{{Diemer}}{{Diemer}}{2018}]{diemer}
{Diemer} B.,  2018, \mn@doi [\apjs] {10.3847/1538-4365/aaee8c}, \href
  {https://ui.adsabs.harvard.edu/abs/2018ApJS..239...35D} {239, 35}

\bibitem[\protect\citeauthoryear{{El-Badry}, {Quataert}, {Weisz}, {Choksi}  \&
  {Boylan-Kolchin}}{{El-Badry} et~al.}{2019}]{elbadry}
{El-Badry} K.,  {Quataert} E.,  {Weisz} D.~R.,  {Choksi} N.,   {Boylan-Kolchin}
  M.,  2019, \mn@doi [\mnras] {10.1093/mnras/sty3007}, \href
  {https://ui.adsabs.harvard.edu/abs/2019MNRAS.482.4528E} {482, 4528}

\bibitem[\protect\citeauthoryear{{Erkal} et~al.,}{{Erkal} et~al.}{2019}]{erkal}
{Erkal} D.,  et~al., 2019, \mn@doi [\mnras] {10.1093/mnras/stz1371}, \href
  {https://ui.adsabs.harvard.edu/abs/2019MNRAS.487.2685E} {487, 2685}

\bibitem[\protect\citeauthoryear{{Ferr{\'e}-Mateu} et~al.,}{{Ferr{\'e}-Mateu}
  et~al.}{2018}]{anna}
{Ferr{\'e}-Mateu} A.,  et~al., 2018, \mn@doi [\mnras] {10.1093/mnras/sty1597},
  \href {https://ui.adsabs.harvard.edu/abs/2018MNRAS.479.4891F} {479, 4891}

\bibitem[\protect\citeauthoryear{{Fliri} \& {Trujillo}}{{Fliri} \&
  {Trujillo}}{2016}]{2016MNRAS.456.1359F}
{Fliri} J.,  {Trujillo} I.,  2016, \mn@doi [\mnras] {10.1093/mnras/stv2686},
  \href {https://ui.adsabs.harvard.edu/abs/2016MNRAS.456.1359F} {456, 1359}

\bibitem[\protect\citeauthoryear{{Forbes}, {Read}, {Gieles}  \&
  {Collins}}{{Forbes} et~al.}{2018}]{forbes2018}
{Forbes} D.~A.,  {Read} J.~I.,  {Gieles} M.,   {Collins} M. L.~M.,  2018,
  \mn@doi [\mnras] {10.1093/mnras/sty2584}, \href
  {https://ui.adsabs.harvard.edu/abs/2018MNRAS.481.5592F} {481, 5592}

\bibitem[\protect\citeauthoryear{{Forbes}, {Gannon}, {Romanowsky}, {Alabi},
  {Brodie}, {Couch}  \& {Ferr{\'e}-Mateu}}{{Forbes} et~al.}{2021}]{forbes1}
{Forbes} D.~A.,  {Gannon} J.~S.,  {Romanowsky} A.~J.,  {Alabi} A.,  {Brodie}
  J.~P.,  {Couch} W.~J.,   {Ferr{\'e}-Mateu} A.,  2021, \mn@doi [\mnras]
  {10.1093/mnras/staa3289}, \href
  {https://ui.adsabs.harvard.edu/abs/2021MNRAS.500.1279F} {500, 1279}

\bibitem[\protect\citeauthoryear{{Freundlich}, {Dekel}, {Jiang}, {Ishai},
  {Cornuault}, {Lapiner}, {Dutton}  \& {Macci{\`o}}}{{Freundlich}
  et~al.}{2020}]{freundlich}
{Freundlich} J.,  {Dekel} A.,  {Jiang} F.,  {Ishai} G.,  {Cornuault} N.,
  {Lapiner} S.,  {Dutton} A.~A.,   {Macci{\`o}} A.~V.,  2020, \mn@doi [\mnras]
  {10.1093/mnras/stz3306}, \href
  {https://ui.adsabs.harvard.edu/abs/2020MNRAS.491.4523F} {491, 4523}

\bibitem[\protect\citeauthoryear{{Gannon}, {Forbes}, {Romanowsky},
  {Ferr{\'e}-Mateu}, {Couch}  \& {Brodie}}{{Gannon} et~al.}{2020}]{gannon}
{Gannon} J.~S.,  {Forbes} D.~A.,  {Romanowsky} A.~J.,  {Ferr{\'e}-Mateu} A.,
  {Couch} W.~J.,   {Brodie} J.~P.,  2020, \mn@doi [\mnras]
  {10.1093/mnras/staa1282}, \href
  {https://ui.adsabs.harvard.edu/abs/2020MNRAS.495.2582G} {495, 2582}

\bibitem[\protect\citeauthoryear{{Gannon} et~al.,}{{Gannon}
  et~al.}{2021}]{gannon2021}
{Gannon} J.~S.,  et~al., 2021, \mn@doi [\mnras] {10.1093/mnras/stab3297}, \href
  {https://ui.adsabs.harvard.edu/abs/2021MNRAS.tmp.3015G} {}

\bibitem[\protect\citeauthoryear{{Gennaro, M., et al.}}{{Gennaro, M., et
  al.}}{2018}]{wfc3}
{Gennaro, M., et al.} 2018, {WFC3 Data Handbook, Version 4.0, (Baltimore:
  STScI)}

\bibitem[\protect\citeauthoryear{{Godwin}, {Metcalfe}  \& {Peach}}{{Godwin}
  et~al.}{1983}]{dfx1}
{Godwin} J.~G.,  {Metcalfe} N.,   {Peach} J.~V.,  1983, \mn@doi [\mnras]
  {10.1093/mnras/202.1.113}, \href
  {https://ui.adsabs.harvard.edu/abs/1983MNRAS.202..113G} {202, 113}

\bibitem[\protect\citeauthoryear{{G{\'o}mez}, {Richtler}, {Infante}  \&
  {Drenkhahn}}{{G{\'o}mez} et~al.}{2001}]{gomez1}
{G{\'o}mez} M.,  {Richtler} T.,  {Infante} L.,   {Drenkhahn} G.,  2001, \mn@doi
  [\aap] {10.1051/0004-6361:20010457}, \href
  {https://ui.adsabs.harvard.edu/abs/2001A&A...371..875G} {371, 875}

\bibitem[\protect\citeauthoryear{{Gu} et~al.,}{{Gu} et~al.}{2018}]{gu}
{Gu} M.,  et~al., 2018, \mn@doi [\apj] {10.3847/1538-4357/aabbae}, \href
  {https://ui.adsabs.harvard.edu/abs/2018ApJ...859...37G} {859, 37}

\bibitem[\protect\citeauthoryear{{Habas} et~al.,}{{Habas}
  et~al.}{2020}]{matlas}
{Habas} R.,  et~al., 2020, \mn@doi [\mnras] {10.1093/mnras/stz3045}, \href
  {https://ui.adsabs.harvard.edu/abs/2020MNRAS.491.1901H} {491, 1901}

\bibitem[\protect\citeauthoryear{{Harris}}{{Harris}}{2018}]{harris2018}
{Harris} W.~E.,  2018, \mn@doi [\aj] {10.3847/1538-3881/aaedb8}, \href
  {https://ui.adsabs.harvard.edu/abs/2018AJ....156..296H} {156, 296}

\bibitem[\protect\citeauthoryear{{Harris}, {Kavelaars}, {Hanes}, {Pritchet}  \&
  {Baum}}{{Harris} et~al.}{2009}]{harris2009}
{Harris} W.~E.,  {Kavelaars} J.~J.,  {Hanes} D.~A.,  {Pritchet} C.~J.,   {Baum}
  W.~A.,  2009, \mn@doi [\aj] {10.1088/0004-6256/137/2/3314}, \href
  {https://ui.adsabs.harvard.edu/abs/2009AJ....137.3314H} {137, 3314}

\bibitem[\protect\citeauthoryear{{Harris}, {Harris}  \& {Alessi}}{{Harris}
  et~al.}{2013}]{harris2013}
{Harris} W.~E.,  {Harris} G. L.~H.,   {Alessi} M.,  2013, \mn@doi [\apj]
  {10.1088/0004-637X/772/2/82}, \href
  {https://ui.adsabs.harvard.edu/abs/2013ApJ...772...82H} {772, 82}

\bibitem[\protect\citeauthoryear{{Harris} et~al.,}{{Harris}
  et~al.}{2014}]{harris2014}
{Harris} W.~E.,  et~al., 2014, \mn@doi [\apj] {10.1088/0004-637X/797/2/128},
  \href {https://ui.adsabs.harvard.edu/abs/2014ApJ...797..128H} {797, 128}

\bibitem[\protect\citeauthoryear{{Harris}, {Blakeslee}  \& {Harris}}{{Harris}
  et~al.}{2017}]{harris2017}
{Harris} W.~E.,  {Blakeslee} J.~P.,   {Harris} G. L.~H.,  2017, \mn@doi [\apj]
  {10.3847/1538-4357/836/1/67}, \href
  {https://ui.adsabs.harvard.edu/abs/2017ApJ...836...67H} {836, 67}

\bibitem[\protect\citeauthoryear{{Impey} \& {Bothun}}{{Impey} \&
  {Bothun}}{1997}]{impey}
{Impey} C.,  {Bothun} G.,  1997, \mn@doi [\araa]
  {10.1146/annurev.astro.35.1.267}, \href
  {https://ui.adsabs.harvard.edu/abs/1997ARA&A..35..267I} {35, 267}

\bibitem[\protect\citeauthoryear{{Into} \& {Portinari}}{{Into} \&
  {Portinari}}{2013}]{tom}
{Into} T.,  {Portinari} L.,  2013, \mn@doi [\mnras] {10.1093/mnras/stt071},
  \href {https://ui.adsabs.harvard.edu/abs/2013MNRAS.430.2715I} {430, 2715}

\bibitem[\protect\citeauthoryear{{Iodice} et~al.,}{{Iodice}
  et~al.}{2020}]{iodice}
{Iodice} E.,  et~al., 2020, \mn@doi [\aap] {10.1051/0004-6361/202038523}, \href
  {https://ui.adsabs.harvard.edu/abs/2020A&A...642A..48I} {642, A48}

\bibitem[\protect\citeauthoryear{{Jensen}, {Tonry}  \& {Luppino}}{{Jensen}
  et~al.}{1999}]{distcoma2}
{Jensen} J.~B.,  {Tonry} J.~L.,   {Luppino} G.~A.,  1999, \mn@doi [\apj]
  {10.1086/306569}, \href
  {https://ui.adsabs.harvard.edu/abs/1999ApJ...510...71J} {510, 71}

\bibitem[\protect\citeauthoryear{{Jiang}, {Dekel}, {Freundlich}, {Romanowsky},
  {Dutton}, {Macci{\`o}}  \& {Di Cintio}}{{Jiang} et~al.}{2019}]{jiang}
{Jiang} F.,  {Dekel} A.,  {Freundlich} J.,  {Romanowsky} A.~J.,  {Dutton}
  A.~A.,  {Macci{\`o}} A.~V.,   {Di Cintio} A.,  2019, \mn@doi [\mnras]
  {10.1093/mnras/stz1499}, \href
  {https://ui.adsabs.harvard.edu/abs/2019MNRAS.487.5272J} {487, 5272}

\bibitem[\protect\citeauthoryear{{Jones}, {Papastergis}, {Pandya}, {Leisman},
  {Romanowsky}, {Yung}, {Somerville}  \& {Adams}}{{Jones}
  et~al.}{2018}]{jones2018}
{Jones} M.~G.,  {Papastergis} E.,  {Pandya} V.,  {Leisman} L.,  {Romanowsky}
  A.~J.,  {Yung} L.~Y.~A.,  {Somerville} R.~S.,   {Adams} E.~A.~K.,  2018,
  \mn@doi [\aap] {10.1051/0004-6361/201732409}, \href
  {https://ui.adsabs.harvard.edu/abs/2018A&A...614A..21J} {614, A21}

\bibitem[\protect\citeauthoryear{{Jones}, {Bennet}, {Mutlu-Pakdil}, {Sand},
  {Spekkens}, {Crnojevi{\'c}}, {Karunakaran}  \& {Zaritsky}}{{Jones}
  et~al.}{2021}]{jones2021}
{Jones} M.~G.,  {Bennet} P.,  {Mutlu-Pakdil} B.,  {Sand} D.~J.,  {Spekkens} K.,
   {Crnojevi{\'c}} D.,  {Karunakaran} A.,   {Zaritsky} D.,  2021, \mn@doi
  [\apj] {10.3847/1538-4357/ac0975}, \href
  {https://ui.adsabs.harvard.edu/abs/2021ApJ...919...72J} {919, 72}

\bibitem[\protect\citeauthoryear{{Jord{\'a}n} et~al.,}{{Jord{\'a}n}
  et~al.}{2007}]{jordan2007}
{Jord{\'a}n} A.,  et~al., 2007, \mn@doi [\apjs] {10.1086/516840}, \href
  {https://ui.adsabs.harvard.edu/abs/2007ApJS..171..101J} {171, 101}

\bibitem[\protect\citeauthoryear{{Jord{\'a}n}, {Peng}, {Blakeslee},
  {C{\^o}t{\'e}}, {Eyheramendy}  \& {Ferrarese}}{{Jord{\'a}n}
  et~al.}{2015}]{jordan2015}
{Jord{\'a}n} A.,  {Peng} E.~W.,  {Blakeslee} J.~P.,  {C{\^o}t{\'e}} P.,
  {Eyheramendy} S.,   {Ferrarese} L.,  2015, \mn@doi [\apjs]
  {10.1088/0067-0049/221/1/13}, \href
  {https://ui.adsabs.harvard.edu/abs/2015ApJS..221...13J} {221, 13}

\bibitem[\protect\citeauthoryear{{Joye} \& {Mandel}}{{Joye} \&
  {Mandel}}{2003}]{ds9}
{Joye} W.~A.,  {Mandel} E.,  2003, in {Payne} H.~E.,  {Jedrzejewski} R.~I.,
  {Hook} R.~N.,  eds,  Astronomical Society of the Pacific Conference Series
  Vol. 295, Astronomical Data Analysis Software and Systems XII. p.~489

\bibitem[\protect\citeauthoryear{{Kado-Fong} et~al.,}{{Kado-Fong}
  et~al.}{2021}]{kado}
{Kado-Fong} E.,  et~al., 2021, \mn@doi [\apj] {10.3847/1538-4357/ac15f0}, \href
  {https://ui.adsabs.harvard.edu/abs/2021ApJ...920...72K} {920, 72}

\bibitem[\protect\citeauthoryear{{Kadowaki}, {Zaritsky}  \&
  {Donnerstein}}{{Kadowaki} et~al.}{2017}]{kadowaski}
{Kadowaki} J.,  {Zaritsky} D.,   {Donnerstein} R.~L.,  2017, \mn@doi [\apjl]
  {10.3847/2041-8213/aa653d}, \href
  {https://ui.adsabs.harvard.edu/abs/2017ApJ...838L..21K} {838, L21}

\bibitem[\protect\citeauthoryear{{Kadowaki}, {Zaritsky}, {Donnerstein}, {RS},
  {Karunakaran}  \& {Spekkens}}{{Kadowaki} et~al.}{2021}]{kadowaski2}
{Kadowaki} J.,  {Zaritsky} D.,  {Donnerstein} R.~L.,  {RS} P.,  {Karunakaran}
  A.,   {Spekkens} K.,  2021, \mn@doi [\apj] {10.3847/1538-4357/ac2948}, \href
  {https://ui.adsabs.harvard.edu/abs/2021ApJ...923..257K} {923, 257}

\bibitem[\protect\citeauthoryear{{Karunakaran}, {Spekkens}, {Zaritsky},
  {Donnerstein}, {Kadowaki}  \& {Dey}}{{Karunakaran}
  et~al.}{2020}]{karunakaran}
{Karunakaran} A.,  {Spekkens} K.,  {Zaritsky} D.,  {Donnerstein} R.~L.,
  {Kadowaki} J.,   {Dey} A.,  2020, \mn@doi [\apj] {10.3847/1538-4357/abb464},
  \href {https://ui.adsabs.harvard.edu/abs/2020ApJ...902...39K} {902, 39}

\bibitem[\protect\citeauthoryear{{Kavelaars}}{{Kavelaars}}{1998}]{kave}
{Kavelaars} J.~J.,  1998, \mn@doi [\pasp] {10.1086/316183}, \href
  {https://ui.adsabs.harvard.edu/abs/1998PASP..110..758K} {110, 758}

\bibitem[\protect\citeauthoryear{{Kissler-Patig}, {Richtler}, {Storm}  \&
  {della Valle}}{{Kissler-Patig} et~al.}{1997}]{kissler}
{Kissler-Patig} M.,  {Richtler} T.,  {Storm} J.,   {della Valle} M.,  1997,
  \aap, \href {https://ui.adsabs.harvard.edu/abs/1997A&A...327..503K} {327,
  503}

\bibitem[\protect\citeauthoryear{{Koda}, {Yagi}, {Yamanoi}  \&
  {Komiyama}}{{Koda} et~al.}{2015}]{koda}
{Koda} J.,  {Yagi} M.,  {Yamanoi} H.,   {Komiyama} Y.,  2015, \mn@doi [\apjl]
  {10.1088/2041-8205/807/1/L2}, \href
  {https://ui.adsabs.harvard.edu/abs/2015ApJ...807L...2K} {807, L2}

\bibitem[\protect\citeauthoryear{{Kruijssen}, {Pfeffer}, {Crain}  \&
  {Bastian}}{{Kruijssen} et~al.}{2019}]{kuij}
{Kruijssen} J.~M.~D.,  {Pfeffer} J.~L.,  {Crain} R.~A.,   {Bastian} N.,  2019,
  \mn@doi [\mnras] {10.1093/mnras/stz968}, \href
  {https://ui.adsabs.harvard.edu/abs/2019MNRAS.486.3134K} {486, 3134}

\bibitem[\protect\citeauthoryear{{Kubo}, {Stebbins}, {Annis}, {Dell'Antonio},
  {Lin}, {Khiabanian}  \& {Frieman}}{{Kubo} et~al.}{2007}]{coma-virial}
{Kubo} J.~M.,  {Stebbins} A.,  {Annis} J.,  {Dell'Antonio} I.~P.,  {Lin} H.,
  {Khiabanian} H.,   {Frieman} J.~A.,  2007, \mn@doi [\apj] {10.1086/523101},
  \href {https://ui.adsabs.harvard.edu/abs/2007ApJ...671.1466K} {671, 1466}

\bibitem[\protect\citeauthoryear{{Kundu} \& {Whitmore}}{{Kundu} \&
  {Whitmore}}{2001}]{kundu}
{Kundu} A.,  {Whitmore} B.~C.,  2001, \mn@doi [\aj] {10.1086/321073}, \href
  {https://ui.adsabs.harvard.edu/abs/2001AJ....121.2950K} {121, 2950}

\bibitem[\protect\citeauthoryear{{Lan{\c{c}}on} et~al.,}{{Lan{\c{c}}on}
  et~al.}{2021}]{ariane}
{Lan{\c{c}}on} A.,  et~al., 2021, in {Siebert} A.,  et~al., eds, SF2A-2021:
  Proceedings of the Annual meeting of the French Society of Astronomy and
  Astrophysics. pp 447--450 (\mn@eprint {arXiv} {2110.13783})

\bibitem[\protect\citeauthoryear{{Lee}, {Hodges-Kluck}  \& {Gallo}}{{Lee}
  et~al.}{2020}]{lee}
{Lee} C.~H.,  {Hodges-Kluck} E.,   {Gallo} E.,  2020, \mn@doi [\mnras]
  {10.1093/mnras/staa1955}, \href
  {https://ui.adsabs.harvard.edu/abs/2020MNRAS.497.2759L} {497, 2759}

\bibitem[\protect\citeauthoryear{{Leisman} et~al.,}{{Leisman}
  et~al.}{2017}]{leisman}
{Leisman} L.,  et~al., 2017, \mn@doi [\apj] {10.3847/1538-4357/aa7575}, \href
  {https://ui.adsabs.harvard.edu/abs/2017ApJ...842..133L} {842, 133}

\bibitem[\protect\citeauthoryear{{Liao} et~al.,}{{Liao} et~al.}{2019}]{liao}
{Liao} S.,  et~al., 2019, \mn@doi [\mnras] {10.1093/mnras/stz2969}, \href
  {https://ui.adsabs.harvard.edu/abs/2019MNRAS.490.5182L} {490, 5182}

\bibitem[\protect\citeauthoryear{{Lim}, {Peng}, {C{\^o}t{\'e}}, {Sales}, {den
  Brok}, {Blakeslee}  \& {Guhathakurta}}{{Lim} et~al.}{2018}]{lim2018}
{Lim} S.,  {Peng} E.~W.,  {C{\^o}t{\'e}} P.,  {Sales} L.~V.,  {den Brok} M.,
  {Blakeslee} J.~P.,   {Guhathakurta} P.,  2018, \mn@doi [\apj]
  {10.3847/1538-4357/aacb81}, \href
  {https://ui.adsabs.harvard.edu/abs/2018ApJ...862...82L} {862, 82}

\bibitem[\protect\citeauthoryear{{Lim} et~al.,}{{Lim} et~al.}{2020}]{lim2020}
{Lim} S.,  et~al., 2020, \mn@doi [\apj] {10.3847/1538-4357/aba433}, \href
  {https://ui.adsabs.harvard.edu/abs/2020ApJ...899...69L} {899, 69}

\bibitem[\protect\citeauthoryear{{Liu}, {Peng}, {Jord{\'a}n}, {Blakeslee},
  {C{\^o}t{\'e}}, {Ferrarese}  \& {Puzia}}{{Liu} et~al.}{2019}]{liu2019}
{Liu} Y.,  {Peng} E.~W.,  {Jord{\'a}n} A.,  {Blakeslee} J.~P.,  {C{\^o}t{\'e}}
  P.,  {Ferrarese} L.,   {Puzia} T.~H.,  2019, \mn@doi [\apj]
  {10.3847/1538-4357/ab12d9}, \href
  {https://ui.adsabs.harvard.edu/abs/2019ApJ...875..156L} {875, 156}

\bibitem[\protect\citeauthoryear{{Mancera Pi{\~n}a}, {Aguerri}, {Peletier},
  {Venhola}, {Trager}  \& {Choque Challapa}}{{Mancera Pi{\~n}a}
  et~al.}{2019}]{pavel2019}
{Mancera Pi{\~n}a} P.~E.,  {Aguerri} J.~A.~L.,  {Peletier} R.~F.,  {Venhola}
  A.,  {Trager} S.,   {Choque Challapa} N.,  2019, \mn@doi [\mnras]
  {10.1093/mnras/stz238}, \href
  {https://ui.adsabs.harvard.edu/abs/2019MNRAS.485.1036M} {485, 1036}

\bibitem[\protect\citeauthoryear{{Mancera Pi{\~n}a} et~al.,}{{Mancera Pi{\~n}a}
  et~al.}{2020}]{pavel2020}
{Mancera Pi{\~n}a} P.~E.,  et~al., 2020, \mn@doi [\mnras]
  {10.1093/mnras/staa1256}, \href
  {https://ui.adsabs.harvard.edu/abs/2020MNRAS.495.3636M} {495, 3636}

\bibitem[\protect\citeauthoryear{{Marleau} et~al.,}{{Marleau}
  et~al.}{2021}]{francine}
{Marleau} F.~R.,  et~al., 2021, \mn@doi [\aap] {10.1051/0004-6361/202141432},
  \href {https://ui.adsabs.harvard.edu/abs/2021A&A...654A.105M} {654, A105}

\bibitem[\protect\citeauthoryear{{Martin} et~al.,}{{Martin}
  et~al.}{2019}]{martin}
{Martin} G.,  et~al., 2019, \mn@doi [\mnras] {10.1093/mnras/stz356}, \href
  {https://ui.adsabs.harvard.edu/abs/2019MNRAS.485..796M} {485, 796}

\bibitem[\protect\citeauthoryear{{Miller} \& {Lotz}}{{Miller} \&
  {Lotz}}{2007}]{miller}
{Miller} B.~W.,  {Lotz} J.~M.,  2007, \mn@doi [\apj] {10.1086/522323}, \href
  {https://ui.adsabs.harvard.edu/abs/2007ApJ...670.1074M} {670, 1074}

\bibitem[\protect\citeauthoryear{{Montes}, {Trujillo}, {Infante-Sainz},
  {Monelli}  \& {Borlaff}}{{Montes} et~al.}{2021}]{montes2}
{Montes} M.,  {Trujillo} I.,  {Infante-Sainz} R.,  {Monelli} M.,   {Borlaff}
  A.~S.,  2021, \mn@doi [\apj] {10.3847/1538-4357/ac0d55}, \href
  {https://ui.adsabs.harvard.edu/abs/2021ApJ...919...56M} {919, 56}

\bibitem[\protect\citeauthoryear{{M{\"u}ller} et~al.,}{{M{\"u}ller}
  et~al.}{2020}]{oliver2020}
{M{\"u}ller} O.,  et~al., 2020, \mn@doi [\aap] {10.1051/0004-6361/202038351},
  \href {https://ui.adsabs.harvard.edu/abs/2020A&A...640A.106M} {640, A106}

\bibitem[\protect\citeauthoryear{{M{\"u}ller} et~al.,}{{M{\"u}ller}
  et~al.}{2021}]{muller21}
{M{\"u}ller} O.,  et~al., 2021, \mn@doi [\apj] {10.3847/1538-4357/ac2831},
  \href {https://ui.adsabs.harvard.edu/abs/2021ApJ...923....9M} {923, 9}

\bibitem[\protect\citeauthoryear{{Navarro}, {Eke}  \& {Frenk}}{{Navarro}
  et~al.}{1996}]{navarro96}
{Navarro} J.~F.,  {Eke} V.~R.,   {Frenk} C.~S.,  1996, \mn@doi [\mnras]
  {10.1093/mnras/283.3.L72}, \href
  {https://ui.adsabs.harvard.edu/abs/1996MNRAS.283L..72N} {283, L72}

\bibitem[\protect\citeauthoryear{{Ochsenbein}, {Bauer}  \&
  {Marcout}}{{Ochsenbein} et~al.}{2000}]{vizier}
{Ochsenbein} F.,  {Bauer} P.,   {Marcout} J.,  2000, \mn@doi [\aaps]
  {10.1051/aas:2000169}, \href
  {https://ui.adsabs.harvard.edu/abs/2000A&AS..143...23O} {143, 23}

\bibitem[\protect\citeauthoryear{{Pandya} et~al.,}{{Pandya}
  et~al.}{2018}]{pandaya}
{Pandya} V.,  et~al., 2018, \mn@doi [\apj] {10.3847/1538-4357/aab498}, \href
  {https://ui.adsabs.harvard.edu/abs/2018ApJ...858...29P} {858, 29}

\bibitem[\protect\citeauthoryear{{Papastergis}, {Adams}  \&
  {Romanowsky}}{{Papastergis} et~al.}{2017}]{manolis}
{Papastergis} E.,  {Adams} E.~A.~K.,   {Romanowsky} A.~J.,  2017, \mn@doi
  [\aap] {10.1051/0004-6361/201730795}, \href
  {https://ui.adsabs.harvard.edu/abs/2017A&A...601L..10P} {601, L10}

\bibitem[\protect\citeauthoryear{Pedregosa et~al.,}{Pedregosa
  et~al.}{2011}]{scikit-learn}
Pedregosa F.,  et~al., 2011, Journal of Machine Learning Research, 12, 2825

\bibitem[\protect\citeauthoryear{{Peng} \& {Lim}}{{Peng} \&
  {Lim}}{2016}]{peng2016}
{Peng} E.~W.,  {Lim} S.,  2016, \mn@doi [\apjl] {10.3847/2041-8205/822/2/L31},
  \href {https://ui.adsabs.harvard.edu/abs/2016ApJ...822L..31P} {822, L31}

\bibitem[\protect\citeauthoryear{{Peng}, {Ho}, {Impey}  \& {Rix}}{{Peng}
  et~al.}{2002}]{galfit1}
{Peng} C.~Y.,  {Ho} L.~C.,  {Impey} C.~D.,   {Rix} H.-W.,  2002, \mn@doi [\aj]
  {10.1086/340952}, \href
  {https://ui.adsabs.harvard.edu/abs/2002AJ....124..266P} {124, 266}

\bibitem[\protect\citeauthoryear{{Peng} et~al.,}{{Peng}
  et~al.}{2006}]{peng2006}
{Peng} E.~W.,  et~al., 2006, \mn@doi [\apj] {10.1086/498210}, \href
  {https://ui.adsabs.harvard.edu/abs/2006ApJ...639...95P} {639, 95}

\bibitem[\protect\citeauthoryear{{Peng} et~al.,}{{Peng}
  et~al.}{2008}]{peng2008}
{Peng} E.~W.,  et~al., 2008, \mn@doi [\apj] {10.1086/587951}, \href
  {https://ui.adsabs.harvard.edu/abs/2008ApJ...681..197P} {681, 197}

\bibitem[\protect\citeauthoryear{{Peng}, {Ho}, {Impey}  \& {Rix}}{{Peng}
  et~al.}{2010}]{galfit2}
{Peng} C.~Y.,  {Ho} L.~C.,  {Impey} C.~D.,   {Rix} H.-W.,  2010, \mn@doi [\aj]
  {10.1088/0004-6256/139/6/2097}, \href
  {https://ui.adsabs.harvard.edu/abs/2010AJ....139.2097P} {139, 2097}

\bibitem[\protect\citeauthoryear{{Pfeffer}, {Kruijssen}, {Crain}  \&
  {Bastian}}{{Pfeffer} et~al.}{2018}]{Pfeffer-2018}
{Pfeffer} J.,  {Kruijssen} J.~M.~D.,  {Crain} R.~A.,   {Bastian} N.,  2018,
  \mn@doi [\mnras] {10.1093/mnras/stx3124}, \href
  {https://ui.adsabs.harvard.edu/abs/2018MNRAS.475.4309P} {475, 4309}

\bibitem[\protect\citeauthoryear{{Pontzen} \& {Governato}}{{Pontzen} \&
  {Governato}}{2012}]{pontzen12}
{Pontzen} A.,  {Governato} F.,  2012, \mn@doi [\mnras]
  {10.1111/j.1365-2966.2012.20571.x}, \href
  {https://ui.adsabs.harvard.edu/abs/2012MNRAS.421.3464P} {421, 3464}

\bibitem[\protect\citeauthoryear{{Poulain} et~al.,}{{Poulain}
  et~al.}{2021}]{poulain}
{Poulain} M.,  et~al., 2021, arXiv e-prints, \href
  {https://ui.adsabs.harvard.edu/abs/2021arXiv211114491P} {p. arXiv:2111.14491}

\bibitem[\protect\citeauthoryear{{Prole}, {Davies}, {Keenan}  \&
  {Davies}}{{Prole} et~al.}{2018}]{prole2018}
{Prole} D.~J.,  {Davies} J.~I.,  {Keenan} O.~C.,   {Davies} L.~J.~M.,  2018,
  \mn@doi [\mnras] {10.1093/mnras/sty1021}, \href
  {https://ui.adsabs.harvard.edu/abs/2018MNRAS.478..667P} {478, 667}

\bibitem[\protect\citeauthoryear{{Prole}, {van der Burg}, {Hilker}  \&
  {Davies}}{{Prole} et~al.}{2019}]{prole2019}
{Prole} D.~J.,  {van der Burg} R.~F.~J.,  {Hilker} M.,   {Davies} J.~I.,  2019,
  \mn@doi [\mnras] {10.1093/mnras/stz1843}, \href
  {https://ui.adsabs.harvard.edu/abs/2019MNRAS.488.2143P} {488, 2143}

\bibitem[\protect\citeauthoryear{{Prole}, {van der Burg}, {Hilker}  \&
  {Spitler}}{{Prole} et~al.}{2021}]{prole2021}
{Prole} D.~J.,  {van der Burg} R.~F.~J.,  {Hilker} M.,   {Spitler} L.~R.,
  2021, \mn@doi [\mnras] {10.1093/mnras/staa3296}, \href
  {https://ui.adsabs.harvard.edu/abs/2021MNRAS.500.2049P} {500, 2049}

\bibitem[\protect\citeauthoryear{{Rejkuba}}{{Rejkuba}}{2012}]{rejkuba2012}
{Rejkuba} M.,  2012, \mn@doi [\apss] {10.1007/s10509-012-0986-9}, \href
  {https://ui.adsabs.harvard.edu/abs/2012Ap&SS.341..195R} {341, 195}

\bibitem[\protect\citeauthoryear{{Rom{\'a}n} \& {Trujillo}}{{Rom{\'a}n} \&
  {Trujillo}}{2017}]{javier}
{Rom{\'a}n} J.,  {Trujillo} I.,  2017, \mn@doi [\mnras] {10.1093/mnras/stx694},
  \href {https://ui.adsabs.harvard.edu/abs/2017MNRAS.468.4039R} {468, 4039}

\bibitem[\protect\citeauthoryear{{Rom{\'a}n}, {Beasley}, {Ruiz-Lara}  \&
  {Valls-Gabaud}}{{Rom{\'a}n} et~al.}{2019}]{javier+red}
{Rom{\'a}n} J.,  {Beasley} M.~A.,  {Ruiz-Lara} T.,   {Valls-Gabaud} D.,  2019,
  \mn@doi [\mnras] {10.1093/mnras/stz835}, \href
  {https://ui.adsabs.harvard.edu/abs/2019MNRAS.486..823R} {486, 823}

\bibitem[\protect\citeauthoryear{{Rom{\'a}n}, {Castilla}  \&
  {Pascual-Granado}}{{Rom{\'a}n} et~al.}{2021}]{javier2021}
{Rom{\'a}n} J.,  {Castilla} A.,   {Pascual-Granado} J.,  2021, \mn@doi [\aap]
  {10.1051/0004-6361/202142161}, \href
  {https://ui.adsabs.harvard.edu/abs/2021A&A...656A..44R} {656, A44}

\bibitem[\protect\citeauthoryear{{Rong} et~al.,}{{Rong}
  et~al.}{2020}]{rong2020_2}
{Rong} Y.,  et~al., 2020, \mn@doi [\apj] {10.3847/1538-4357/aba74a}, \href
  {https://ui.adsabs.harvard.edu/abs/2020ApJ...899...78R} {899, 78}

\bibitem[\protect\citeauthoryear{{Ruiz-Lara} et~al.,}{{Ruiz-Lara}
  et~al.}{2018}]{ruizlara2018}
{Ruiz-Lara} T.,  et~al., 2018, \mn@doi [\mnras] {10.1093/mnras/sty1112}, \href
  {https://ui.adsabs.harvard.edu/abs/2018MNRAS.478.2034R} {478, 2034}

\bibitem[\protect\citeauthoryear{{Saifollahi}, {Trujillo}, {Beasley},
  {Peletier}  \& {Knapen}}{{Saifollahi} et~al.}{2021}]{saifollahi2020}
{Saifollahi} T.,  {Trujillo} I.,  {Beasley} M.~A.,  {Peletier} R.~F.,
  {Knapen} J.~H.,  2021, \mn@doi [\mnras] {10.1093/mnras/staa3016}, \href
  {https://ui.adsabs.harvard.edu/abs/2021MNRAS.502.5921S} {502, 5921}

\bibitem[\protect\citeauthoryear{{Sales}, {Navarro}, {Pe{\~n}afiel}, {Peng},
  {Lim}  \& {Hernquist}}{{Sales} et~al.}{2020}]{Sales20}
{Sales} L.~V.,  {Navarro} J.~F.,  {Pe{\~n}afiel} L.,  {Peng} E.~W.,  {Lim} S.,
   {Hernquist} L.,  2020, \mn@doi [\mnras] {10.1093/mnras/staa854}, \href
  {https://ui.adsabs.harvard.edu/abs/2020MNRAS.494.1848S} {494, 1848}

\bibitem[\protect\citeauthoryear{{Secker} \& {Harris}}{{Secker} \&
  {Harris}}{1993}]{secker}
{Secker} J.,  {Harris} W.~E.,  1993, \mn@doi [\aj] {10.1086/116515}, \href
  {https://ui.adsabs.harvard.edu/abs/1993AJ....105.1358S} {105, 1358}

\bibitem[\protect\citeauthoryear{{Sengupta}, {Scott}, {Chung}  \&
  {Wong}}{{Sengupta} et~al.}{2019}]{sengupta}
{Sengupta} C.,  {Scott} T.~C.,  {Chung} A.,   {Wong} O.~I.,  2019, \mn@doi
  [\mnras] {10.1093/mnras/stz1884}, \href
  {https://ui.adsabs.harvard.edu/abs/2019MNRAS.488.3222S} {488, 3222}

\bibitem[\protect\citeauthoryear{{Sif{\'o}n}, {van der Burg}, {Hoekstra},
  {Muzzin}  \& {Herbonnet}}{{Sif{\'o}n} et~al.}{2018}]{sifon}
{Sif{\'o}n} C.,  {van der Burg} R. F.~J.,  {Hoekstra} H.,  {Muzzin} A.,
  {Herbonnet} R.,  2018, \mn@doi [\mnras] {10.1093/mnras/stx2648}, \href
  {https://ui.adsabs.harvard.edu/abs/2018MNRAS.473.3747S} {473, 3747}

\bibitem[\protect\citeauthoryear{{Singh}, {Zaritsky}, {Donnerstein}  \&
  {Spekkens}}{{Singh} et~al.}{2019}]{singh}
{Singh} P.~R.,  {Zaritsky} D.,  {Donnerstein} R.,   {Spekkens} K.,  2019,
  \mn@doi [\aj] {10.3847/1538-3881/ab16f2}, \href
  {https://ui.adsabs.harvard.edu/abs/2019AJ....157..212S} {157, 212}

\bibitem[\protect\citeauthoryear{{Somalwar}, {Greene}, {Greco}, {Huang},
  {Beaton}, {Goulding}  \& {Lancaster}}{{Somalwar} et~al.}{2020}]{somalwar}
{Somalwar} J.~J.,  {Greene} J.~E.,  {Greco} J.~P.,  {Huang} S.,  {Beaton}
  R.~L.,  {Goulding} A.~D.,   {Lancaster} L.,  2020, \mn@doi [\apj]
  {10.3847/1538-4357/abb1b2}, \href
  {https://ui.adsabs.harvard.edu/abs/2020ApJ...902...45S} {902, 45}

\bibitem[\protect\citeauthoryear{{Spekkens} \& {Karunakaran}}{{Spekkens} \&
  {Karunakaran}}{2018}]{spekkens}
{Spekkens} K.,  {Karunakaran} A.,  2018, \mn@doi [\apj]
  {10.3847/1538-4357/aa94be}, \href
  {https://ui.adsabs.harvard.edu/abs/2018ApJ...855...28S} {855, 28}

\bibitem[\protect\citeauthoryear{{Struble} \& {Rood}}{{Struble} \&
  {Rood}}{1999}]{coma-vel}
{Struble} M.~F.,  {Rood} H.~J.,  1999, \mn@doi [\apjs] {10.1086/313274}, \href
  {https://ui.adsabs.harvard.edu/abs/1999ApJS..125...35S} {125, 35}

\bibitem[\protect\citeauthoryear{{Thomsen}, {Baum}, {Hammergren}  \&
  {Worthey}}{{Thomsen} et~al.}{1997}]{distcoma1}
{Thomsen} B.,  {Baum} W.~A.,  {Hammergren} M.,   {Worthey} G.,  1997, \mn@doi
  [\apjl] {10.1086/310735}, \href
  {https://ui.adsabs.harvard.edu/abs/1997ApJ...483L..37T} {483, L37}

\bibitem[\protect\citeauthoryear{{Toloba} et~al.,}{{Toloba}
  et~al.}{2018}]{toloba}
{Toloba} E.,  et~al., 2018, \mn@doi [\apjl] {10.3847/2041-8213/aab603}, \href
  {https://ui.adsabs.harvard.edu/abs/2018ApJ...856L..31T} {856, L31}

\bibitem[\protect\citeauthoryear{{Tremmel}, {Wright}, {Brooks}, {Munshi},
  {Nagai}  \& {Quinn}}{{Tremmel} et~al.}{2020}]{tremmel}
{Tremmel} M.,  {Wright} A.~C.,  {Brooks} A.~M.,  {Munshi} F.,  {Nagai} D.,
  {Quinn} T.~R.,  2020, \mn@doi [\mnras] {10.1093/mnras/staa2015}, \href
  {https://ui.adsabs.harvard.edu/abs/2020MNRAS.497.2786T} {497, 2786}

\bibitem[\protect\citeauthoryear{{Trujillo-Gomez}, {Kruijssen}  \&
  {Reina-Campos}}{{Trujillo-Gomez} et~al.}{2022}]{gomez}
{Trujillo-Gomez} S.,  {Kruijssen} J.~M.~D.,   {Reina-Campos} M.,  2022, \mn@doi
  [\mnras] {10.1093/mnras/stab3401}, \href
  {https://ui.adsabs.harvard.edu/abs/2022MNRAS.510.3356T} {510, 3356}

\bibitem[\protect\citeauthoryear{{Trujillo}, {Graham}  \& {Caon}}{{Trujillo}
  et~al.}{2001}]{trujilo2001}
{Trujillo} I.,  {Graham} A.~W.,   {Caon} N.,  2001, \mn@doi [\mnras]
  {10.1046/j.1365-8711.2001.04471.x}, \href
  {https://ui.adsabs.harvard.edu/abs/2001MNRAS.326..869T} {326, 869}

\bibitem[\protect\citeauthoryear{{Trujillo}, {Roman}, {Filho}  \& {S{\'a}nchez
  Almeida}}{{Trujillo} et~al.}{2017}]{udgsfr1}
{Trujillo} I.,  {Roman} J.,  {Filho} M.,   {S{\'a}nchez Almeida} J.,  2017,
  \mn@doi [\apj] {10.3847/1538-4357/aa5cbb}, \href
  {https://ui.adsabs.harvard.edu/abs/2017ApJ...836..191T} {836, 191}

\bibitem[\protect\citeauthoryear{{Trujillo} et~al.,}{{Trujillo}
  et~al.}{2021}]{2021A&A...654A..40T}
{Trujillo} I.,  et~al., 2021, \mn@doi [\aap] {10.1051/0004-6361/202141603},
  \href {https://ui.adsabs.harvard.edu/abs/2021A&A...654A..40T} {654, A40}

\bibitem[\protect\citeauthoryear{{Valenzuela}, {Moster}, {Remus}, {O'Leary}  \&
  {Burkert}}{{Valenzuela} et~al.}{2021}]{lucas}
{Valenzuela} L.~M.,  {Moster} B.~P.,  {Remus} R.-S.,  {O'Leary} J.~A.,
  {Burkert} A.,  2021, \mn@doi [\mnras] {10.1093/mnras/stab1701}, \href
  {https://ui.adsabs.harvard.edu/abs/2021MNRAS.505.5815V} {505, 5815}

\bibitem[\protect\citeauthoryear{{Van Nest}, {Munshi}, {Wright}, {Tremmel},
  {Brooks}, {Nagai}  \& {Quinn}}{{Van Nest} et~al.}{2021}]{vannest}
{Van Nest} J.~D.,  {Munshi} F.,  {Wright} A.~C.,  {Tremmel} M.,  {Brooks}
  A.~M.,  {Nagai} D.,   {Quinn} T.,  2021, arXiv e-prints, \href
  {https://ui.adsabs.harvard.edu/abs/2021arXiv210812985V} {p. arXiv:2108.12985}

\bibitem[\protect\citeauthoryear{{Vazdekis}, {S{\'a}nchez-Bl{\'a}zquez},
  {Falc{\'o}n-Barroso}, {Cenarro}, {Beasley}, {Cardiel}, {Gorgas}  \&
  {Peletier}}{{Vazdekis} et~al.}{2010}]{vazdekis}
{Vazdekis} A.,  {S{\'a}nchez-Bl{\'a}zquez} P.,  {Falc{\'o}n-Barroso} J.,
  {Cenarro} A.~J.,  {Beasley} M.~A.,  {Cardiel} N.,  {Gorgas} J.,   {Peletier}
  R.~F.,  2010, \mn@doi [\mnras] {10.1111/j.1365-2966.2010.16407.x}, \href
  {https://ui.adsabs.harvard.edu/abs/2010MNRAS.404.1639V} {404, 1639}

\bibitem[\protect\citeauthoryear{{Vazdekis}, {Koleva}, {Ricciardelli},
  {R{\"o}ck}  \& {Falc{\'o}n-Barroso}}{{Vazdekis} et~al.}{2016}]{vazdekis2016}
{Vazdekis} A.,  {Koleva} M.,  {Ricciardelli} E.,  {R{\"o}ck} B.,
  {Falc{\'o}n-Barroso} J.,  2016, \mn@doi [\mnras] {10.1093/mnras/stw2231},
  \href {https://ui.adsabs.harvard.edu/abs/2016MNRAS.463.3409V} {463, 3409}

\bibitem[\protect\citeauthoryear{{Venhola} et~al.,}{{Venhola}
  et~al.}{2017}]{venhola2017}
{Venhola} A.,  et~al., 2017, \mn@doi [\aap] {10.1051/0004-6361/201730696},
  \href {https://ui.adsabs.harvard.edu/abs/2017A&A...608A.142V} {608, A142}

\bibitem[\protect\citeauthoryear{{Villaume} et~al.,}{{Villaume}
  et~al.}{2022}]{alexa}
{Villaume} A.,  et~al., 2022, \mn@doi [\apj] {10.3847/1538-4357/ac341e}, \href
  {https://ui.adsabs.harvard.edu/abs/2022ApJ...924...32V} {924, 32}

\bibitem[\protect\citeauthoryear{Virtanen et~al.,}{Virtanen
  et~al.}{2020}]{scipy}
Virtanen P.,  et~al., 2020, \mn@doi [Nature Methods]
  {10.1038/s41592-019-0686-2}, \href {https://rdcu.be/b08Wh} {17, 261}

\bibitem[\protect\citeauthoryear{{Wenger} et~al.,}{{Wenger}
  et~al.}{2000}]{simbad}
{Wenger} M.,  et~al., 2000, \mn@doi [\aaps] {10.1051/aas:2000332}, \href
  {https://ui.adsabs.harvard.edu/abs/2000A&AS..143....9W} {143, 9}

\bibitem[\protect\citeauthoryear{{Wright}, {Tremmel}, {Brooks}, {Munshi},
  {Nagai}, {Sharma}  \& {Quinn}}{{Wright} et~al.}{2021}]{wright}
{Wright} A.~C.,  {Tremmel} M.,  {Brooks} A.~M.,  {Munshi} F.,  {Nagai} D.,
  {Sharma} R.~S.,   {Quinn} T.~R.,  2021, \mn@doi [\mnras]
  {10.1093/mnras/stab081}, \href
  {https://ui.adsabs.harvard.edu/abs/2021MNRAS.502.5370W} {502, 5370}

\bibitem[\protect\citeauthoryear{{Yagi}, {Koda}, {Komiyama}  \&
  {Yamanoi}}{{Yagi} et~al.}{2016}]{yagi}
{Yagi} M.,  {Koda} J.,  {Komiyama} Y.,   {Yamanoi} H.,  2016, \mn@doi [\apjs]
  {10.3847/0067-0049/225/1/11}, \href
  {https://ui.adsabs.harvard.edu/abs/2016ApJS..225...11Y} {225, 11}

\bibitem[\protect\citeauthoryear{{Yozin} \& {Bekki}}{{Yozin} \&
  {Bekki}}{2015}]{yozin}
{Yozin} C.,  {Bekki} K.,  2015, \mn@doi [\mnras] {10.1093/mnras/stv1073}, \href
  {https://ui.adsabs.harvard.edu/abs/2015MNRAS.452..937Y} {452, 937}

\bibitem[\protect\citeauthoryear{{Zaritsky}}{{Zaritsky}}{2017}]{dennis}
{Zaritsky} D.,  2017, \mn@doi [\mnras] {10.1093/mnrasl/slw198}, \href
  {https://ui.adsabs.harvard.edu/abs/2017MNRAS.464L.110Z} {464, L110}

\bibitem[\protect\citeauthoryear{{Zaritsky} et~al.,}{{Zaritsky}
  et~al.}{2019}]{smudges1}
{Zaritsky} D.,  et~al., 2019, \mn@doi [\apjs] {10.3847/1538-4365/aaefe9}, \href
  {https://ui.adsabs.harvard.edu/abs/2019ApJS..240....1Z} {240, 1}

\bibitem[\protect\citeauthoryear{{Zaritsky}, {Donnerstein}, {Karunakaran},
  {Barbosa}, {Dey}, {Kadowaki}, {Spekkens}  \& {Zhang}}{{Zaritsky}
  et~al.}{2021}]{dennis2}
{Zaritsky} D.,  {Donnerstein} R.,  {Karunakaran} A.,  {Barbosa} C.~E.,  {Dey}
  A.,  {Kadowaki} J.,  {Spekkens} K.,   {Zhang} H.,  2021, \mn@doi [\apjs]
  {10.3847/1538-4365/ac2607}, \href
  {https://ui.adsabs.harvard.edu/abs/2021ApJS..257...60Z} {257, 60}

\bibitem[\protect\citeauthoryear{{van Dokkum}}{{van Dokkum}}{2001}]{cosmic}
{van Dokkum} P.~G.,  2001, \mn@doi [\pasp] {10.1086/323894}, \href
  {https://ui.adsabs.harvard.edu/abs/2001PASP..113.1420V} {113, 1420}

\bibitem[\protect\citeauthoryear{{van Dokkum}, {Abraham}, {Merritt}, {Zhang},
  {Geha}  \& {Conroy}}{{van Dokkum} et~al.}{2015a}]{vd15}
{van Dokkum} P.~G.,  {Abraham} R.,  {Merritt} A.,  {Zhang} J.,  {Geha} M.,
  {Conroy} C.,  2015a, \mn@doi [\apjl] {10.1088/2041-8205/798/2/L45}, \href
  {https://ui.adsabs.harvard.edu/abs/2015ApJ...798L..45V} {798, L45}

\bibitem[\protect\citeauthoryear{{van Dokkum} et~al.,}{{van Dokkum}
  et~al.}{2015b}]{vd15b}
{van Dokkum} P.~G.,  et~al., 2015b, \mn@doi [\apjl]
  {10.1088/2041-8205/804/1/L26}, \href
  {https://ui.adsabs.harvard.edu/abs/2015ApJ...804L..26V} {804, L26}

\bibitem[\protect\citeauthoryear{{van Dokkum} et~al.,}{{van Dokkum}
  et~al.}{2016}]{vd16}
{van Dokkum} P.,  et~al., 2016, \mn@doi [\apjl] {10.3847/2041-8205/828/1/L6},
  \href {https://ui.adsabs.harvard.edu/abs/2016ApJ...828L...6V} {828, L6}

\bibitem[\protect\citeauthoryear{{van Dokkum} et~al.,}{{van Dokkum}
  et~al.}{2017}]{vd17}
{van Dokkum} P.,  et~al., 2017, \mn@doi [\apjl] {10.3847/2041-8213/aa7ca2},
  \href {https://ui.adsabs.harvard.edu/abs/2017ApJ...844L..11V} {844, L11}

\bibitem[\protect\citeauthoryear{{van Dokkum} et~al.,}{{van Dokkum}
  et~al.}{2019}]{vd19}
{van Dokkum} P.,  et~al., 2019, \mn@doi [\apj] {10.3847/1538-4357/ab2914},
  \href {https://ui.adsabs.harvard.edu/abs/2019ApJ...880...91V} {880, 91}

\bibitem[\protect\citeauthoryear{{van der Walt}, {Colbert}  \&
  {Varoquaux}}{{van der Walt} et~al.}{2011}]{numpy}
{van der Walt} S.,  {Colbert} S.~C.,   {Varoquaux} G.,  2011, \mn@doi
  [Computing in Science and Engineering] {10.1109/MCSE.2011.37}, \href
  {https://ui.adsabs.harvard.edu/abs/2011CSE....13b..22V} {13, 22}

\makeatother
\end{thebibliography}



\appendix
\section{S\'ersic models and profiles, GC selection using compactness and colour, completeness functions}

\begin{figure*}
\includegraphics[width=0.75\linewidth]{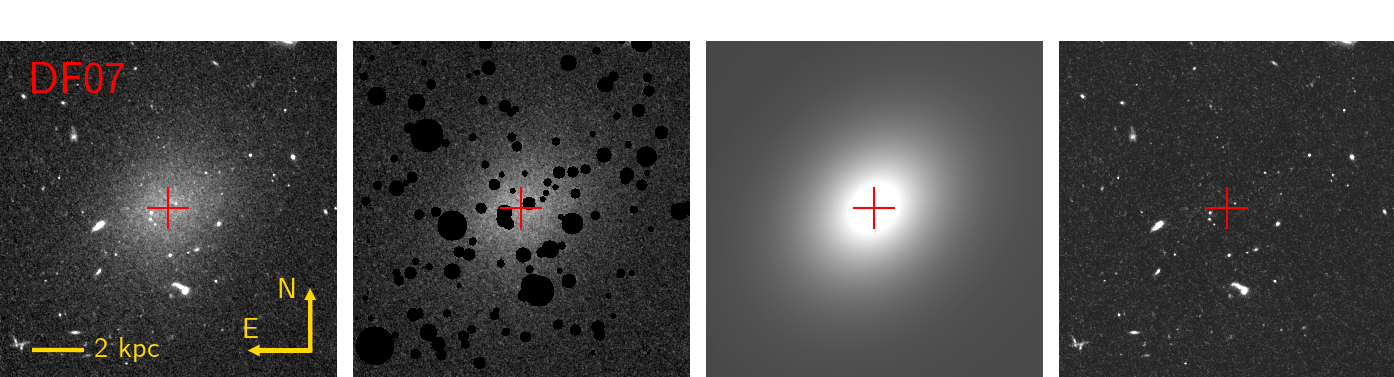}
\includegraphics[width=0.75\linewidth]{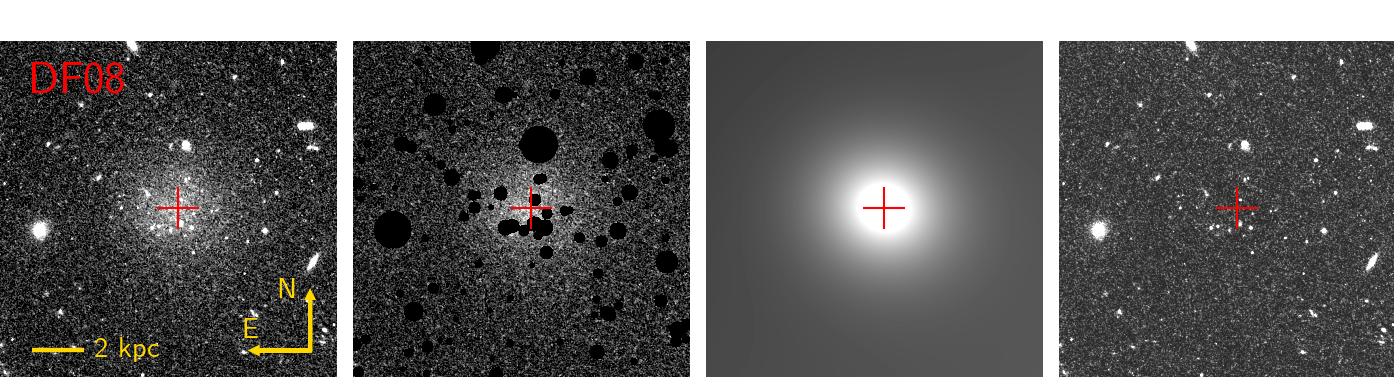}
\includegraphics[width=0.75\linewidth]{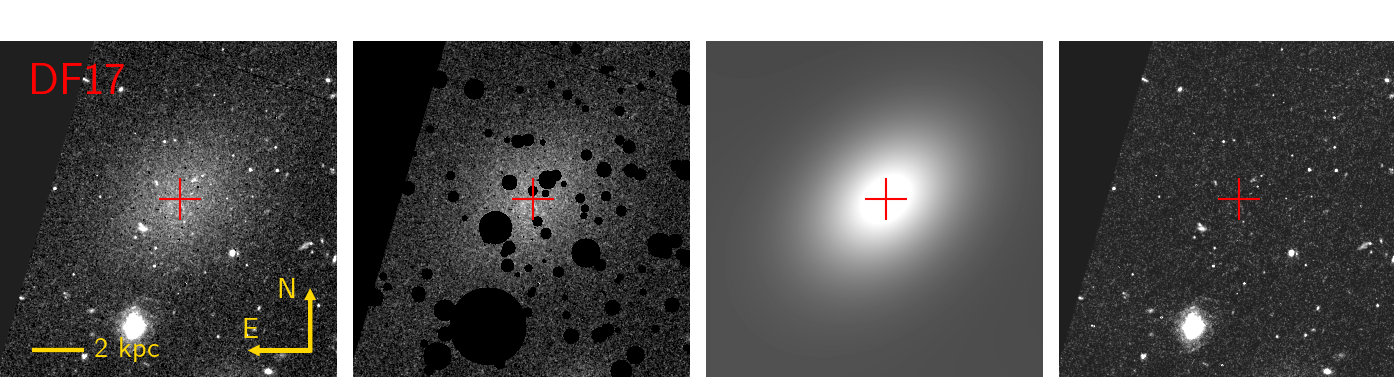}
\includegraphics[width=0.75\linewidth]{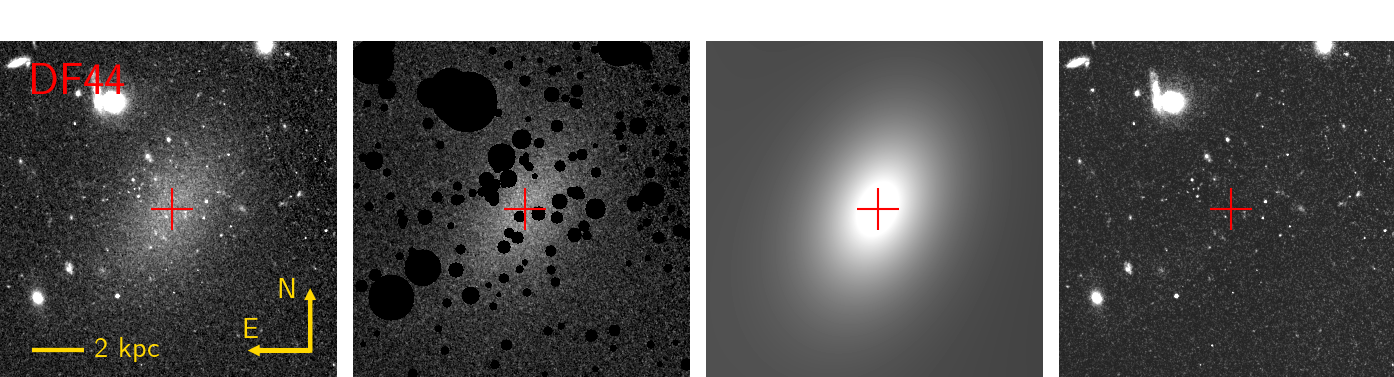}
\includegraphics[width=0.75\linewidth]{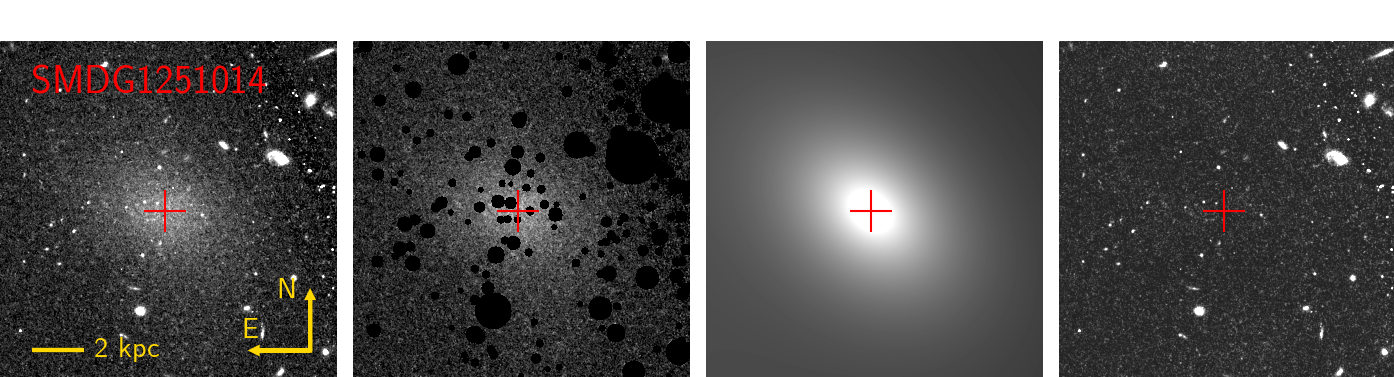}
\includegraphics[width=0.75\linewidth]{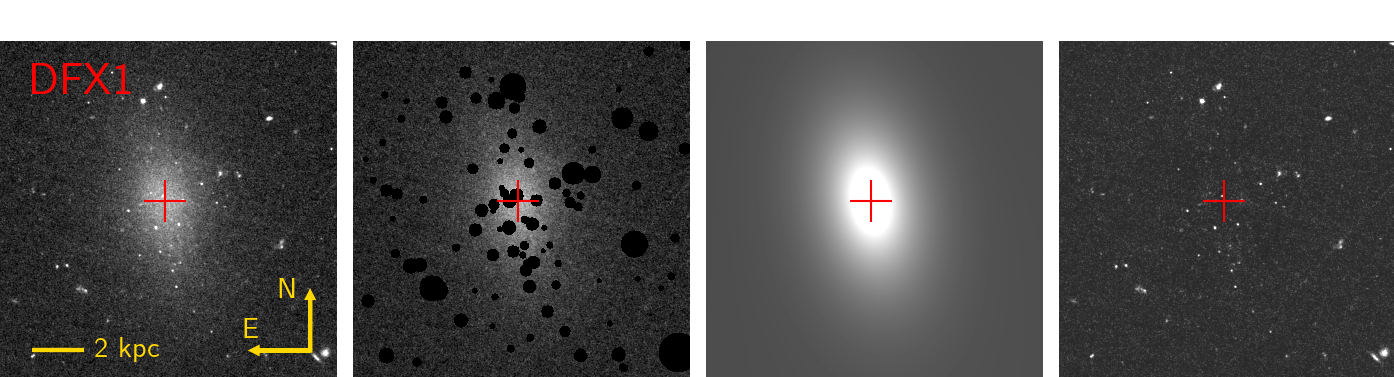}
\caption{Modeling the light profile of the galaxies with a S\'ersic profile in their primary filter. From left to right: the galaxy frame, the mask frame, the galaxy model and the residuals after subtracting the model from the main frame.}
\label{sersicmodel-app}
\end{figure*}

\begin{figure*}
\centering
\includegraphics[width=0.75\linewidth]{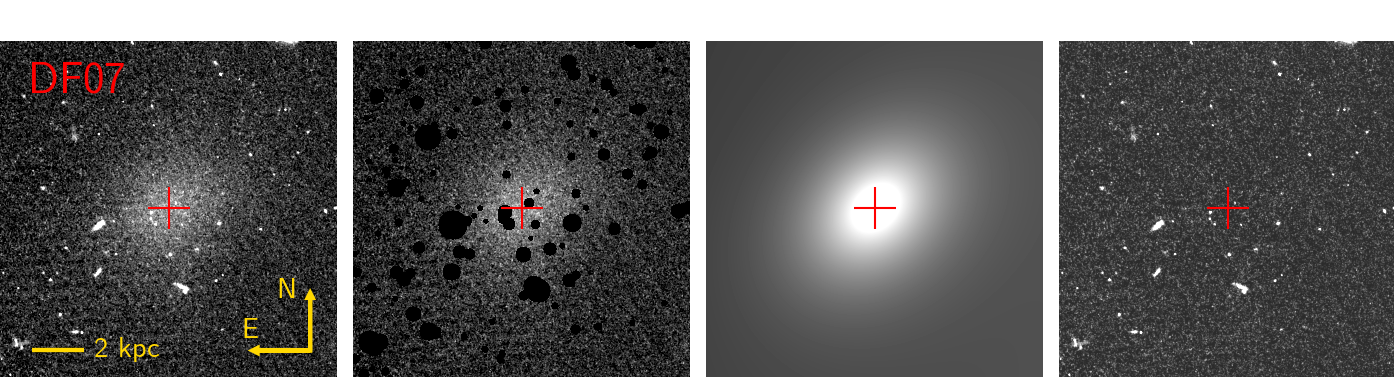}
\includegraphics[width=0.75\linewidth]{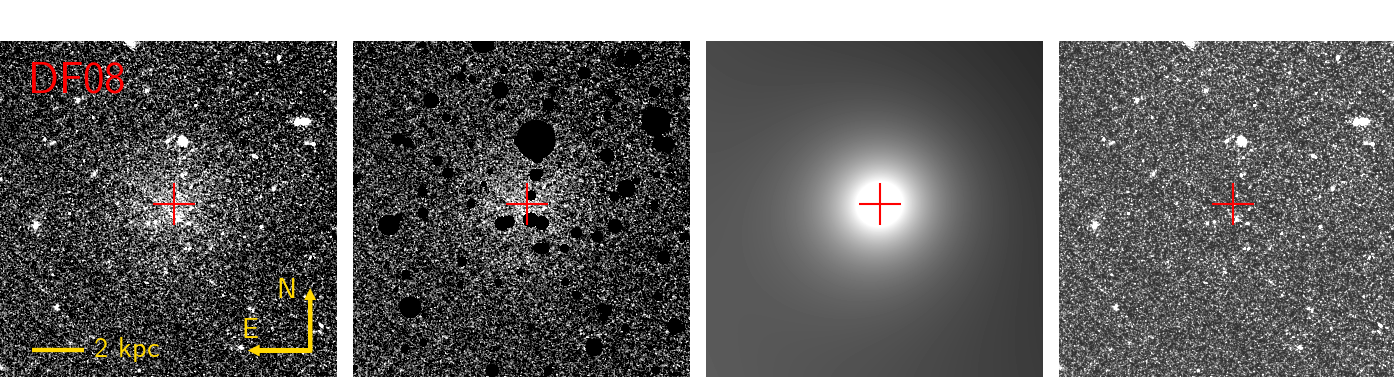}
\includegraphics[width=0.75\linewidth]{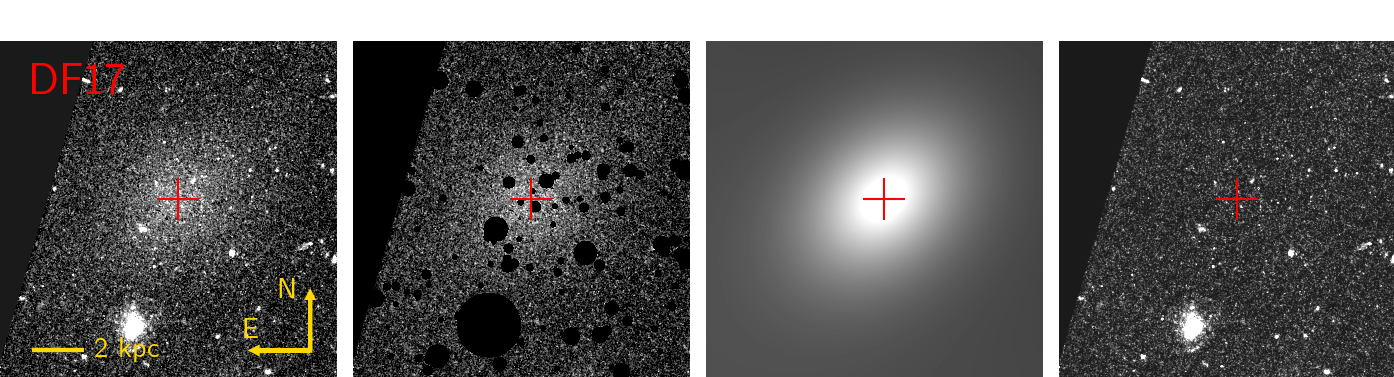}
\includegraphics[width=0.75\linewidth]{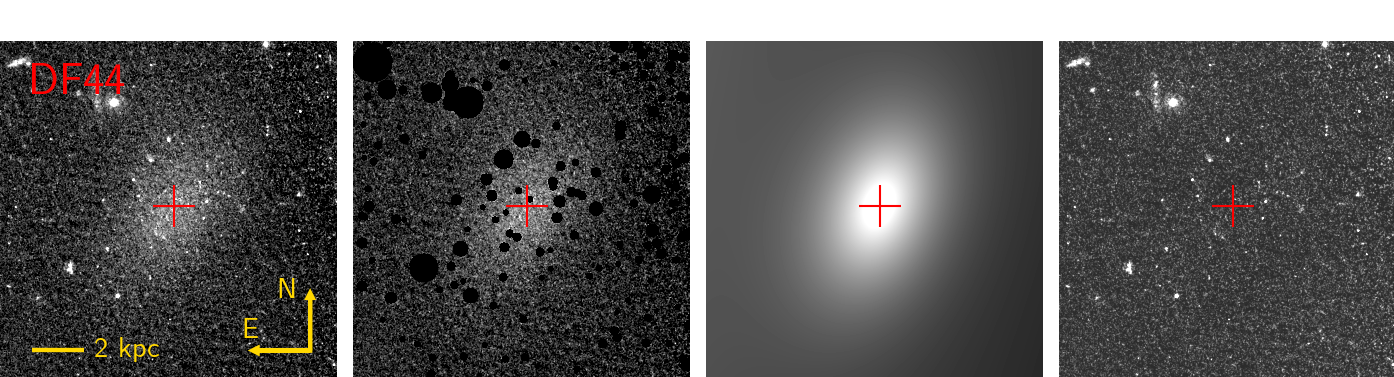}
\includegraphics[width=0.75\linewidth]{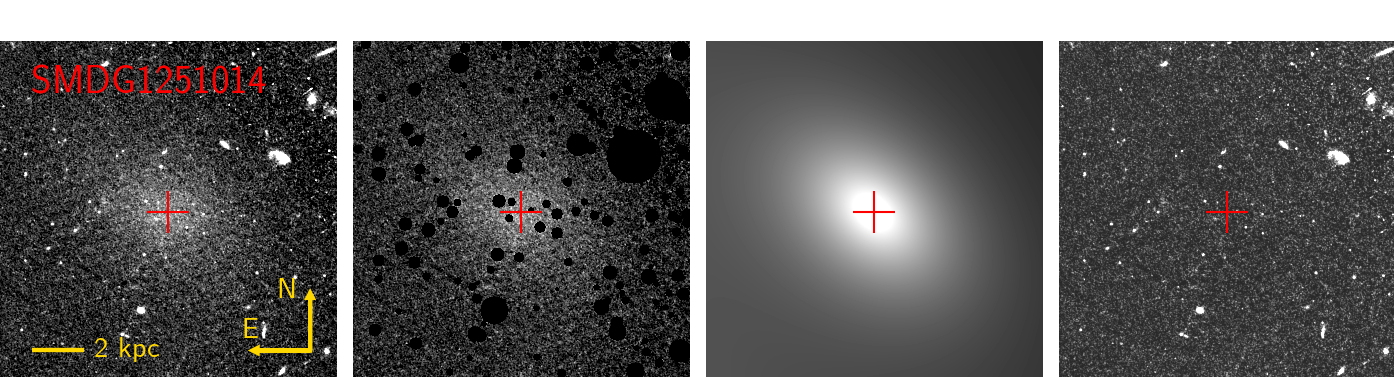}
\includegraphics[width=0.75\linewidth]{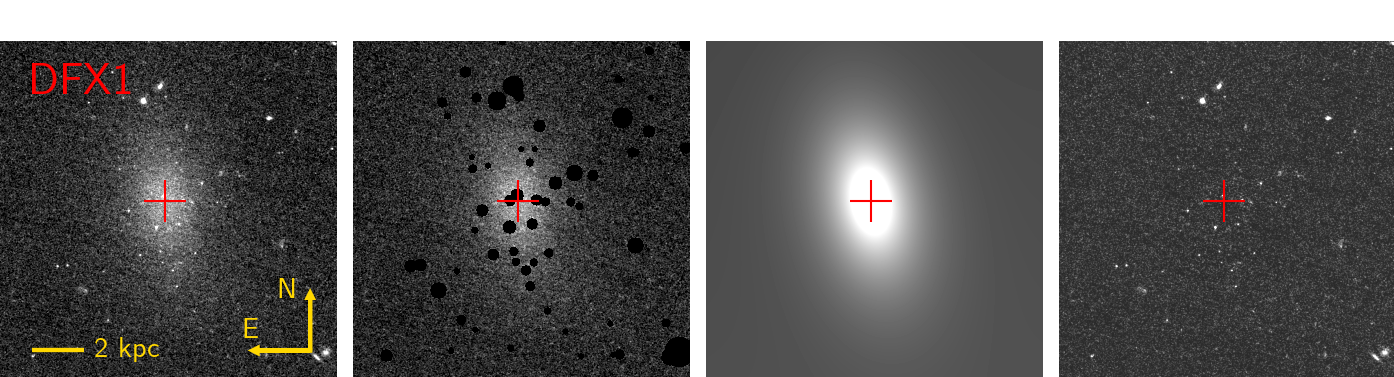}
\caption{As Fig.~\ref{sersicmodel-app}, now for the secondary filter.}
\label{sersicmodel2-app}
\end{figure*}

\begin{figure*}
\centering
\includegraphics[width=0.3\linewidth]{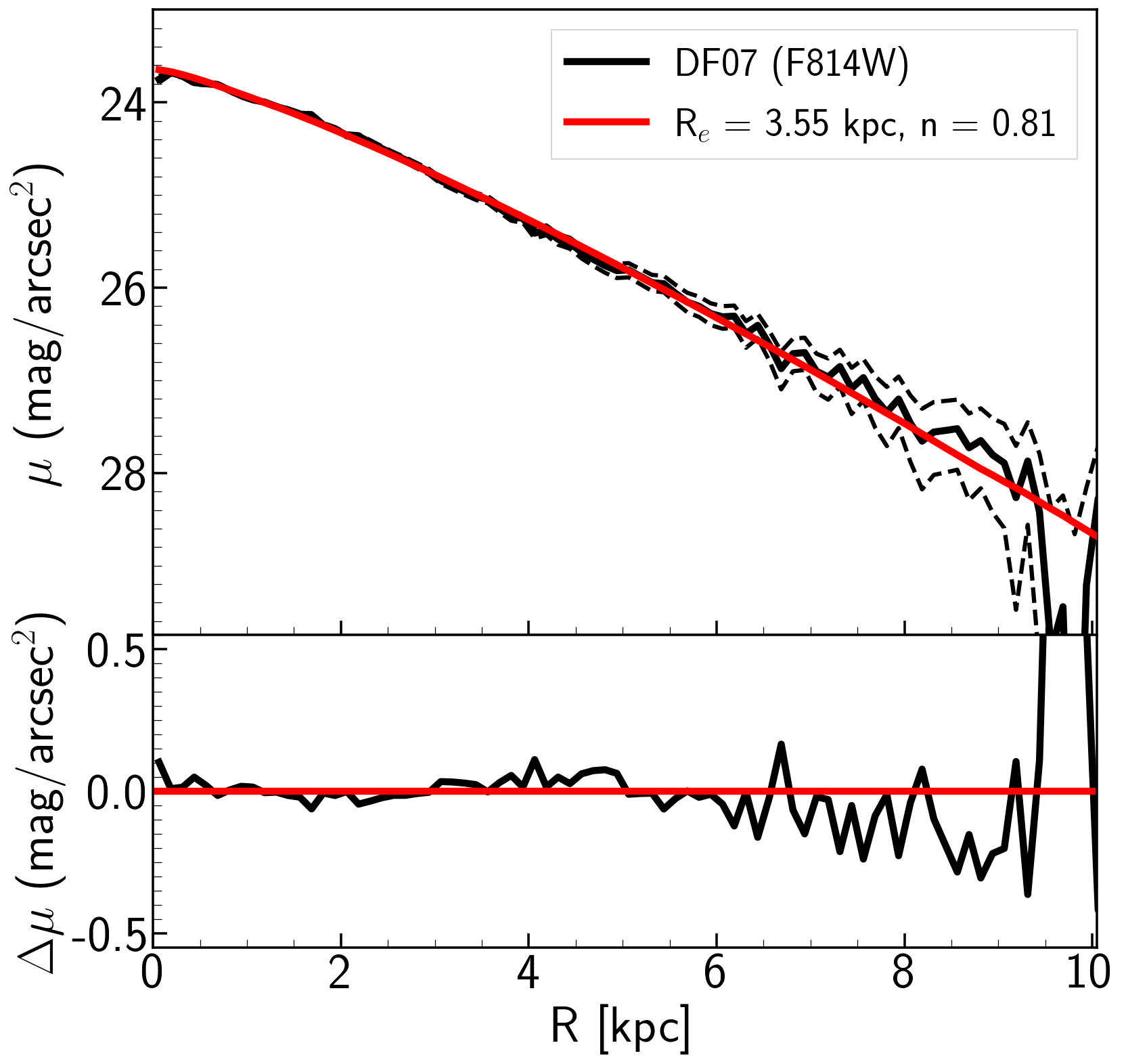}
\includegraphics[width=0.3\linewidth]{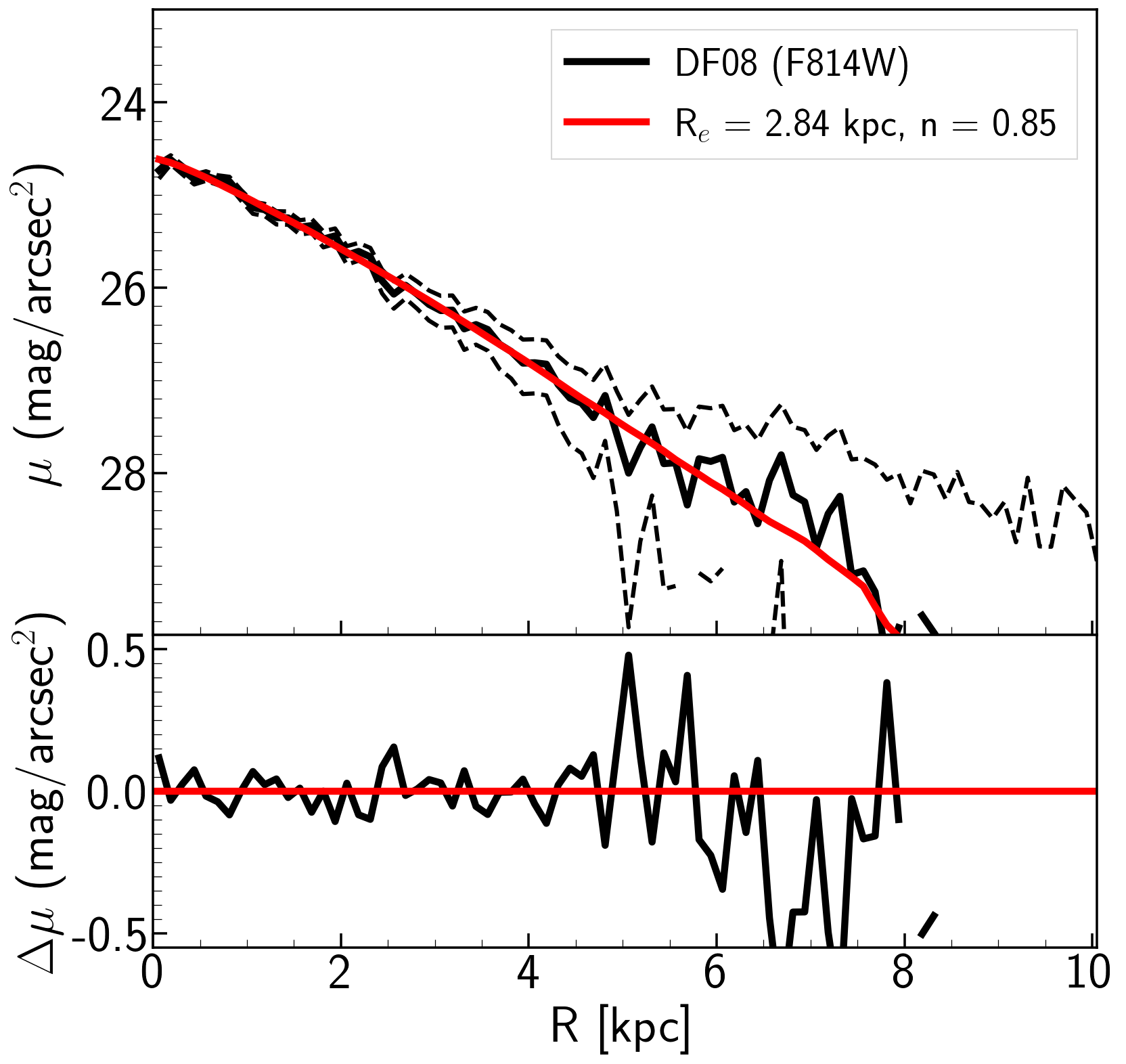}
\includegraphics[width=0.3\linewidth]{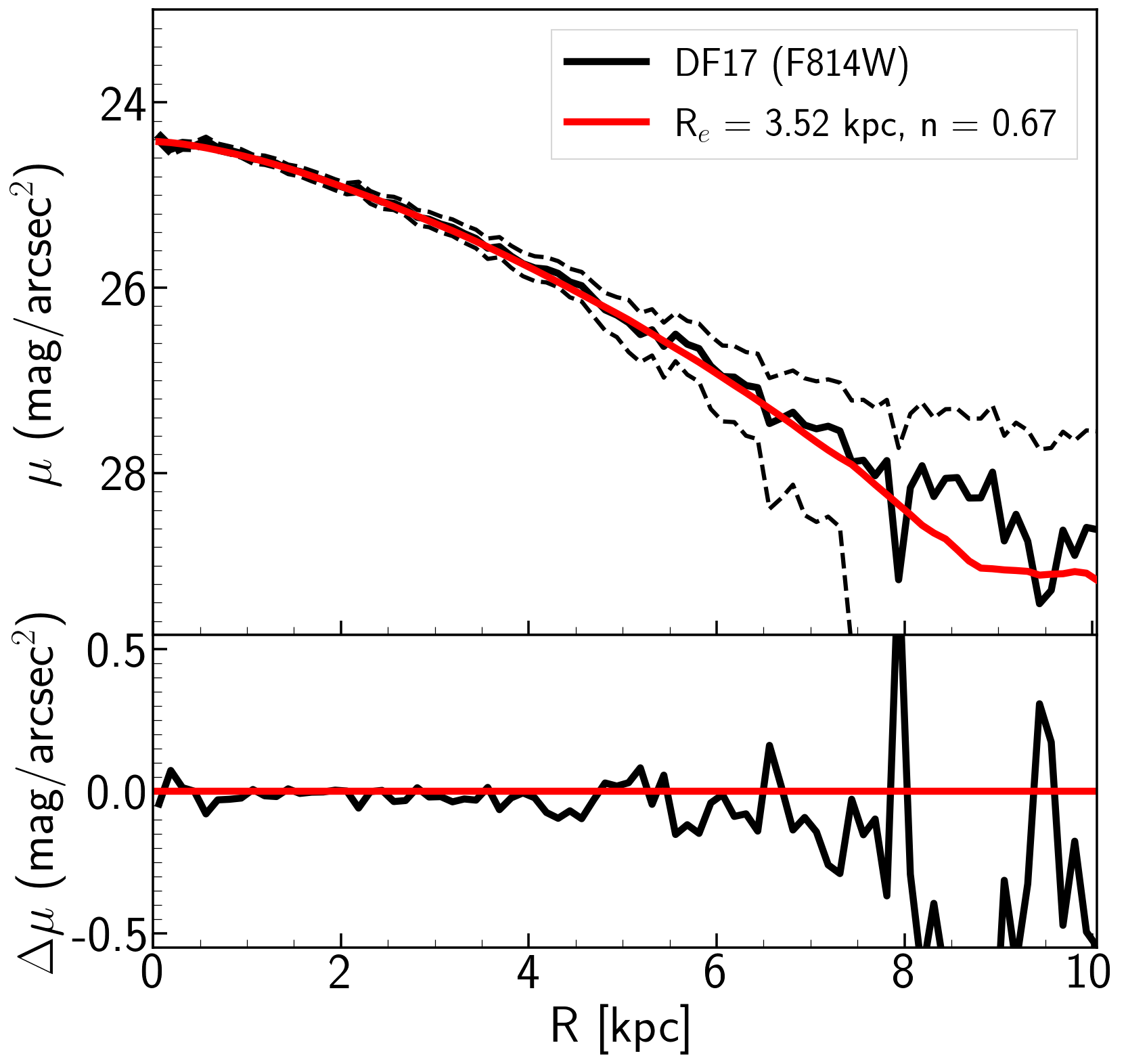}
\includegraphics[width=0.3\linewidth]{DF44_814_sersic_profile.png}
\includegraphics[width=0.3\linewidth]{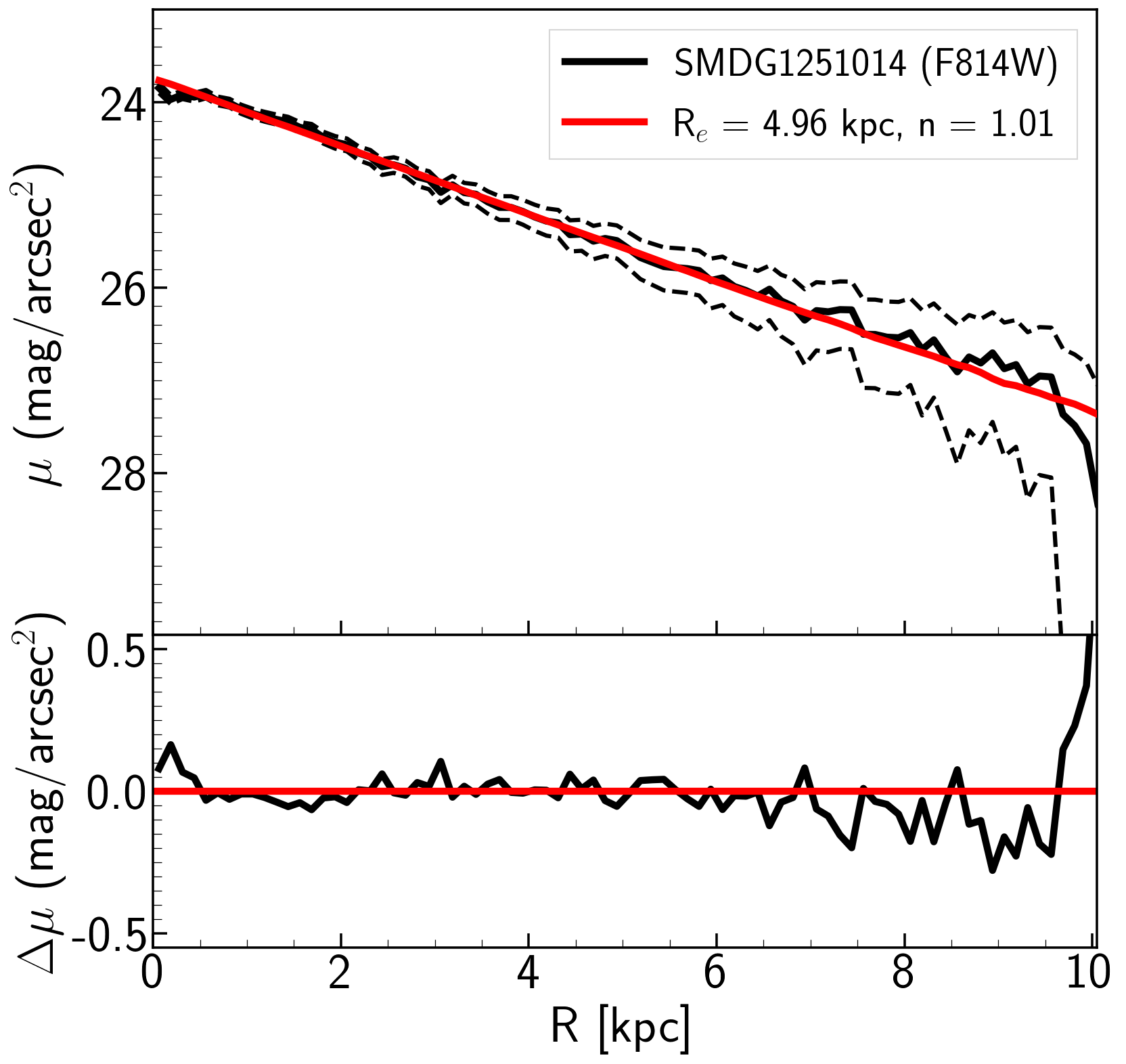}
\includegraphics[width=0.3\linewidth]{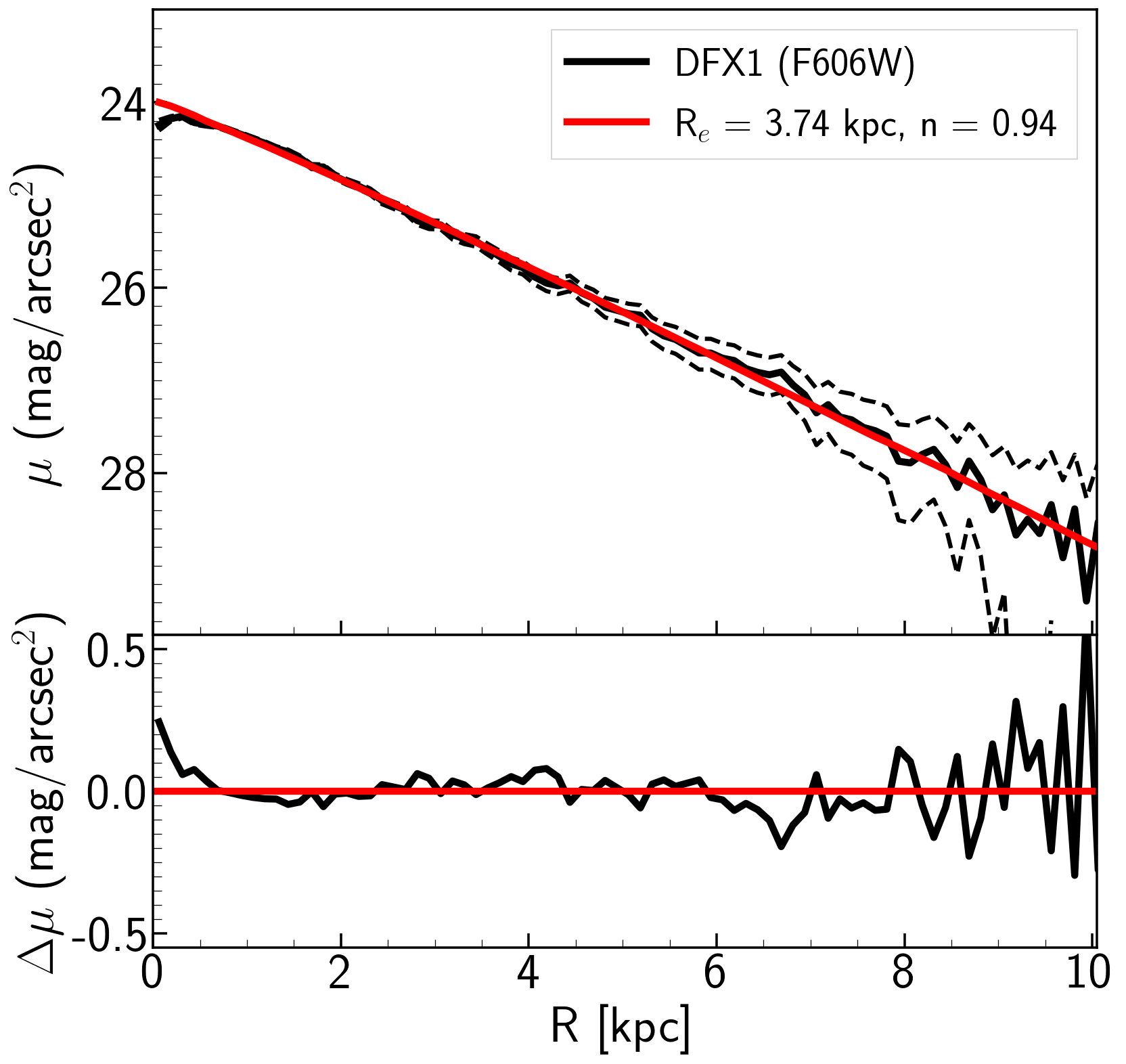}
\caption{S\'ersic profiles of the UDG sample in their primary filter. The light profile, best-fit S\'ersic function for UDGs and fitting residuals. The black dashed lines indicate the 1-sigma uncertainty of the light profile.}
\label{sersicprofile-app}
\end{figure*}

\newpage

\begin{figure*}
\centering
\includegraphics[width=0.3\linewidth]{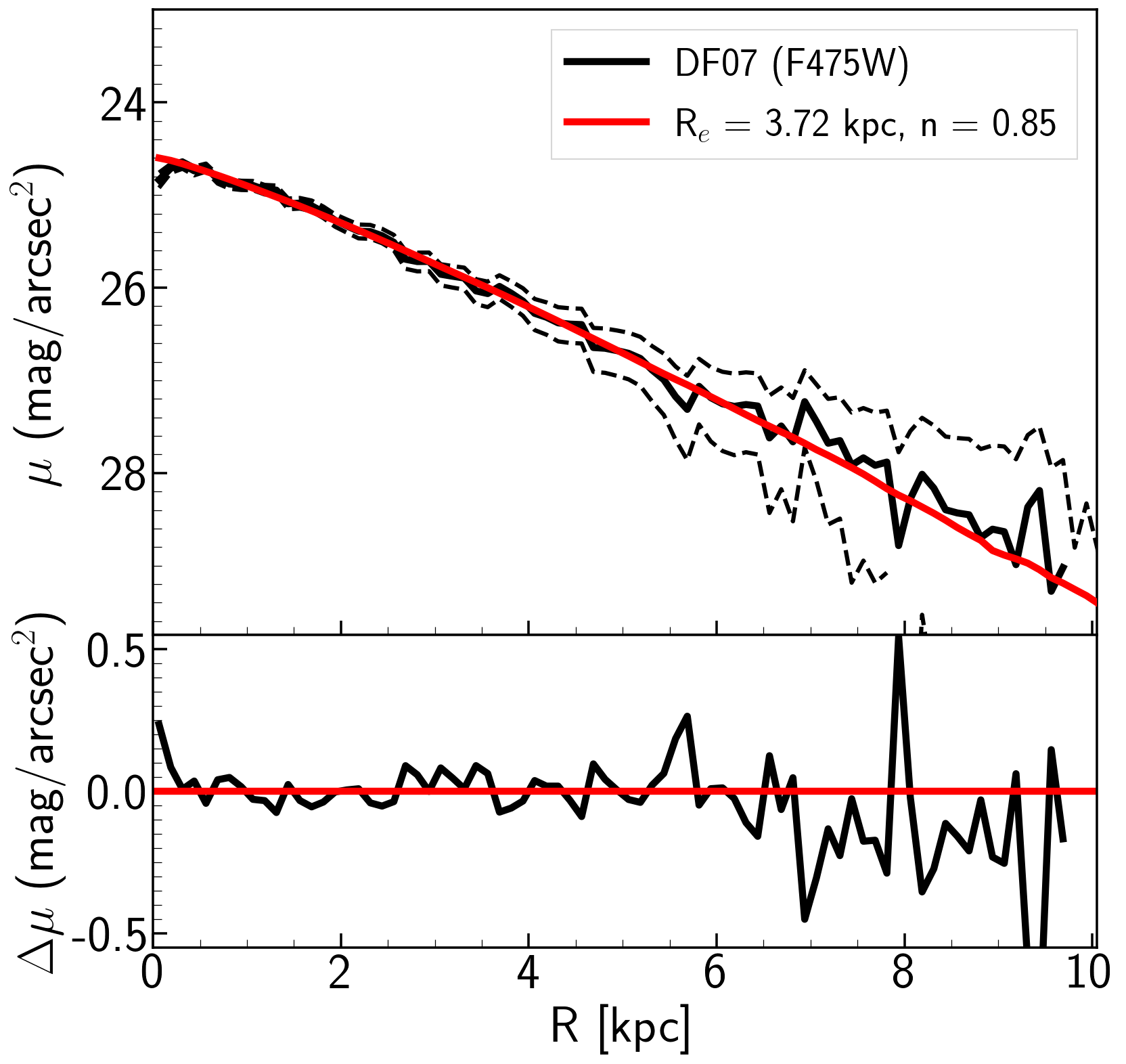}
\includegraphics[width=0.3\linewidth]{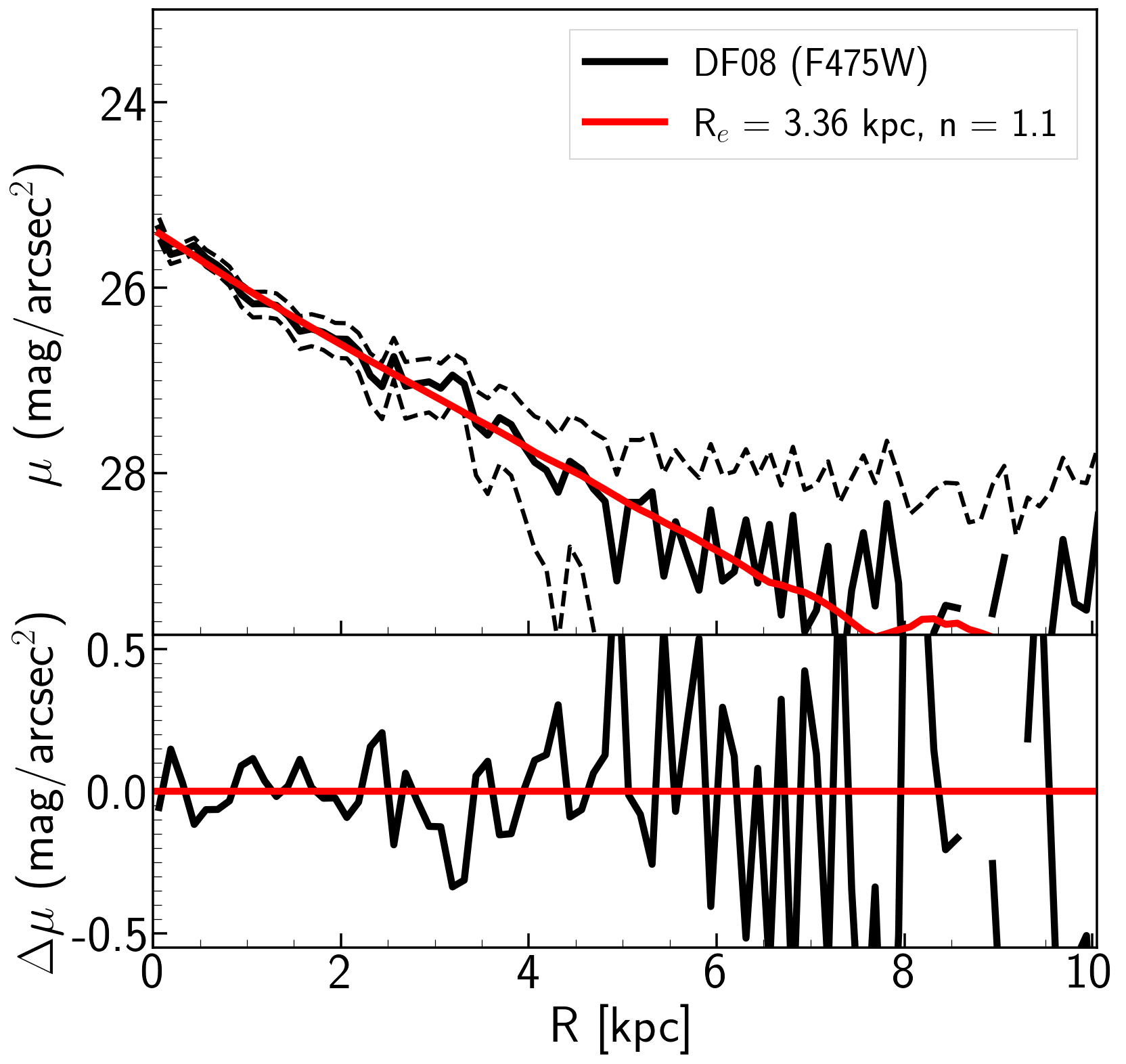}
\includegraphics[width=0.3\linewidth]{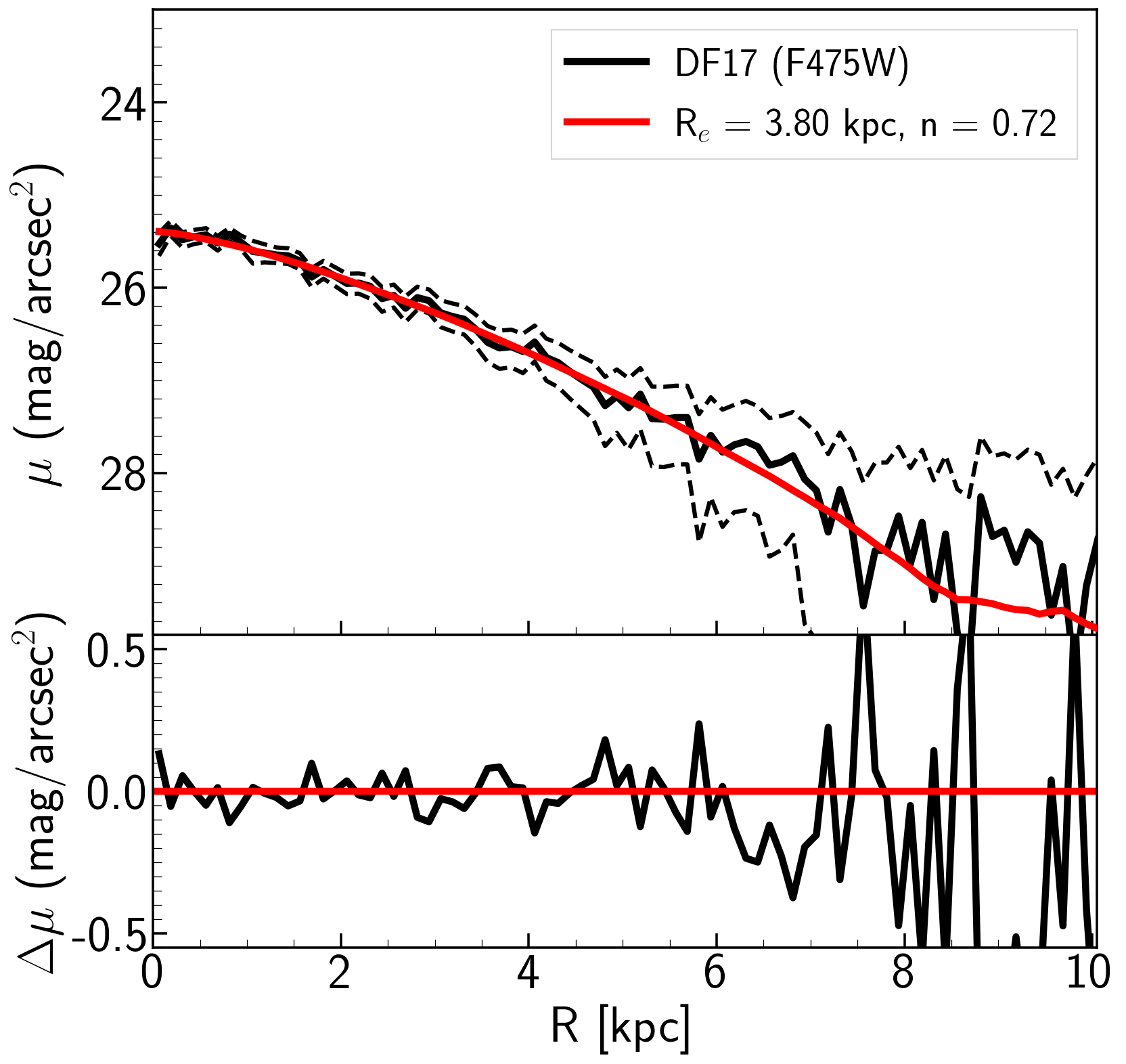}
\includegraphics[width=0.3\linewidth]{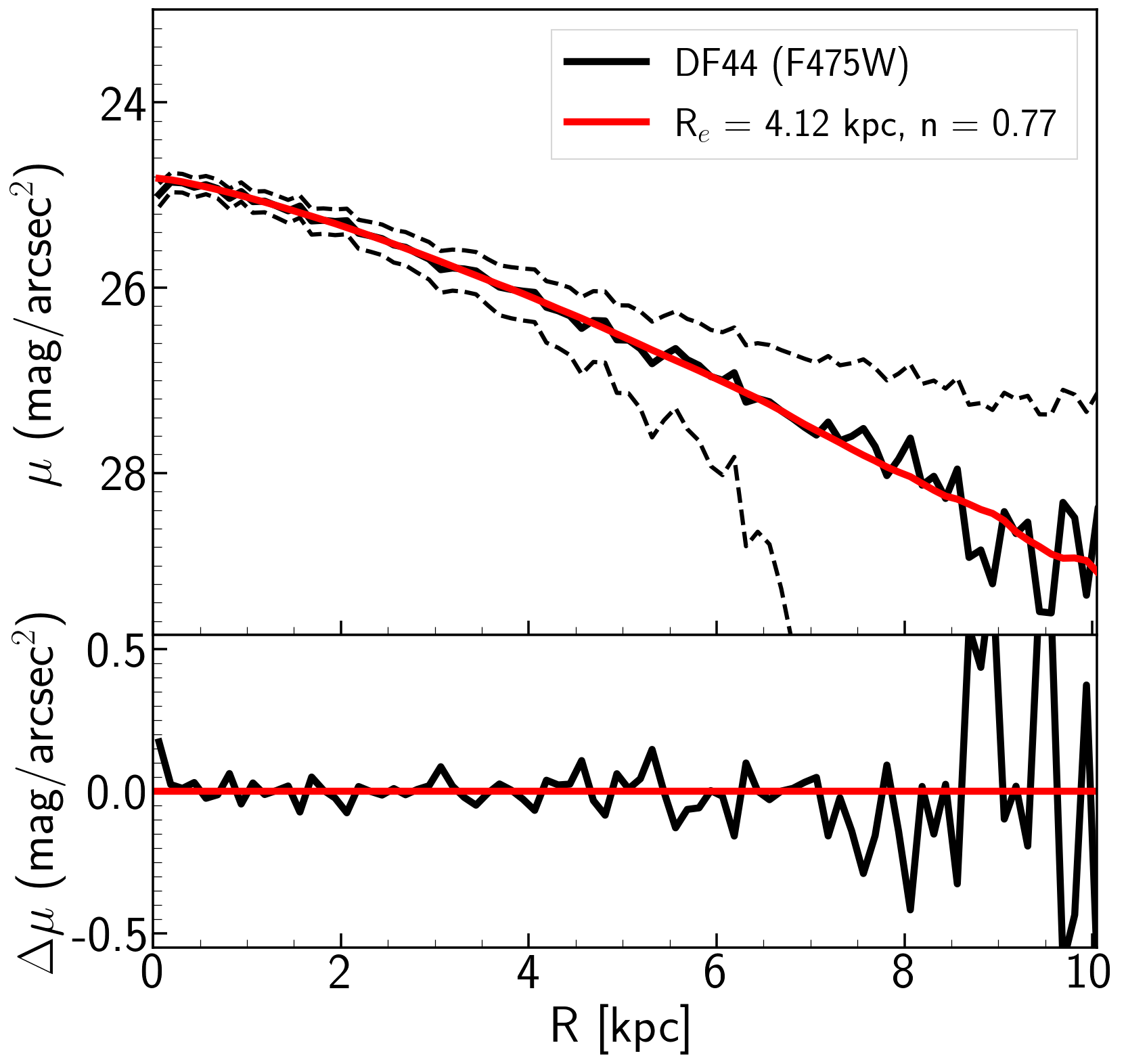}
\includegraphics[width=0.3\linewidth]{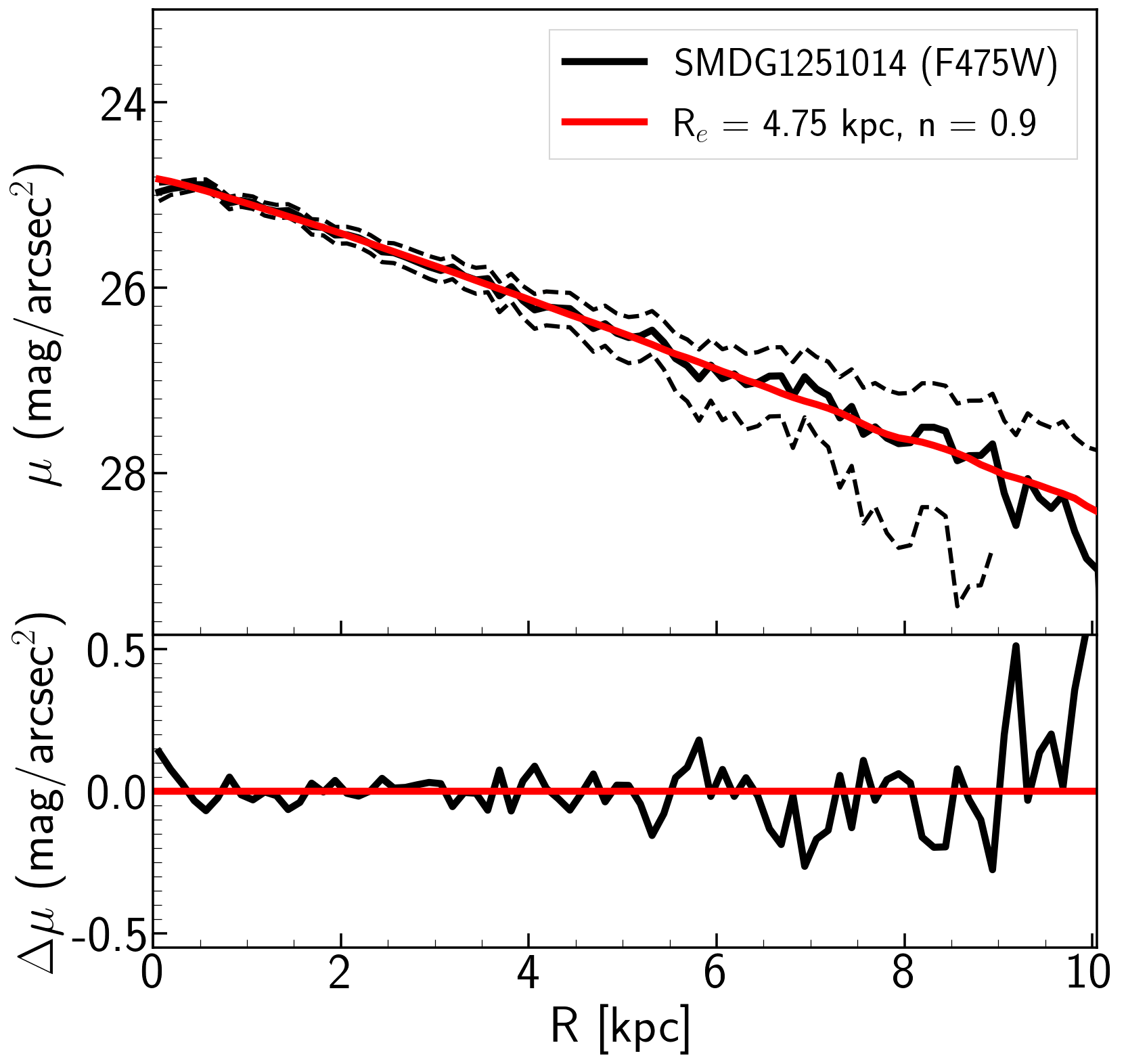}
\includegraphics[width=0.3\linewidth]{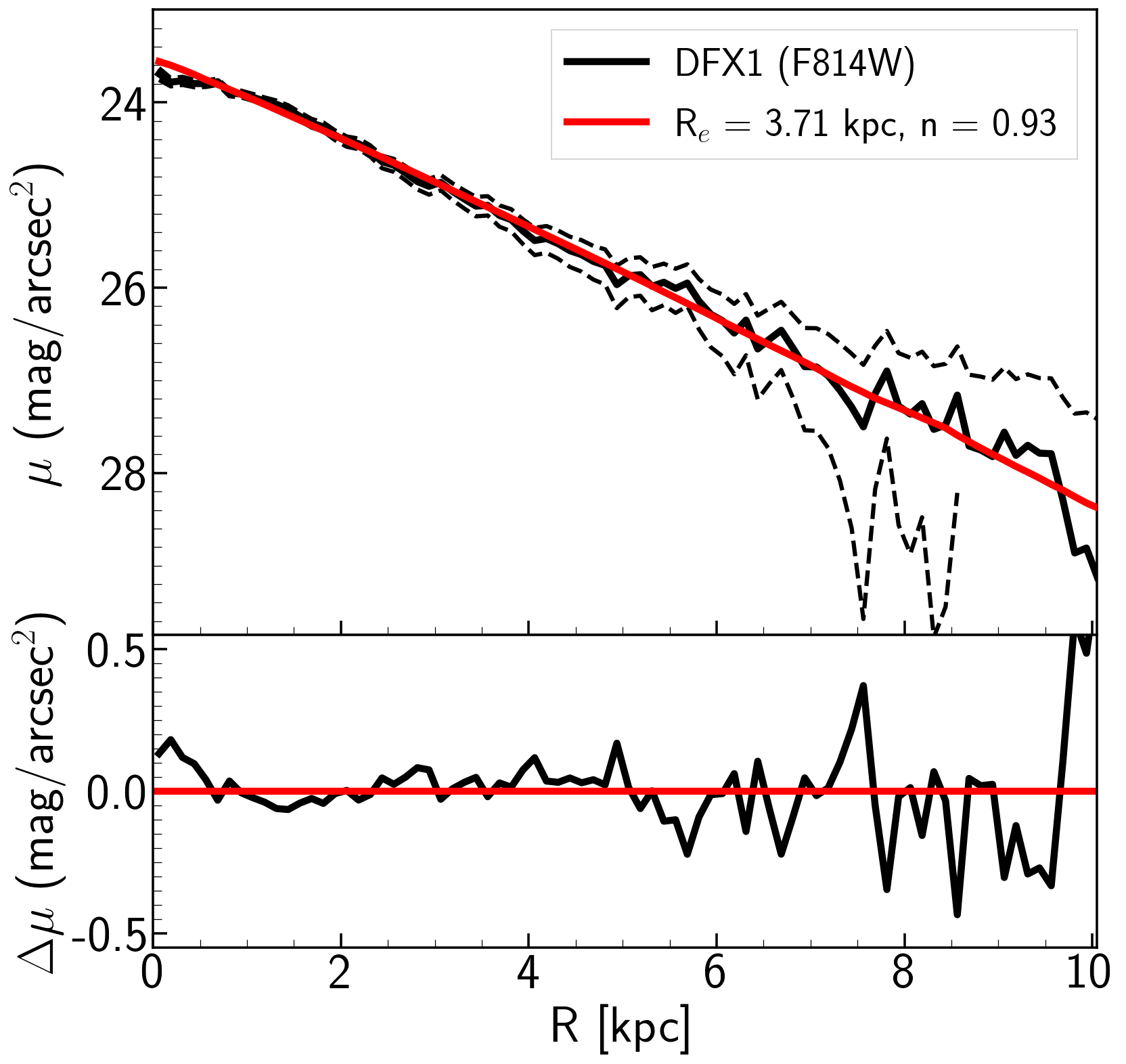}
\caption{As Fig.~\ref{sersicprofile-app}, now for the secondary filter.}
\label{sersicprofile2-app}
\end{figure*}

\newpage

\begin{figure*}
\centering
\includegraphics[width=0.49\linewidth]{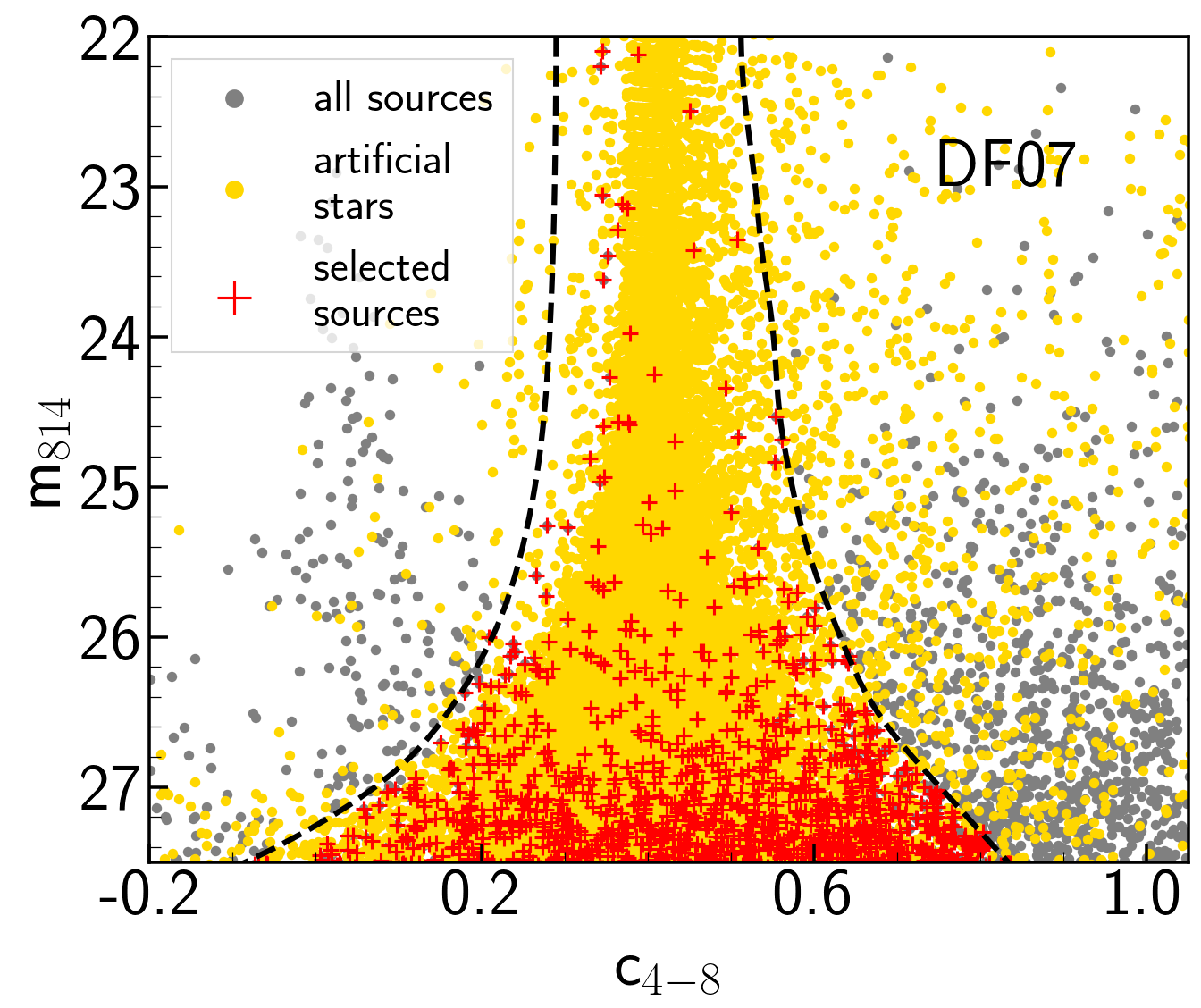}
\includegraphics[width=0.49\linewidth]{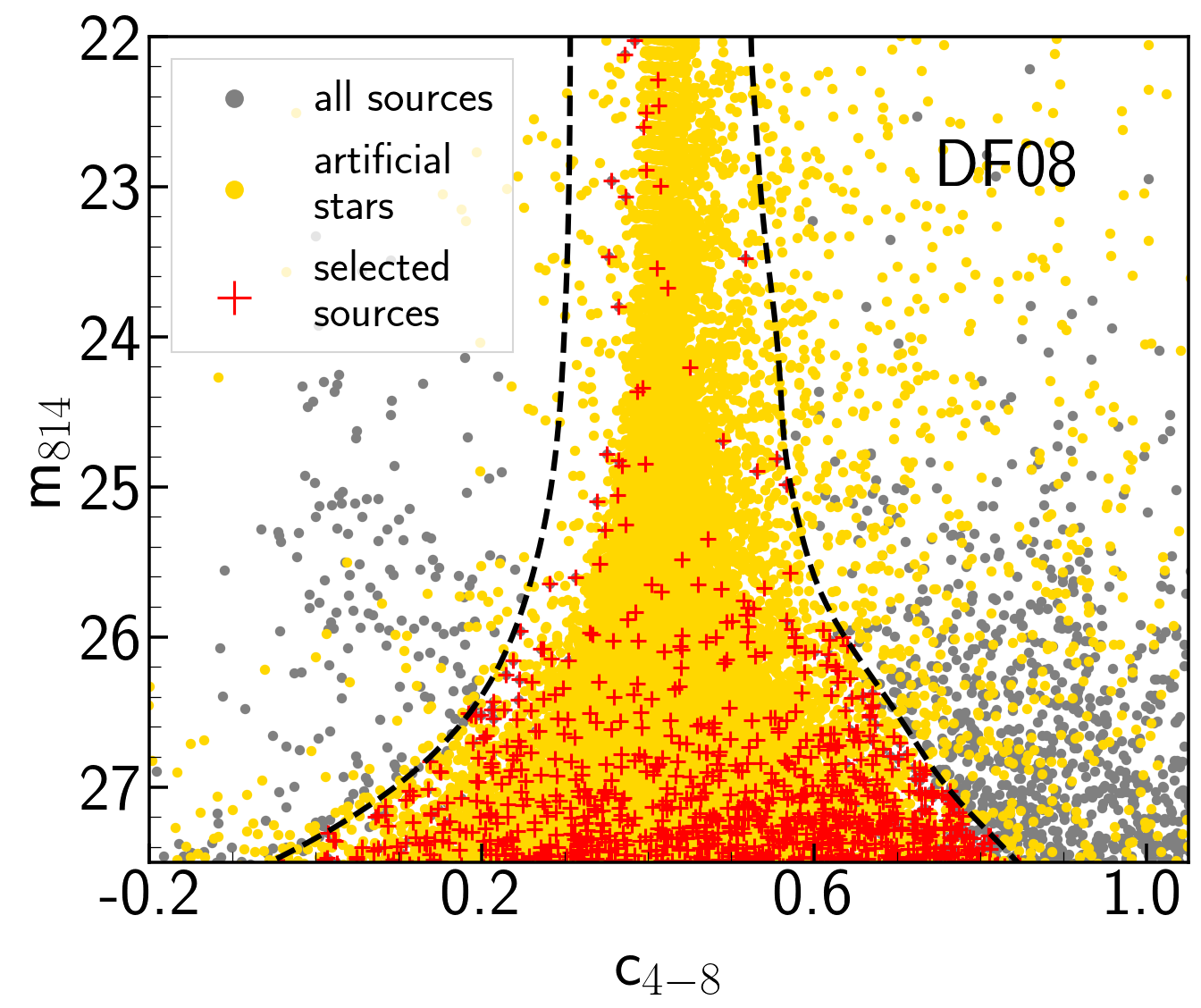}
\includegraphics[width=0.49\linewidth]{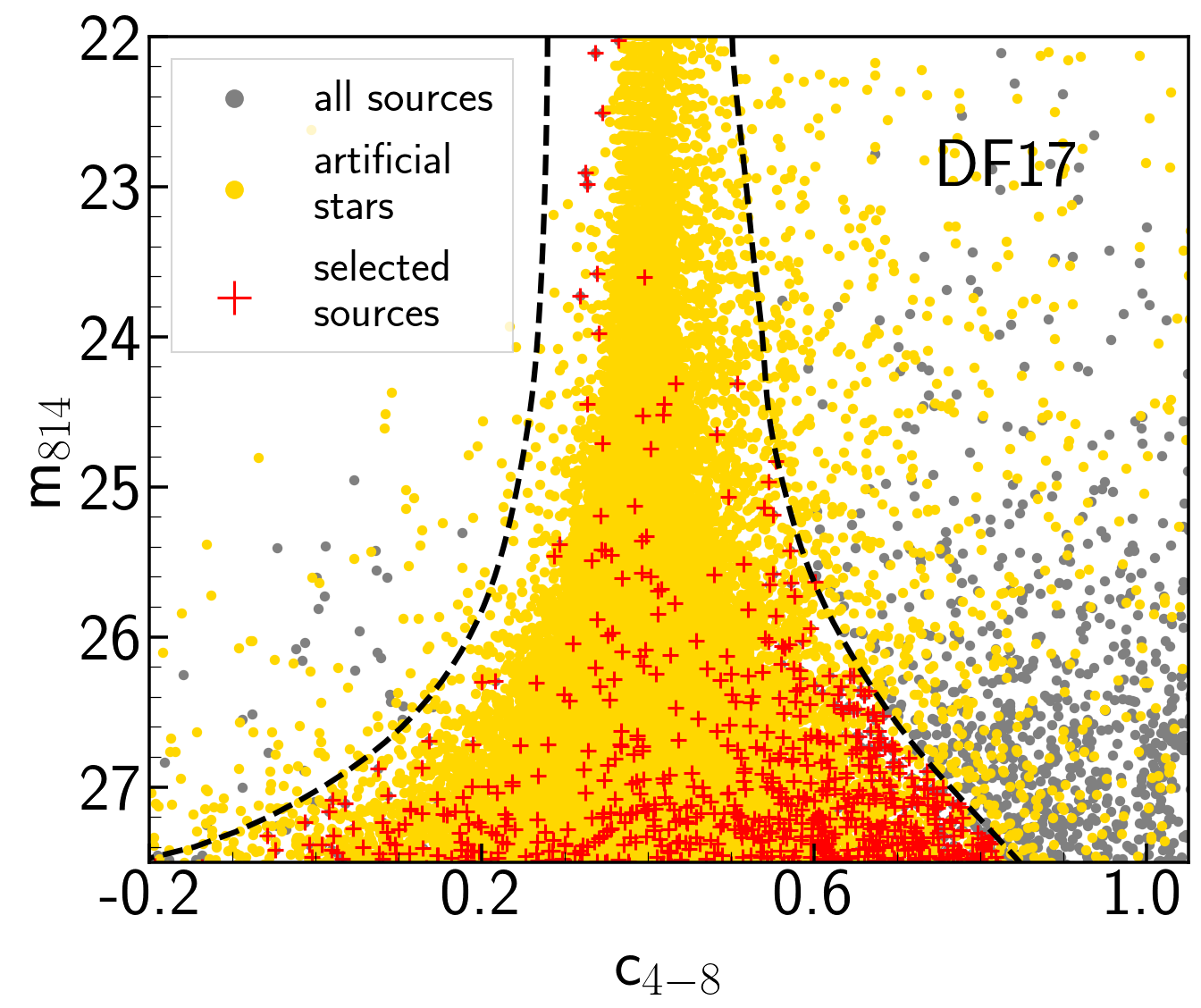}
\includegraphics[width=0.49\linewidth]{selected_sources_compactness_plot_DF44_814.png}
\includegraphics[width=0.49\linewidth]{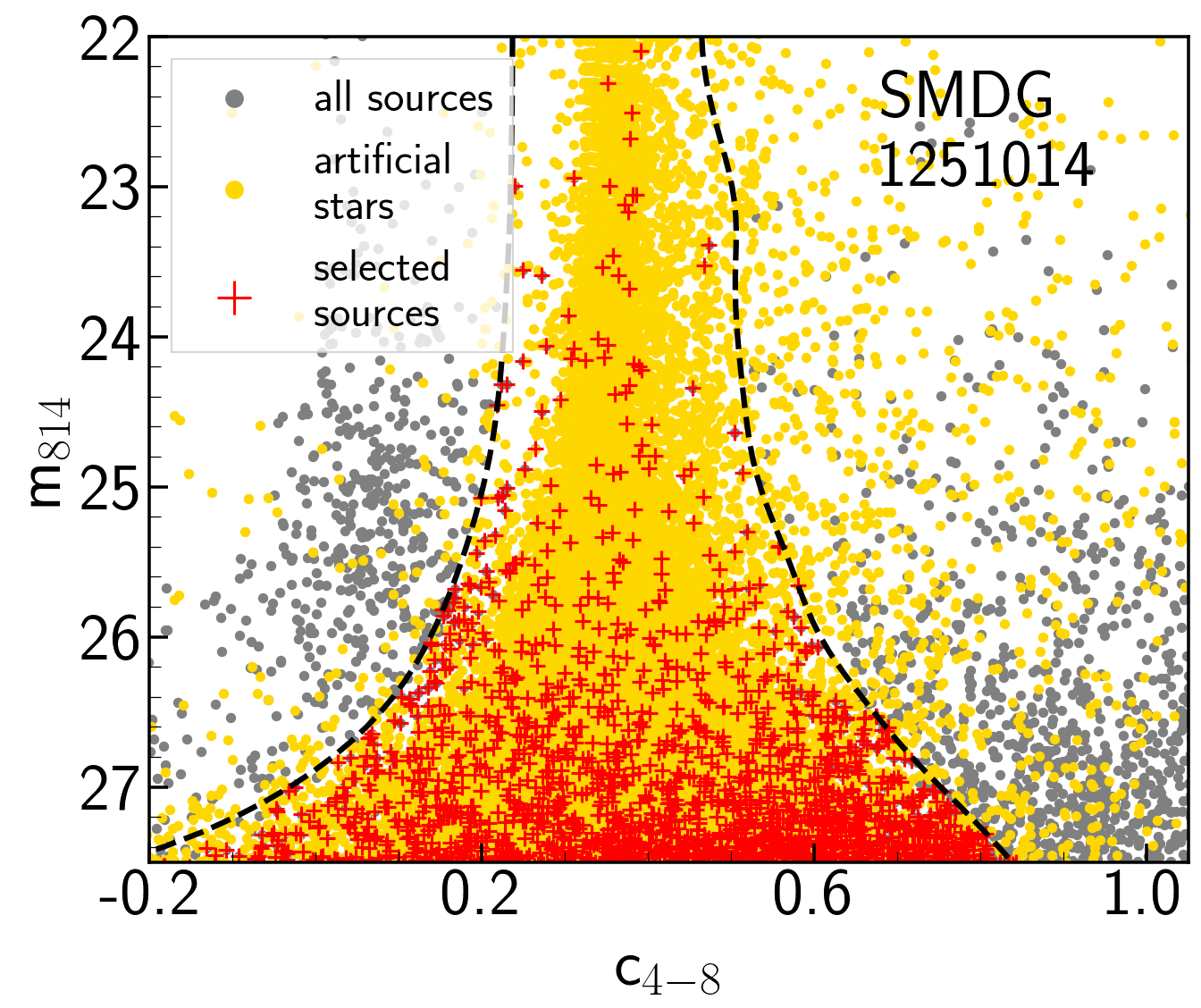}
\includegraphics[width=0.49\linewidth]{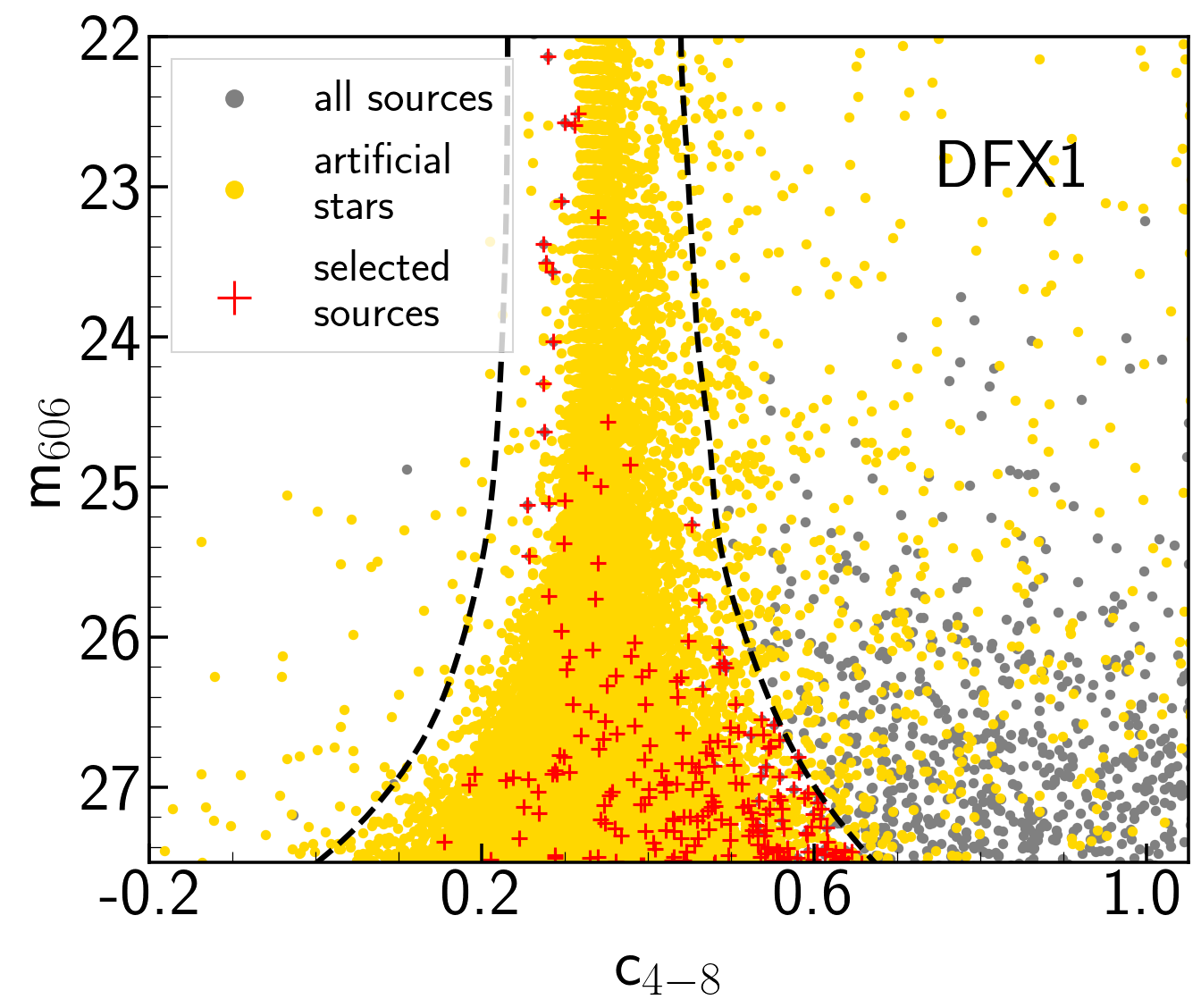}
\caption{Compactness index ($c_{4-8}$) - magnitude diagram for the primary frames (six primary frames for six galaxies). The boundaries of point source selection are indicated with black dashed lines. All the sources in the primary frames, simulated stars and selected point sources are shown in grey dots, yellow dots and red crosses, respectively.}
\label{compactness-app}
\end{figure*}

\begin{figure*}
\centering
\includegraphics[width=0.49\linewidth]{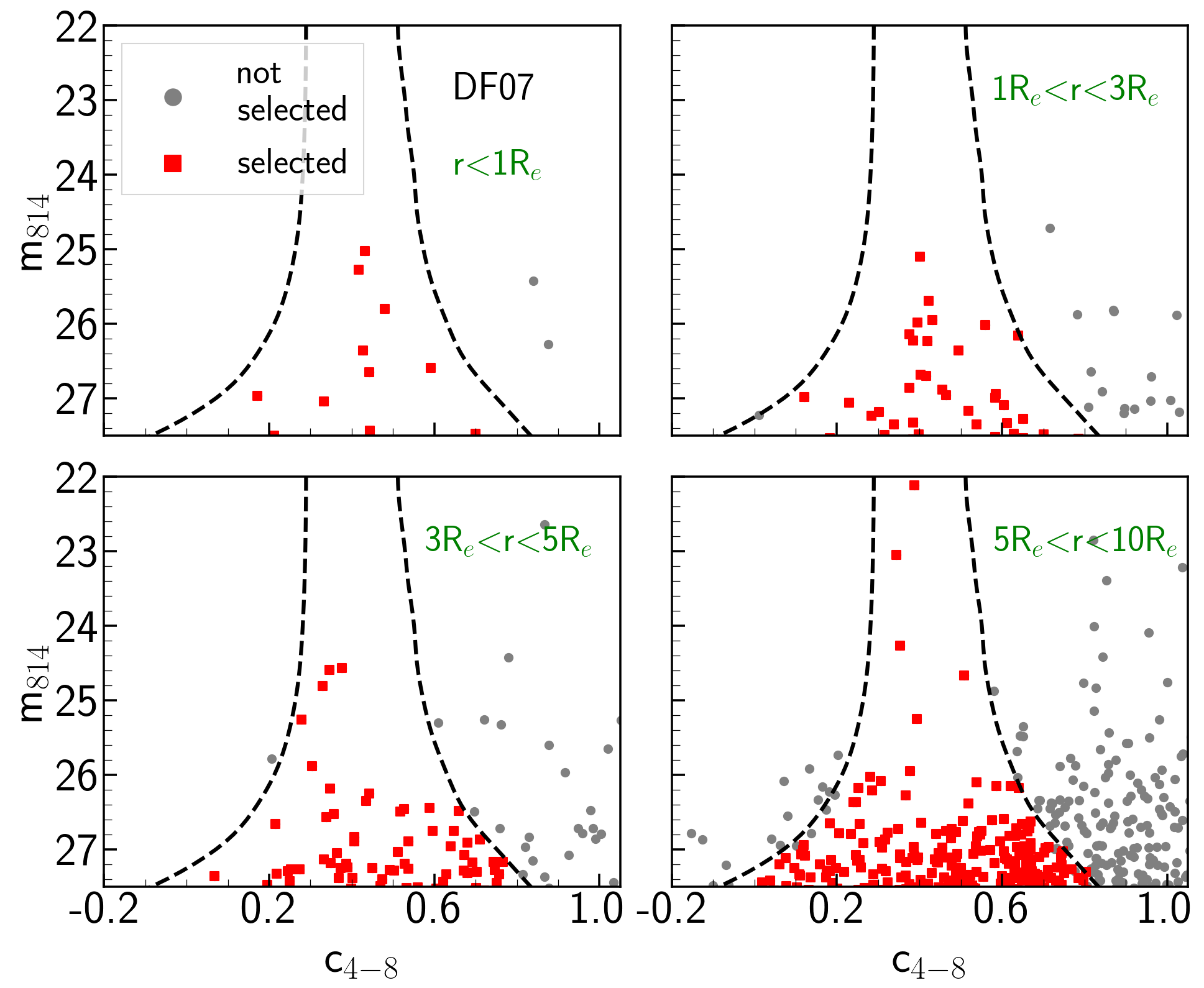}
\includegraphics[width=0.49\linewidth]{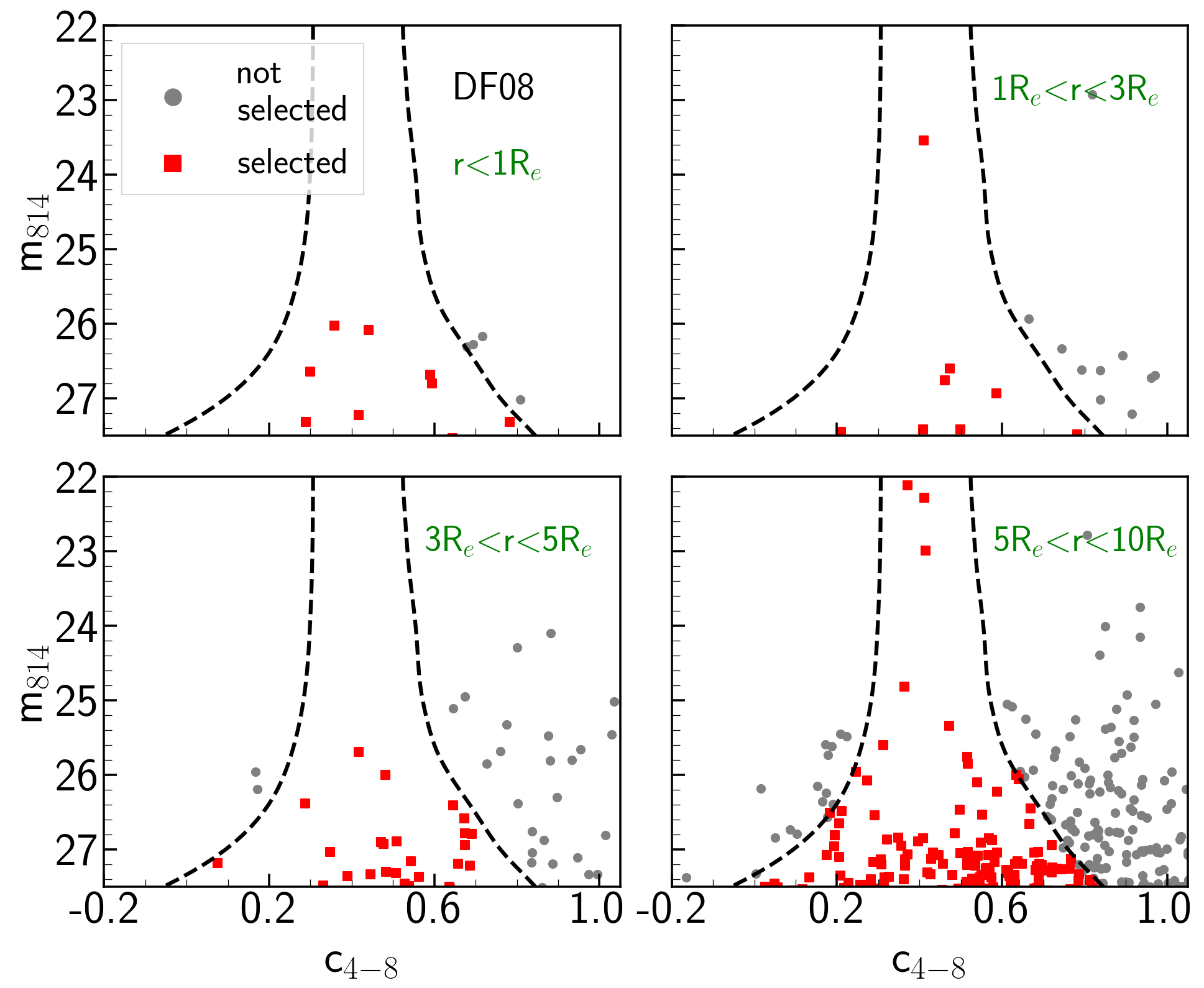}
\includegraphics[width=0.49\linewidth]{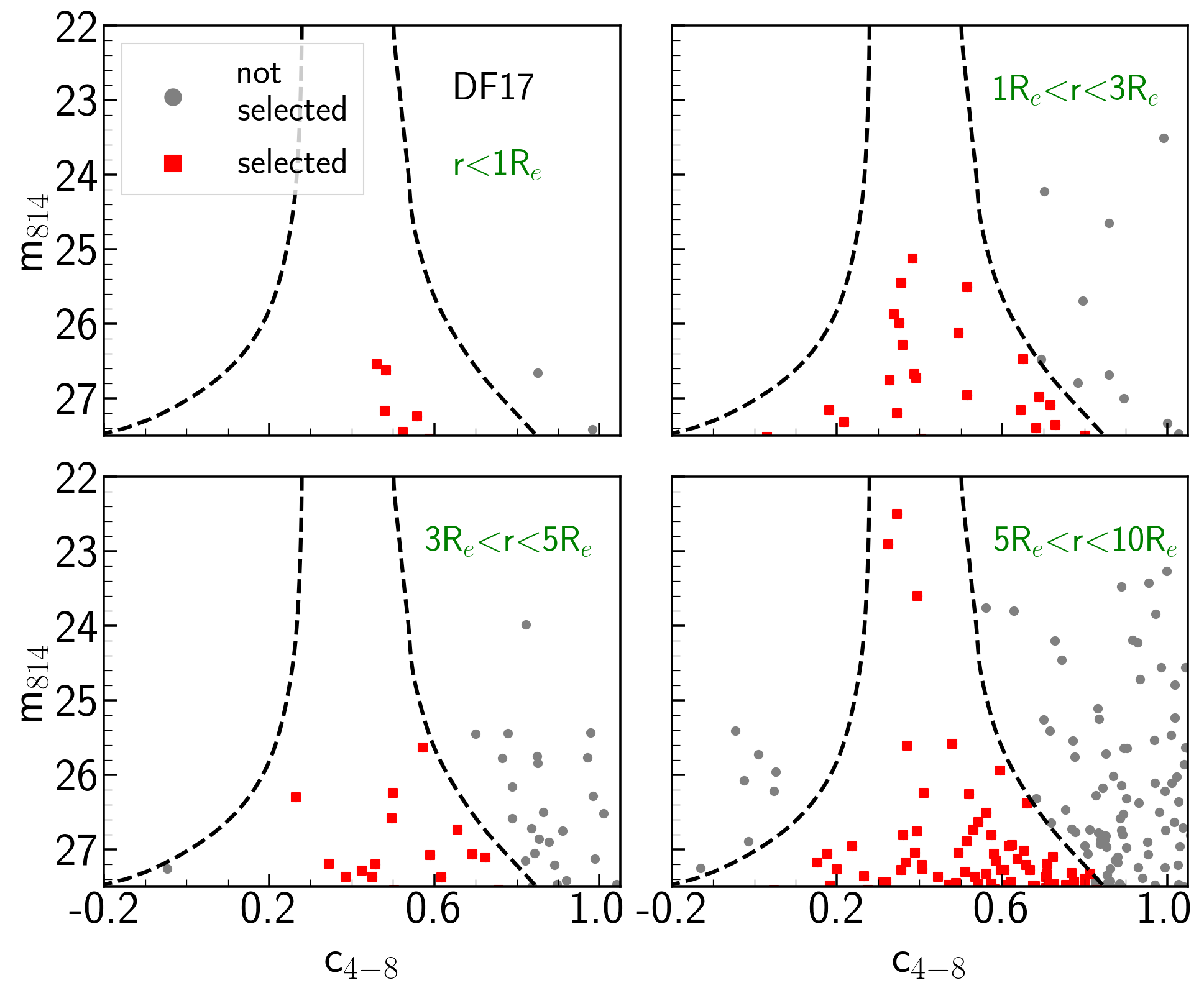}
\includegraphics[width=0.49\linewidth]{selected_sources_compactness_plot_advanced_DF44_814.png}
\includegraphics[width=0.49\linewidth]{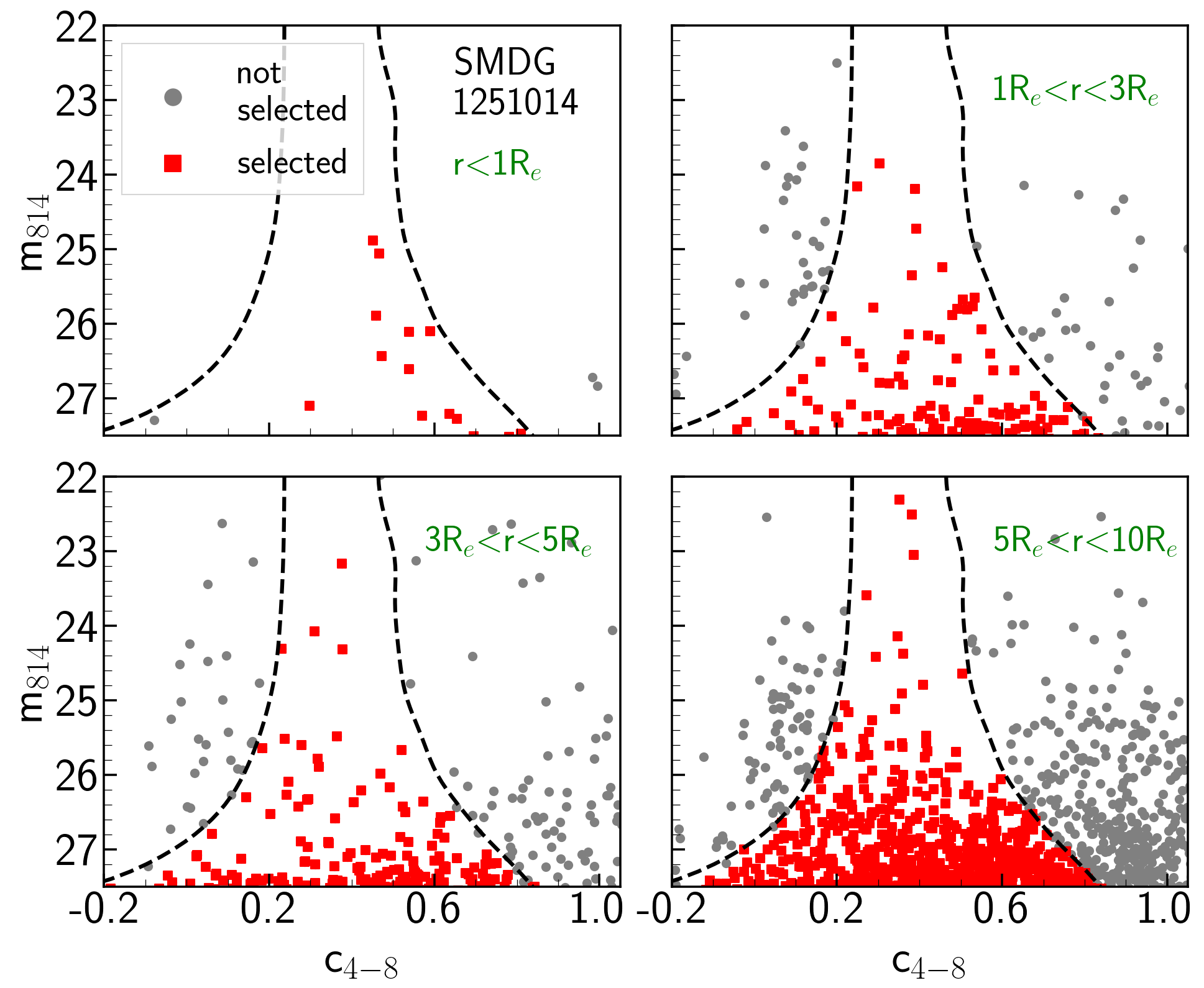}
\includegraphics[width=0.49\linewidth]{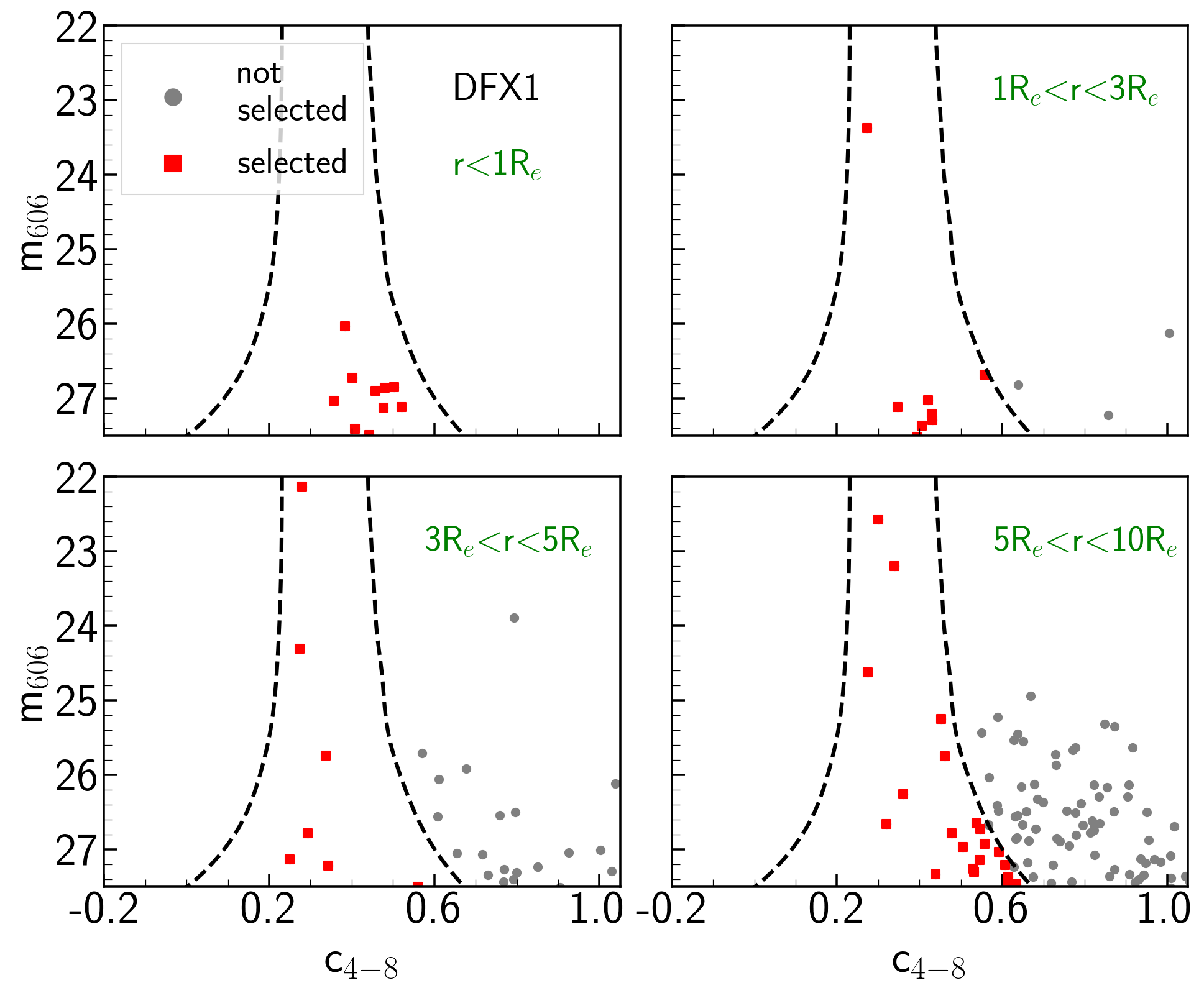}
\caption{Similar diagram as the Fig.~\ref{compactness-app} for sources within different radial distances. The boundaries of point source selection are indicated with black dashed lines. Selected and not selected (discarded) sources are shown in red and grey dots, respectively.}
\label{compactness2-app}
\end{figure*}

\begin{figure*}
\centering
\includegraphics[width=0.49\linewidth]{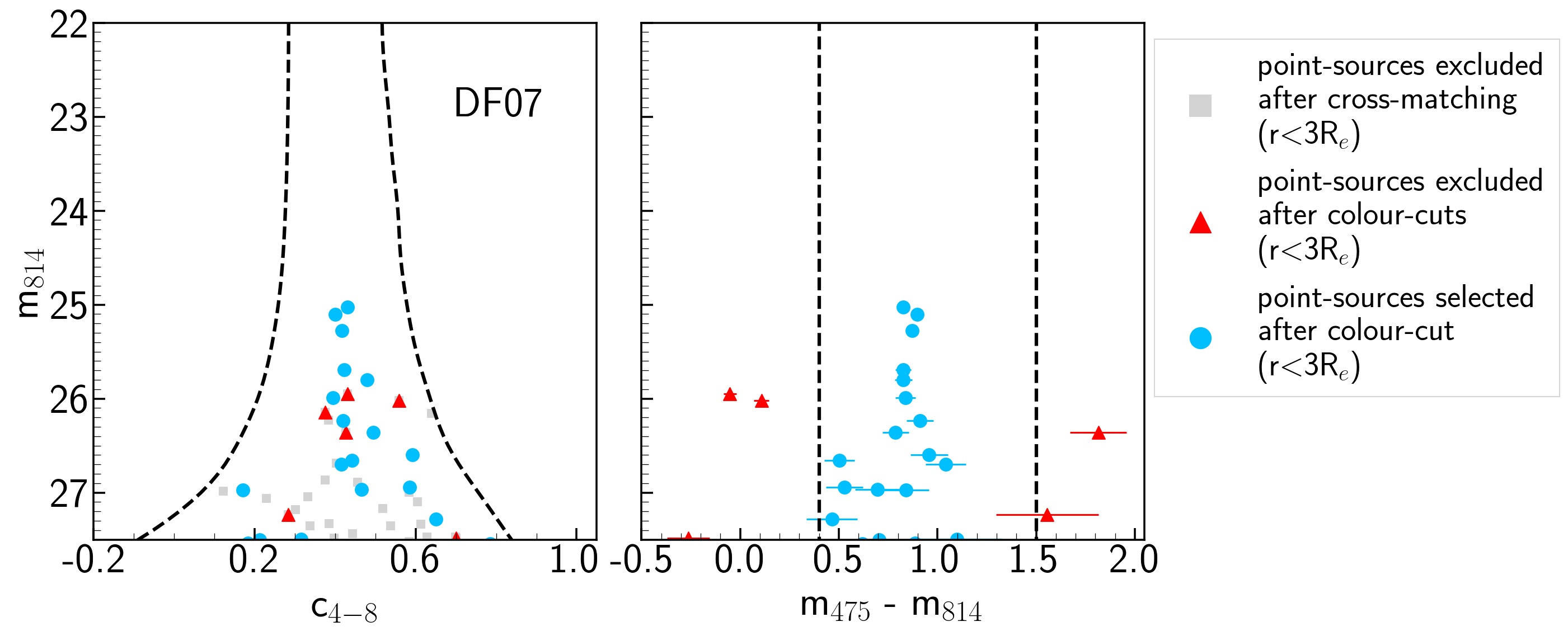}
\includegraphics[width=0.49\linewidth]{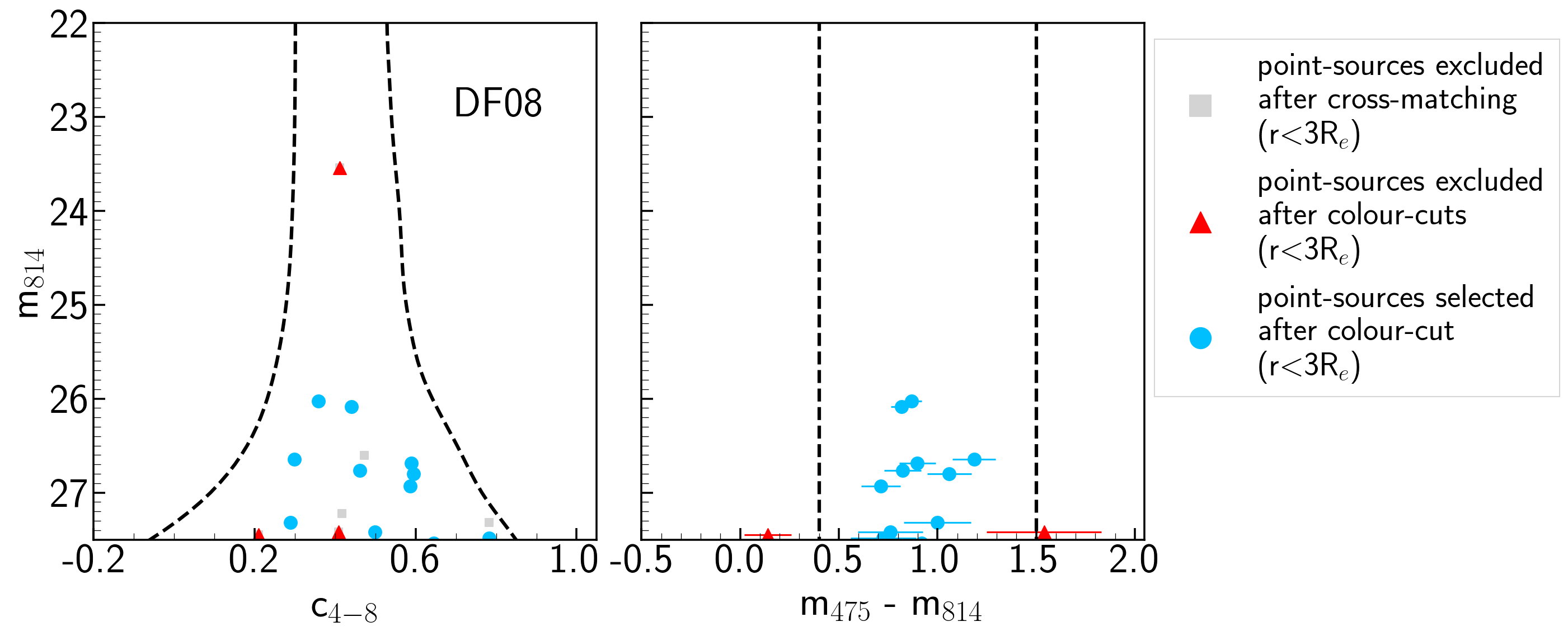}
\includegraphics[width=0.49\linewidth]{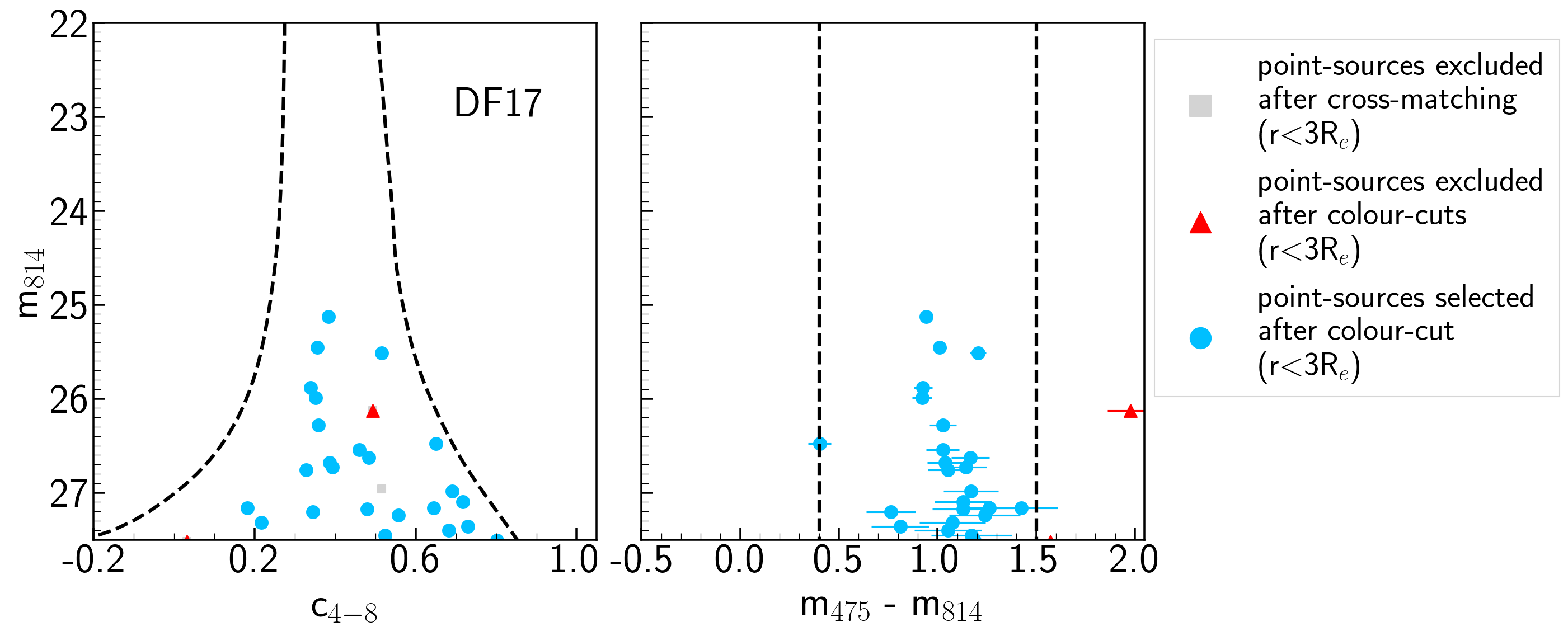}
\includegraphics[width=0.49\linewidth]{selected_sources_color_plot_DF44_814.png}
\includegraphics[width=0.49\linewidth]{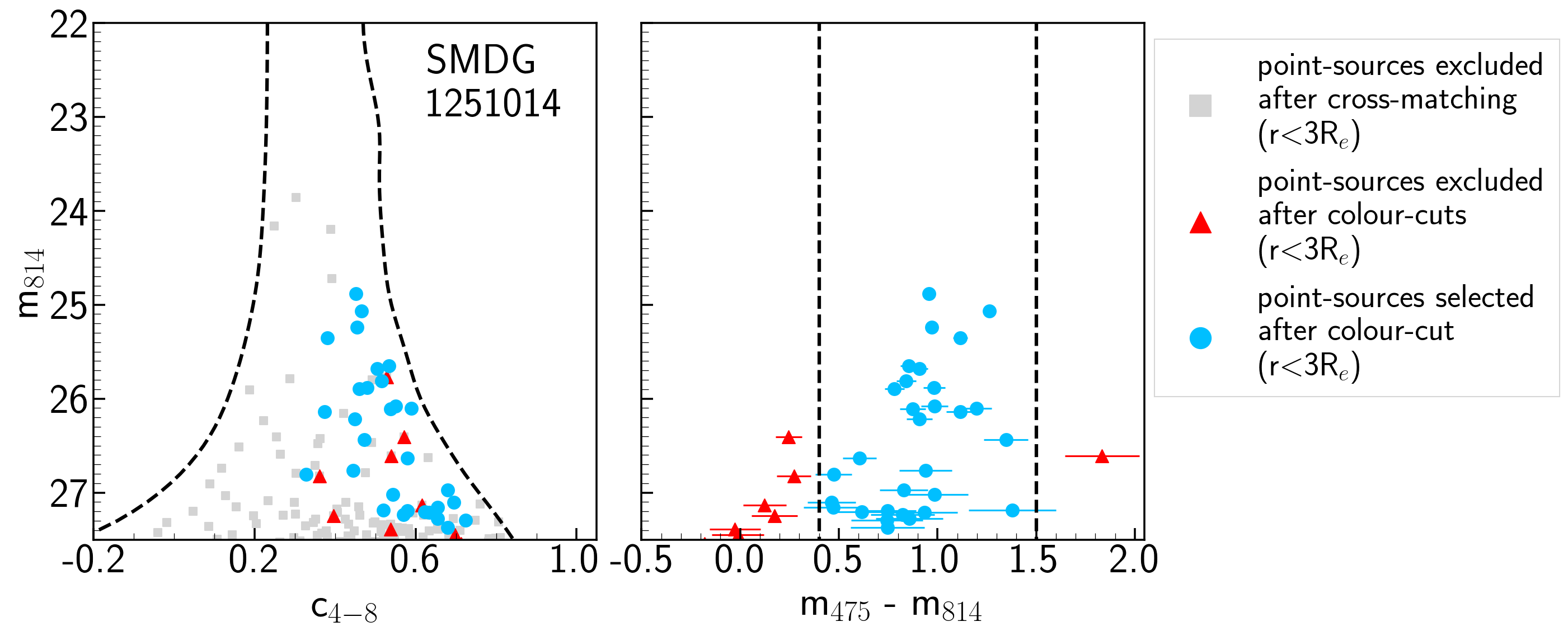}
\includegraphics[width=0.49\linewidth]{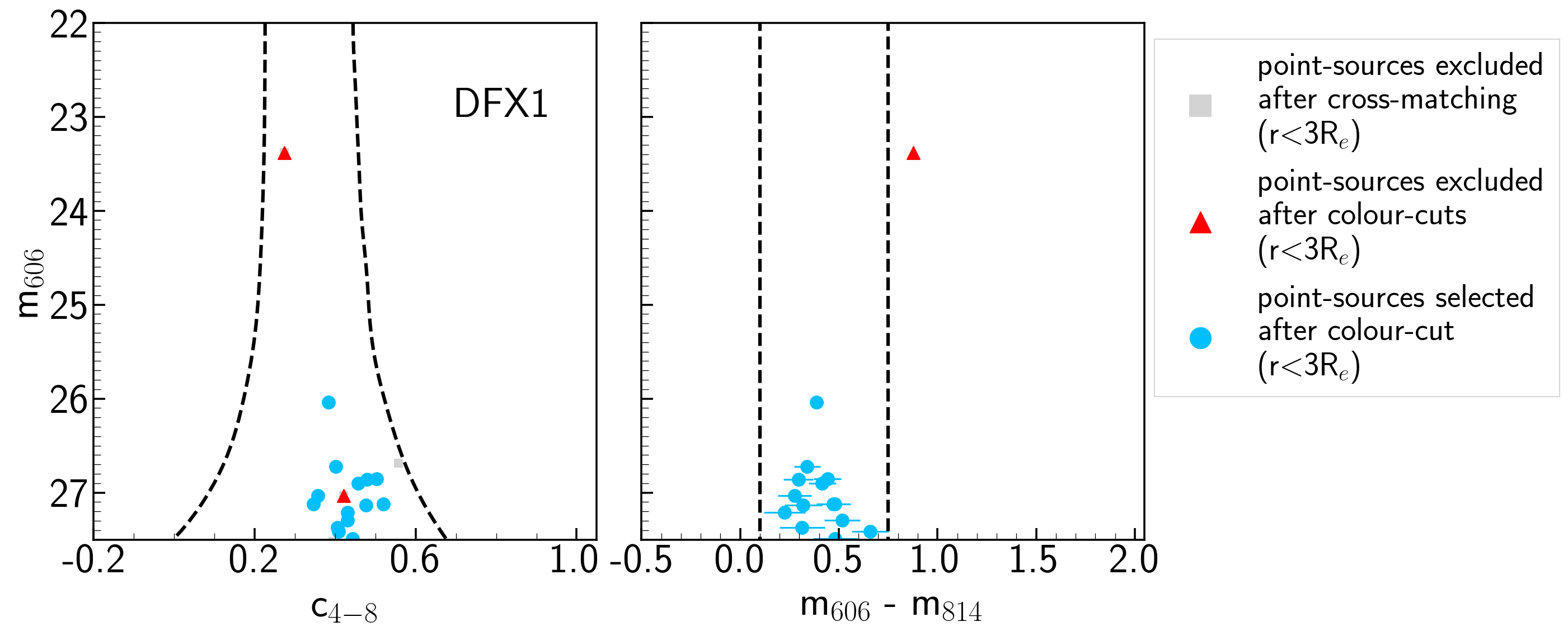}
\caption{Compactness index ($c_{4-8}$) - magnitude diagram (left) and colour-magnitude diagram (right) of the selected point sources around the UDG samples and within 3$R_{\rm e}$ from the host galaxies.}
\label{colour-select-app}
\end{figure*}

\begin{figure*}
\centering
\includegraphics[width=0.3\linewidth]{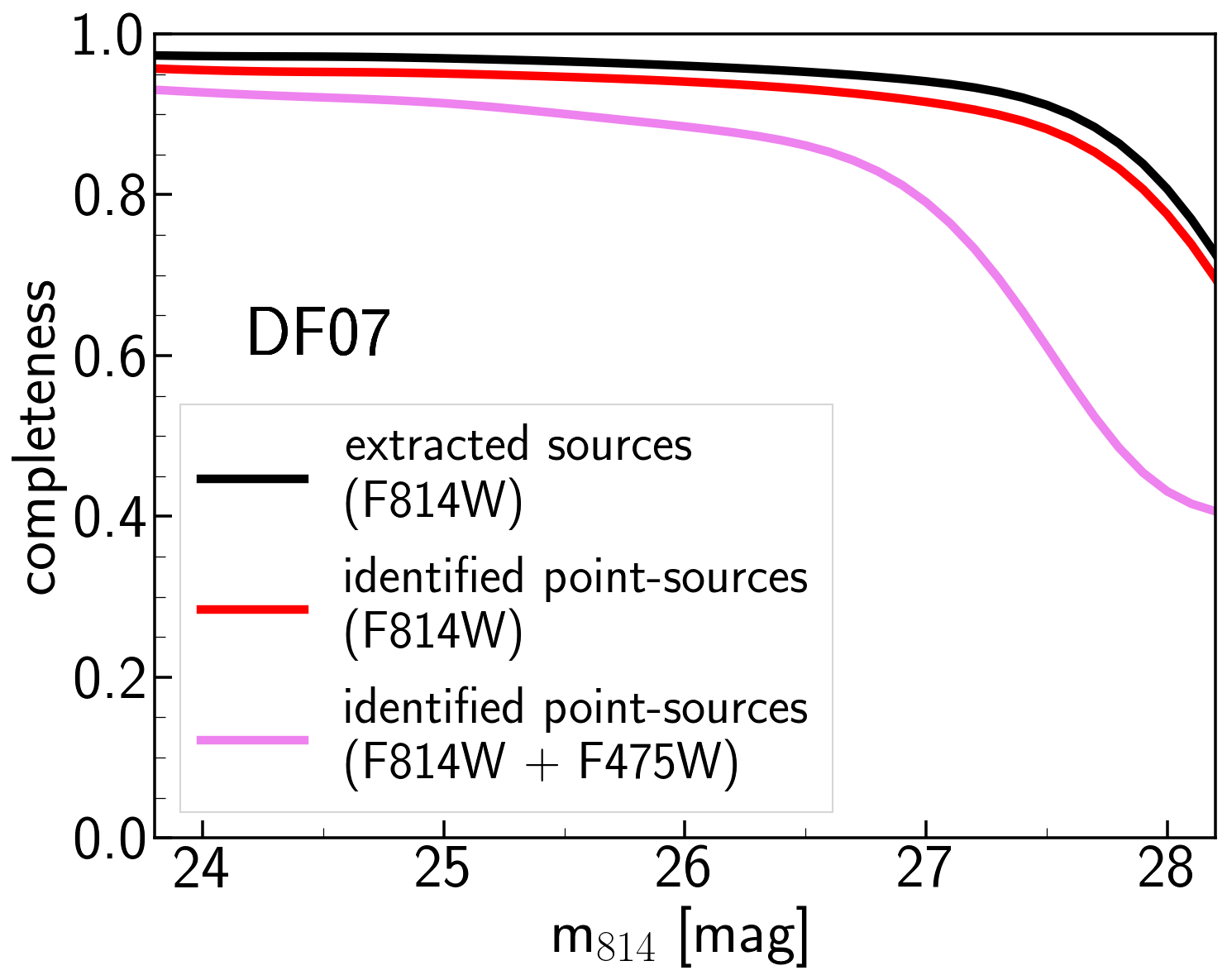}
\includegraphics[width=0.3\linewidth]{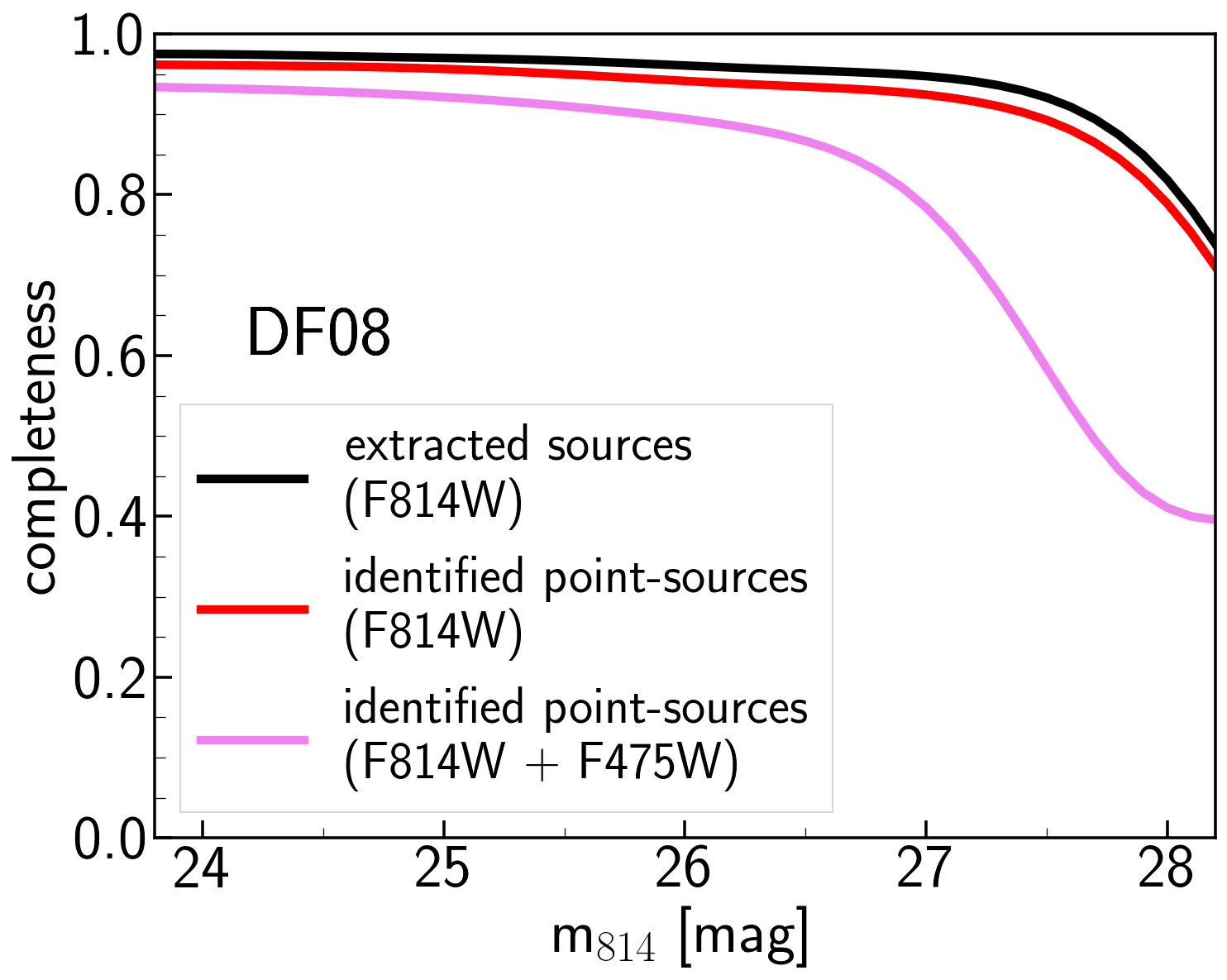}
\includegraphics[width=0.3\linewidth]{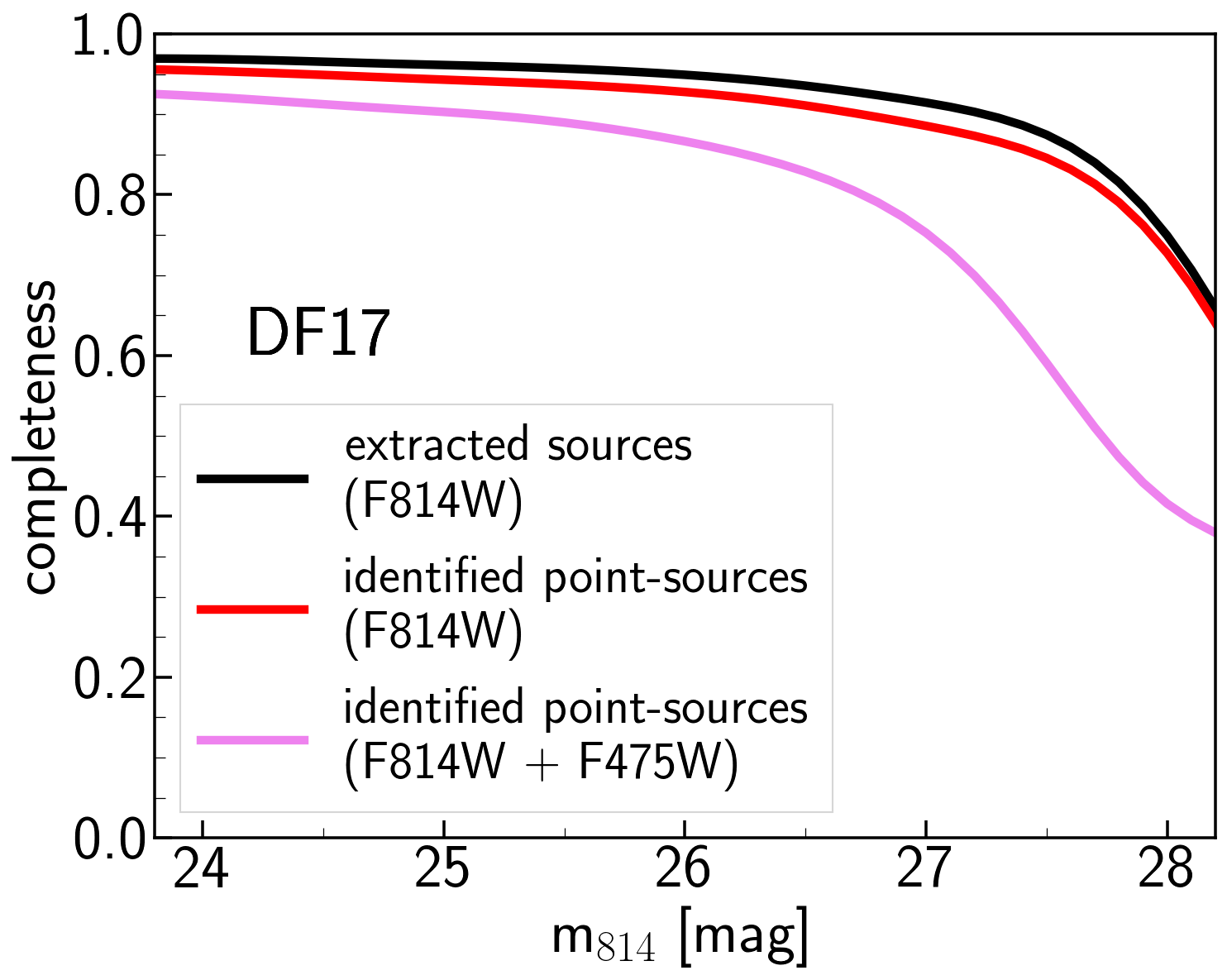}
\includegraphics[width=0.3\linewidth]{completeness_DF44_814.png}
\includegraphics[width=0.3\linewidth]{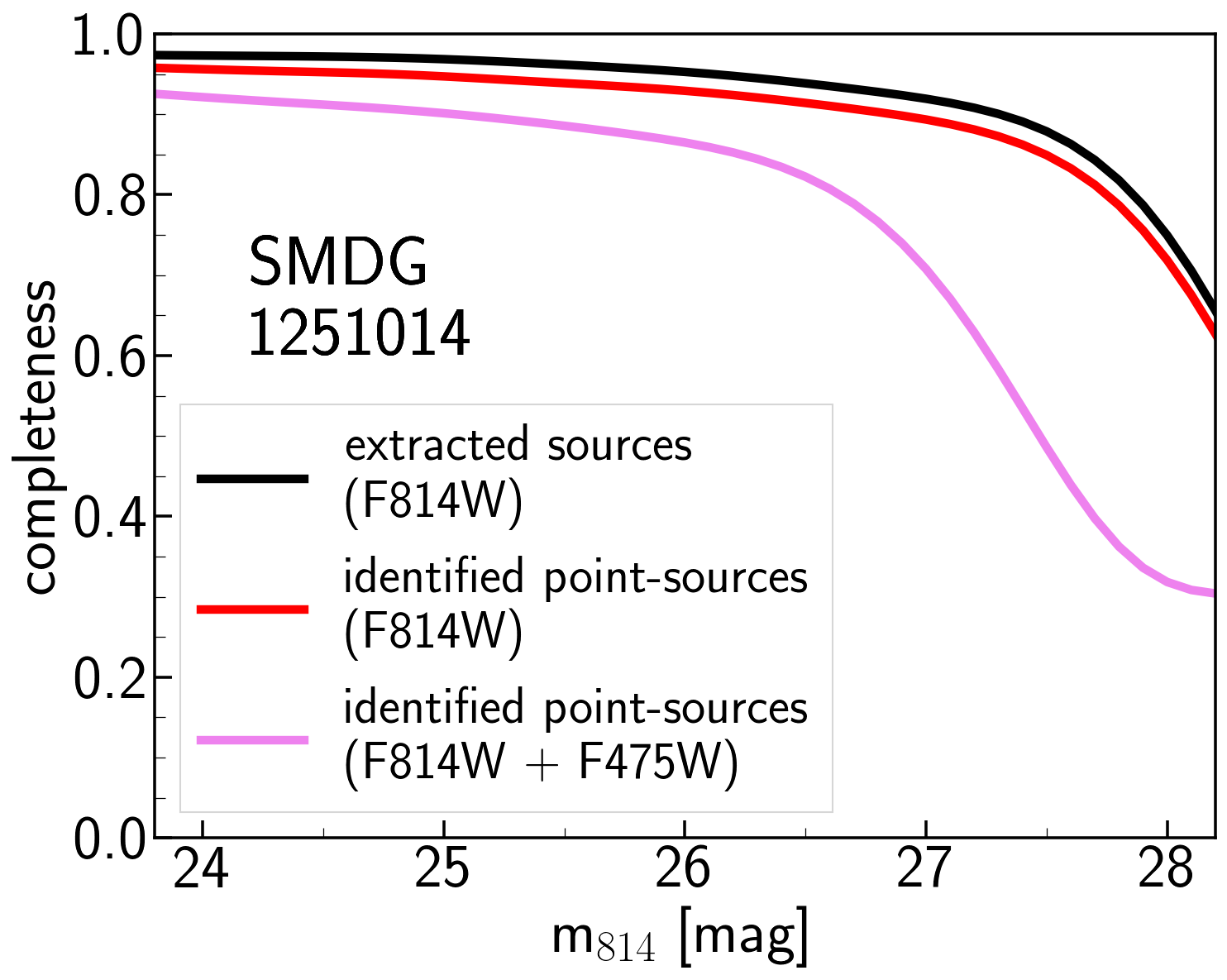}
\includegraphics[width=0.3\linewidth]{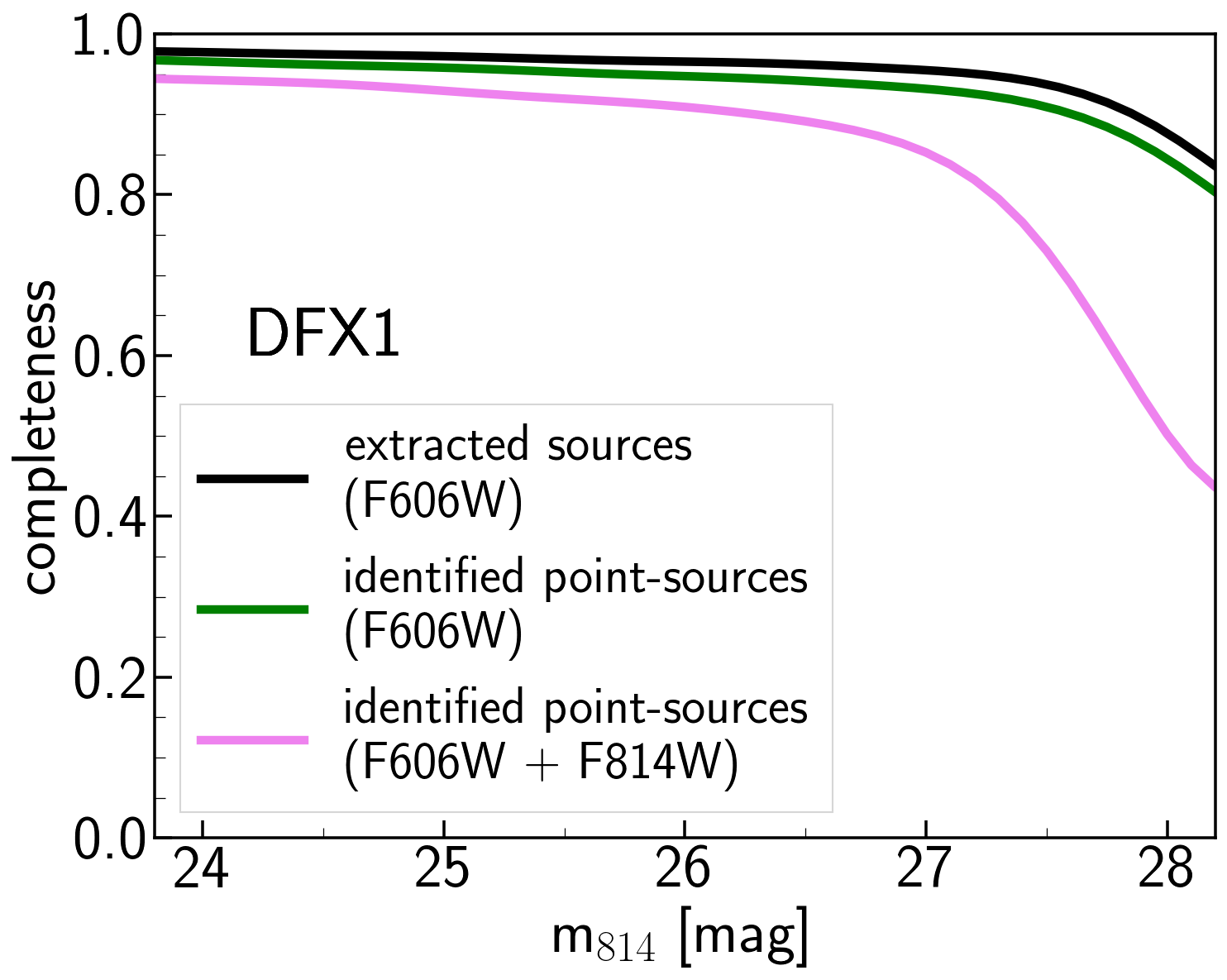}
\caption{The completeness of source extraction (black), point source selection (red for $F814W$ primary filter and green for $F606W$ primary filter) and source selection after including the secondary filter (shallower data, purple), for the dataset corresponding to each galaxy.}
\label{comp-app}
\end{figure*}

\clearpage
\onecolumn

\section{Catalogues of GC candidates}

\begin{longtable}{lcccccc}
\caption{GC candidates around UDGs. Columns from left to right represent host galaxy name, R.A., declination, distance from the host galaxy, magnitude in $F475W$ ($F606W$ for DFX1) and $F814W$, and compactness index in the primary filter. Quoted errors of magnitudes and compactness are statistical ones, systematic errors are not included.} \\ \hline  
\label{gc-candidates}
Host Galaxy & RA & DEC & R & $m_{475}$ & $m_{814}$ & $c_{4-8}$ \\ 
- & hms & dms & arcsec & AB mag & AB mag & AB mag \\ 
\hline
DF07 & 12h 57m 01.692s & +28d 23m 20.530s & 4.47 & 25.85$\pm$0.02 & 25.03$\pm$0.01 & 0.43$\pm$0.02 \\
DF07 & 12h 57m 01.113s & +28d 23m 30.033s & 10.14 & 26.00$\pm$0.02 & 25.10$\pm$0.01 & 0.40$\pm$0.02 \\ 
DF07 & 12h 57m 01.826s & +28d 23m 24.568s & 1.94 & 26.15$\pm$0.03 & 25.28$\pm$0.02 & 0.42$\pm$0.03 \\ 
DF07 & 12h 57m 01.361s & +28d 23m 18.061s & 8.60 & 26.52$\pm$0.03 & 25.69$\pm$0.02 & 0.42$\pm$0.03 \\ 
DF07 & 12h 57m 01.831s & +28d 23m 23.933s & 2.23 & 26.63$\pm$0.04 & 25.80$\pm$0.02 & 0.48$\pm$0.04 \\ 
DF07 & 12h 57m 00.862s & +28d 23m 34.287s & 15.62 & 26.83$\pm$0.04 & 25.99$\pm$0.03 & 0.40$\pm$0.04 \\ 
DF07 & 12h 57m 02.080s & +28d 23m 20.593s & 7.21 & 27.15$\pm$0.06 & 26.23$\pm$0.03 & 0.42$\pm$0.06 \\ 
DF07 & 12h 57m 00.942s & +28d 23m 28.155s & 11.78 & 27.15$\pm$0.06 & 26.36$\pm$0.04 & 0.49$\pm$0.06 \\ 
DF07 & 12h 57m 01.648s & +28d 23m 25.442s & 0.88 & 27.55$\pm$0.08 & 26.60$\pm$0.05 & 0.59$\pm$0.07 \\ 
DF07 & 12h 57m 01.797s & +28d 23m 23.448s & 2.13 & 27.16$\pm$0.06 & 26.66$\pm$0.05 & 0.44$\pm$0.08 \\ 
DF07 & 12h 57m 00.836s & +28d 23m 27.157s & 13.12 & 27.74$\pm$0.09 & 26.70$\pm$0.05 & 0.42$\pm$0.08 \\ 
DF07 & 12h 57m 01.314s & +28d 23m 13.434s & 12.92 & 27.47$\pm$0.07 & 26.94$\pm$0.06 & 0.59$\pm$0.09 \\ 
DF07 & 12h 57m 02.503s & +28d 23m 18.804s & 13.55 & 27.66$\pm$0.09 & 26.97$\pm$0.06 & 0.46$\pm$0.10 \\ 
DF07 & 12h 57m 01.317s & +28d 23m 26.361s & 5.89 & 27.81$\pm$0.10 & 26.97$\pm$0.06 & 0.17$\pm$0.12 \\ 
DF07 & 12h 57m 02.155s & +28d 23m 20.797s & 8.01 & 27.75$\pm$0.10 & 27.28$\pm$0.08 & 0.65$\pm$0.12 \\ 
DF07 & 12h 57m 02.252s & +28d 23m 31.150s & 10.32 & 28.60$\pm$0.20 & 27.50$\pm$0.10 & 0.32$\pm$0.18 \\ 
DF07 & 12h 57m 01.414s & +28d 23m 21.387s & 5.60 & 28.21$\pm$0.14 & 27.50$\pm$0.10 & 0.21$\pm$0.19 \\ 
\hline 
\\\\\\
\hline


Host galaxy & RA & DEC & R & $m_{475}$ & $m_{814}$ & $c_{4-8}$ \\ - & hms & dms & arcsec & AB mag & AB mag & AB mag \\ 
\hline
DF08 & 13h 01m 30.226s & +28d 22m 23.748s & 4.98 & 26.90$\pm$0.04 & 26.03$\pm$0.03 & 0.36$\pm$0.04  \\
DF08 & 13h 01m 30.473s & +28d 22m 26.255s & 2.06 & 26.91$\pm$0.04 & 26.09$\pm$0.03 & 0.44$\pm$0.05  \\
DF08 & 13h 01m 30.255s & +28d 22m 30.810s & 3.54 & 27.83$\pm$0.10 & 26.65$\pm$0.05 & 0.30$\pm$0.08  \\
DF08 & 13h 01m 30.311s & +28d 22m 28.775s & 1.53 & 27.59$\pm$0.08 & 26.69$\pm$0.05 & 0.59$\pm$0.07  \\
DF08 & 13h 01m 29.987s & +28d 22m 25.159s & 6.80 & 27.59$\pm$0.08 & 26.76$\pm$0.05 & 0.46$\pm$0.08  \\
DF08 & 13h 01m 30.743s & +28d 22m 28.009s & 5.15 & 27.86$\pm$0.10 & 26.80$\pm$0.05 & 0.59$\pm$0.08  \\
DF08 & 13h 01m 29.834s & +28d 22m 24.806s & 9.06 & 27.65$\pm$0.08 & 26.93$\pm$0.06 & 0.59$\pm$0.09  \\
DF08 & 13h 01m 30.756s & +28d 22m 29.126s & 5.46 & 28.32$\pm$0.15 & 27.32$\pm$0.08 & 0.29$\pm$0.14  \\
DF08 & 13h 01m 30.165s & +28d 22m 38.834s & 11.39 & 28.18$\pm$0.14 & 27.42$\pm$0.09 & 0.50$\pm$0.14  \\
DF08 & 13h 01m 29.715s & +28d 22m 28.620s & 10.28 & 28.21$\pm$0.14 & 27.49$\pm$0.09 & 0.78$\pm$0.13  \\
\hline 
\\\\\\
\hline


Host Galaxy & RA & DEC & R & $m_{475}$ & $m_{814}$ & $c_{4-8}$ \\ - & hms & dms & arcsec & AB mag & AB mag & AB mag \\ 
\hline
DF17 & 13h 01m 59.118s & +27d 50m 12.841s & 12.41 & 26.07$\pm$0.02 & 25.13$\pm$0.01 & 0.38$\pm$0.02  \\ 
DF17 & 13h 01m 57.526s & +27d 50m 14.550s & 12.13 & 26.47$\pm$0.03 & 25.45$\pm$0.02 & 0.36$\pm$0.03  \\ 
DF17 & 13h 01m 57.718s & +27d 50m 14.608s & 9.43 & 26.72$\pm$0.04 & 25.51$\pm$0.02 & 0.52$\pm$0.03  \\ 
DF17 & 13h 01m 58.405s & +27d 50m 05.476s & 5.74 & 26.81$\pm$0.04 & 25.88$\pm$0.02 & 0.34$\pm$0.04  \\ 
DF17 & 13h 01m 58.332s & +27d 50m 17.286s & 6.30 & 26.91$\pm$0.04 & 25.99$\pm$0.03 & 0.35$\pm$0.04  \\ 
DF17 & 13h 01m 57.537s & +27d 50m 16.736s & 12.79 & 27.31$\pm$0.06 & 26.29$\pm$0.03 & 0.36$\pm$0.06  \\ 
DF17 & 13h 01m 57.746s & +27d 49m 58.905s & 14.67 & 26.88$\pm$0.04 & 26.48$\pm$0.04 & 0.65$\pm$0.06  \\ 
DF17 & 13h 01m 58.397s & +27d 50m 07.170s & 4.09 & 27.57$\pm$0.07 & 26.54$\pm$0.04 & 0.46$\pm$0.07  \\ 
DF17 & 13h 01m 58.541s & +27d 50m 08.186s & 4.59 & 27.80$\pm$0.09 & 26.63$\pm$0.04 & 0.48$\pm$0.07  \\ 
DF17 & 13h 01m 58.984s & +27d 50m 04.670s & 12.05 & 27.72$\pm$0.08 & 26.68$\pm$0.05 & 0.39$\pm$0.07  \\ 
DF17 & 13h 01m 57.744s & +27d 50m 05.557s & 9.95 & 27.87$\pm$0.10 & 26.73$\pm$0.05 & 0.39$\pm$0.08  \\ 
DF17 & 13h 01m 57.841s & +27d 50m 15.961s & 8.48 & 27.81$\pm$0.09 & 26.76$\pm$0.05 & 0.33$\pm$0.09  \\ 
DF17 & 13h 01m 57.809s & +27d 50m 06.634s & 8.55 & 28.16$\pm$0.13 & 26.99$\pm$0.06 & 0.69$\pm$0.08  \\ 
DF17 & 13h 01m 57.276s & +27d 50m 06.796s & 15.91 & 28.23$\pm$0.13 & 27.10$\pm$0.07 & 0.72$\pm$0.09  \\ 
DF17 & 13h 01m 57.977s & +27d 50m 21.703s & 11.74 & 28.43$\pm$0.15 & 27.16$\pm$0.07 & 0.18$\pm$0.12  \\ 
DF17 & 13h 01m 58.258s & +27d 50m 17.415s & 6.44 & 28.59$\pm$0.17 & 27.17$\pm$0.07 & 0.65$\pm$0.10  \\ 
DF17 & 13h 01m 58.059s & +27d 50m 14.057s & 4.73 & 28.30$\pm$0.14 & 27.17$\pm$0.07 & 0.48$\pm$0.11  \\ 
DF17 & 13h 01m 57.703s & +27d 50m 12.062s & 9.01 & 27.97$\pm$0.10 & 27.21$\pm$0.07 & 0.34$\pm$0.12  \\ 
DF17 & 13h 01m 58.223s & +27d 50m 12.570s & 1.94 & 28.48$\pm$0.16 & 27.24$\pm$0.08 & 0.56$\pm$0.12  \\ 
DF17 & 13h 01m 57.885s & +27d 50m 11.882s & 6.28 & 28.39$\pm$0.15 & 27.32$\pm$0.08 & 0.22$\pm$0.15  \\ 
DF17 & 13h 01m 58.042s & +27d 50m 04.862s & 7.25 & 28.17$\pm$0.12 & 27.36$\pm$0.08 & 0.73$\pm$0.11  \\ 
DF17 & 13h 01m 59.039s & +27d 50m 02.486s & 13.98 & 28.46$\pm$0.15 & 27.40$\pm$0.08 & 0.68$\pm$0.12  \\ 
DF17 & 13h 01m 58.191s & +27d 50m 09.678s & 2.10 & 28.63$\pm$0.18 & 27.45$\pm$0.09 & 0.52$\pm$0.14  \\ 
DF17 & 13h 01m 58.265s & +27d 50m 01.057s & 9.95 & 28.69$\pm$0.19 & 27.50$\pm$0.09 & 0.80$\pm$0.12  \\ 
\hline 
\\\\\\
\hline


Host Galaxy & RA & DEC & R & $m_{475}$ & $m_{814}$ & $c_{4-8}$ \\ - & hms & dms & arcsec & AB mag & AB mag & AB mag \\ 
\hline
DF44 & 13h 00m 58.365s & +26d 58m 26.645s & 9.99 & 24.20$\pm$0.01 & 23.63$\pm$0.01 & 0.39$\pm$0.01  \\ 
DF44 & 13h 00m 58.417s & +26d 58m 53.848s & 19.86 & 24.92$\pm$0.01 & 24.06$\pm$0.01 & 0.46$\pm$0.01  \\ 
DF44 & 13h 00m 57.734s & +26d 58m 35.624s & 4.03 & 26.54$\pm$0.04 & 25.30$\pm$0.02 & 0.49$\pm$0.03  \\ 
DF44 & 13h 00m 58.255s & +26d 58m 37.673s & 4.66 & 26.40$\pm$0.03 & 25.47$\pm$0.02 & 0.45$\pm$0.03  \\ 
DF44 & 13h 00m 58.217s & +26d 58m 36.120s & 3.44 & 27.21$\pm$0.06 & 26.10$\pm$0.03 & 0.54$\pm$0.05  \\ 
DF44 & 13h 00m 57.628s & +26d 58m 32.418s & 6.13 & 27.74$\pm$0.10 & 26.41$\pm$0.04 & 0.64$\pm$0.06  \\ 
DF44 & 13h 00m 57.750s & +26d 58m 34.088s & 3.84 & 27.07$\pm$0.06 & 26.54$\pm$0.05 & 0.52$\pm$0.07  \\ 
DF44 & 13h 00m 57.756s & +26d 58m 42.759s & 8.57 & 27.96$\pm$0.12 & 26.75$\pm$0.06 & 0.50$\pm$0.09  \\ 
DF44 & 13h 00m 58.018s & +26d 58m 34.187s & 0.85 & 28.26$\pm$0.16 & 26.80$\pm$0.06 & 0.47$\pm$0.09  \\ 
DF44 & 13h 00m 58.130s & +26d 58m 31.265s & 4.21 & 27.90$\pm$0.11 & 26.93$\pm$0.07 & 0.72$\pm$0.09  \\ 
DF44 & 13h 00m 58.211s & +26d 58m 36.856s & 3.66 & 28.47$\pm$0.20 & 27.00$\pm$0.07 & 0.44$\pm$0.11  \\ 
DF44 & 13h 00m 57.729s & +26d 58m 38.936s & 5.65 & 27.75$\pm$0.10 & 27.09$\pm$0.08 & 0.47$\pm$0.12  \\ 
DF44 & 13h 00m 57.798s & +26d 58m 30.938s & 5.06 & 28.15$\pm$0.15 & 27.21$\pm$0.08 & 0.81$\pm$0.11  \\ 
DF44 & 13h 00m 57.988s & +26d 58m 39.473s & 4.47 & 28.32$\pm$0.17 & 27.27$\pm$0.09 & 0.81$\pm$0.13  \\ 
DF44 & 13h 00m 57.344s & +26d 58m 23.378s & 15.22 & 27.91$\pm$0.12 & 27.31$\pm$0.09 & 0.84$\pm$0.12  \\ 
DF44 & 13h 00m 58.175s & +26d 58m 38.665s & 4.51 & 28.29$\pm$0.17 & 27.39$\pm$0.10 & 0.57$\pm$0.15  \\ 
DF44 & 13h 00m 58.097s & +26d 58m 48.583s & 13.66 & 28.28$\pm$0.16 & 27.41$\pm$0.10 & 0.47$\pm$0.16  \\ 
DF44 & 13h 00m 59.127s & +26d 58m 39.782s & 17.57 & 28.11$\pm$0.14 & 27.43$\pm$0.09 & 0.35$\pm$0.16  \\ 
DF44 & 13h 00m 57.765s & +26d 58m 27.948s & 7.88 & 28.40$\pm$0.18 & 27.44$\pm$0.10 & 0.19$\pm$0.19  \\ 
DF44 & 13h 00m 58.199s & +26d 58m 35.660s & 3.06 & 28.36$\pm$0.18 & 27.46$\pm$0.10 & 0.53$\pm$0.16  \\ 
DF44 & 13h 00m 58.536s & +26d 58m 25.958s & 12.10 & 28.76$\pm$0.24 & 27.50$\pm$0.10 & 0.56$\pm$0.15  \\ 
DF44 & 13h 00m 57.537s & +26d 58m 32.334s & 7.42 & 28.50$\pm$0.20 & 27.50$\pm$0.11 & 0.78$\pm$0.15  \\ 
\hline 
\\\\\\
\hline


Host Galaxy & RA & DEC & R & $m_{475}$ & $m_{814}$ & $c_{4-8}$ \\ - & hms & dms & arcsec & AB mag & AB mag & AB mag \\ 
\hline
SMDG1251014 & 12h 51m 01.366s & +27d 47m 56.849s & 3.97 & 25.84$\pm$0.02 & 24.88$\pm$0.01 & 0.45$\pm$0.02 \\ 
SMDG1251014 & 12h 51m 01.486s & +27d 47m 48.510s & 5.29 & 26.33$\pm$0.03 & 25.07$\pm$0.02 & 0.47$\pm$0.02 \\ 
SMDG1251014 & 12h 51m 01.791s & +27d 47m 33.136s & 21.18 & 26.21$\pm$0.03 & 25.24$\pm$0.02 & 0.45$\pm$0.03 \\ 
SMDG1251014 & 12h 51m 02.361s & +27d 47m 41.343s & 19.73 & 26.47$\pm$0.03 & 25.35$\pm$0.02 & 0.38$\pm$0.03 \\ 
SMDG1251014 & 12h 51m 00.677s & +27d 47m 46.748s & 11.23 & 26.51$\pm$0.03 & 25.65$\pm$0.02 & 0.53$\pm$0.04 \\ 
SMDG1251014 & 12h 51m 02.141s & +27d 47m 51.386s & 12.72 & 26.59$\pm$0.03 & 25.68$\pm$0.02 & 0.50$\pm$0.04 \\ 
SMDG1251014 & 12h 51m 00.194s & +27d 47m 46.506s & 17.80 & 26.66$\pm$0.04 & 25.81$\pm$0.03 & 0.52$\pm$0.05 \\ 
SMDG1251014 & 12h 51m 01.874s & +27d 47m 53.512s & 8.62 & 26.87$\pm$0.05 & 25.88$\pm$0.03 & 0.48$\pm$0.04 \\ 
SMDG1251014 & 12h 51m 01.471s & +27d 47m 53.577s & 2.63 & 26.68$\pm$0.04 & 25.90$\pm$0.03 & 0.46$\pm$0.05 \\ 
SMDG1251014 & 12h 51m 00.150s & +27d 47m 44.684s & 19.14 & 27.07$\pm$0.05 & 26.08$\pm$0.04 & 0.55$\pm$0.06 \\ 
SMDG1251014 & 12h 51m 01.059s & +27d 47m 52.332s & 3.66 & 27.30$\pm$0.07 & 26.10$\pm$0.04 & 0.59$\pm$0.05 \\ 
SMDG1251014 & 12h 51m 00.818s & +27d 47m 58.577s & 9.12 & 26.98$\pm$0.05 & 26.11$\pm$0.04 & 0.54$\pm$0.06 \\ 
SMDG1251014 & 12h 51m 01.636s & +27d 47m 58.326s & 7.34 & 27.26$\pm$0.06 & 26.14$\pm$0.04 & 0.37$\pm$0.06 \\ 
SMDG1251014 & 12h 51m 02.334s & +27d 47m 47.917s & 16.33 & 27.12$\pm$0.05 & 26.22$\pm$0.04 & 0.45$\pm$0.06 \\ 
SMDG1251014 & 12h 51m 01.381s & +27d 47m 53.445s & 1.30 & 27.79$\pm$0.10 & 26.44$\pm$0.05 & 0.47$\pm$0.08 \\ 
SMDG1251014 & 12h 51m 00.977s & +27d 48m 04.059s & 12.07 & 27.24$\pm$0.07 & 26.63$\pm$0.05 & 0.58$\pm$0.08 \\ 
SMDG1251014 & 12h 51m 00.608s & +27d 47m 59.469s & 12.22 & 27.71$\pm$0.11 & 26.77$\pm$0.07 & 0.45$\pm$0.12 \\ 
SMDG1251014 & 12h 51m 00.696s & +27d 47m 51.682s & 9.14 & 27.28$\pm$0.07 & 26.81$\pm$0.06 & 0.33$\pm$0.11 \\ 
SMDG1251014 & 12h 51m 02.061s & +27d 47m 36.847s & 19.78 & 27.80$\pm$0.10 & 26.97$\pm$0.07 & 0.68$\pm$0.10 \\ 
SMDG1251014 & 12h 51m 00.659s & +27d 48m 06.908s & 16.90 & 28.01$\pm$0.14 & 27.02$\pm$0.09 & 0.54$\pm$0.14 \\ 
SMDG1251014 & 12h 51m 00.434s & +27d 47m 48.927s & 13.61 & 27.57$\pm$0.08 & 27.10$\pm$0.09 & 0.70$\pm$0.12 \\ 
SMDG1251014 & 12h 51m 00.193s & +27d 48m 02.504s & 19.12 & 27.63$\pm$0.10 & 27.15$\pm$0.11 & 0.66$\pm$0.16 \\ 
SMDG1251014 & 12h 51m 01.293s & +27d 48m 01.670s & 8.67 & 28.57$\pm$0.20 & 27.19$\pm$0.09 & 0.52$\pm$0.14 \\ 
SMDG1251014 & 12h 51m 01.004s & +27d 47m 36.604s & 16.98 & 27.94$\pm$0.12 & 27.20$\pm$0.09 & 0.58$\pm$0.14 \\ 
SMDG1251014 & 12h 51m 01.635s & +27d 47m 44.250s & 10.09 & 27.82$\pm$0.10 & 27.20$\pm$0.09 & 0.62$\pm$0.13 \\ 
SMDG1251014 & 12h 51m 01.200s & +27d 47m 38.733s & 14.34 & 28.15$\pm$0.14 & 27.21$\pm$0.09 & 0.63$\pm$0.14 \\ 
SMDG1251014 & 12h 51m 01.137s & +27d 47m 53.285s & 2.44 & 28.06$\pm$0.13 & 27.24$\pm$0.10 & 0.57$\pm$0.15 \\ 
SMDG1251014 & 12h 51m 01.238s & +27d 47m 52.692s & 0.96 & 28.13$\pm$0.14 & 27.28$\pm$0.10 & 0.65$\pm$0.15 \\ 
SMDG1251014 & 12h 51m 00.665s & +27d 47m 59.074s & 11.28 & 28.04$\pm$0.14 & 27.29$\pm$0.11 & 0.72$\pm$0.16 \\ 
SMDG1251014 & 12h 51m 00.128s & +27d 48m 04.576s & 21.04 & 28.12$\pm$0.14 & 27.37$\pm$0.12 & 0.68$\pm$0.17 \\ 
\hline 
\\\\\\
\hline


Host Galaxy & RA & DEC & R & $m_{606}$ & $m_{814}$ & $c_{4-8}$ \\ - & hms & dms & arcsec & AB mag & AB mag & AB mag \\ 
\hline
DFX1 & 13h 01m 15.894s & +27d 12m 35.144s & 2.33 & 26.04$\pm$0.02 & 25.65$\pm$0.03 & 0.38$\pm$0.03 \\ 
DFX1 & 13h 01m 15.569s & +27d 12m 39.320s & 4.16 & 26.73$\pm$0.03 & 26.39$\pm$0.06 & 0.40$\pm$0.05 \\ 
DFX1 & 13h 01m 15.702s & +27d 12m 37.737s & 1.63 & 26.85$\pm$0.03 & 26.41$\pm$0.06 & 0.50$\pm$0.05 \\ 
DFX1 & 13h 01m 16.090s & +27d 12m 36.765s & 4.36 & 26.90$\pm$0.03 & 26.48$\pm$0.06 & 0.46$\pm$0.05 \\ 
DFX1 & 13h 01m 15.731s & +27d 12m 35.464s & 1.85 & 26.86$\pm$0.03 & 26.57$\pm$0.07 & 0.48$\pm$0.05 \\ 
DFX1 & 13h 01m 16.206s & +27d 12m 49.881s & 14.25 & 27.12$\pm$0.04 & 26.64$\pm$0.07 & 0.35$\pm$0.07 \\ 
DFX1 & 13h 01m 15.890s & +27d 12m 37.688s & 1.51 & 27.12$\pm$0.04 & 26.65$\pm$0.07 & 0.52$\pm$0.06 \\ 
DFX1 & 13h 01m 16.010s & +27d 12m 32.522s & 5.47 & 27.41$\pm$0.05 & 26.75$\pm$0.08 & 0.41$\pm$0.08 \\ 
DFX1 & 13h 01m 15.775s & +27d 12m 31.074s & 5.93 & 27.03$\pm$0.04 & 26.76$\pm$0.08 & 0.36$\pm$0.06 \\ 
DFX1 & 13h 01m 16.374s & +27d 12m 47.035s & 13.22 & 27.29$\pm$0.05 & 26.78$\pm$0.08 & 0.43$\pm$0.07 \\ 
DFX1 & 13h 01m 15.843s & +27d 12m 38.150s & 1.32 & 27.13$\pm$0.04 & 26.81$\pm$0.09 & 0.48$\pm$0.06 \\ 
DFX1 & 13h 01m 15.823s & +27d 12m 47.546s & 10.55 & 27.21$\pm$0.04 & 26.98$\pm$0.10 & 0.43$\pm$0.07 \\ 
DFX1 & 13h 01m 15.864s & +27d 12m 32.869s & 4.24 & 27.49$\pm$0.05 & 27.01$\pm$0.10 & 0.44$\pm$0.09 \\ 
DFX1 & 13h 01m 15.461s & +27d 12m 37.514s & 5.10 & 27.37$\pm$0.05 & 27.06$\pm$0.11 & 0.41$\pm$0.08 \\ 
DFX1 & 13h 01m 16.154s & +27d 12m 40.703s & 6.47 & 27.61$\pm$0.06 & 27.06$\pm$0.10 & 0.40$\pm$0.10 \\ 
DFX1 & 13h 01m 15.623s & +27d 12m 27.205s & 10.14 & 27.87$\pm$0.07 & 27.27$\pm$0.12 & 0.71$\pm$0.10 \\ 
DFX1 & 13h 01m 15.751s & +27d 12m 41.295s & 4.35 & 27.84$\pm$0.07 & 27.29$\pm$0.13 & 0.37$\pm$0.12 \\ 
DFX1 & 13h 01m 15.403s & +27d 12m 37.138s & 5.94 & 27.97$\pm$0.08 & 27.41$\pm$0.14 & 0.41$\pm$0.13 \\ 
DFX1 & 13h 01m 15.313s & +27d 12m 40.390s & 8.04 & 27.75$\pm$0.07 & 27.41$\pm$0.14 & 0.48$\pm$0.11 \\ 
\hline 
\end{longtable}
\clearpage
\twocolumn


\label{lastpage}
\end{document}